\documentclass[review,times,authoryear]{elsarticle}

\usepackage{wrapfig}
\usepackage[font=small]{caption}
\usepackage[hyperref,usenames,svgnames,dvipsnam]{xcolor}
\usepackage[labelformat=empty]{subcaption}
\usepackage{amsmath}
\usepackage{mathtools}
\usepackage{bm}
\usepackage{breqn}
\usepackage{geometry}
\usepackage{booktabs}
\usepackage{multirow}
\usepackage{xfrac}
\usepackage{units}
\usepackage{tabularx}
\usepackage{dblfloatfix}

\usepackage{array}
\newcolumntype{P}[1]{>{\centering\arraybackslash}p{#1}}
\newcolumntype{M}[1]{>{\centering\arraybackslash}m{#1}}

\usepackage{siunitx}
\usepackage[colorlinks]{hyperref}

\biboptions{authoryear}

\journal{Icarus}

\begin{document}

\hypersetup{
     allcolors = MidnightBlue
}

\urlstyle{same}

\begin{frontmatter}

\title{Ocean tidal heating in icy satellites with solid shells}

\author[label1]{Isamu Matsuyama\corref{cor1}}
\cortext[cor1]{Corresponding author}
\ead{isa@lpl.arizona.edu}

\author[label2]{Mikael Beuthe}
\author[label1]{Hamish C. F. C. Hay}
\author[label3]{Francis Nimmo}
\author[label4]{Shunichi Kamata}

\address[label1]{Lunar and Planetary Laboratory, University of Arizona, Tucson, AZ 85719, USA}
\address[label2]{Royal Observatory of Belgium, Brussels, Belgium}
\address[label3]{Department of Earth and Planetary Sciences, University of California, Santa Cruz, CA, USA}
\address[label4]{Creative Research Institution, Hokkaido University, Sapporo, Japan}



\begin{abstract}

As a long-term energy source, tidal heating in subsurface oceans of
icy satellites can influence their thermal, rotational, and orbital
evolution, and the sustainability of oceans. We present a new theoretical
treatment for tidal heating in thin subsurface oceans with overlying
incompressible elastic shells of arbitrary thickness. The stabilizing
effect of an overlying shell damps ocean tides, reducing tidal heating.
This effect is more pronounced on Enceladus than on Europa because
the effective rigidity on a small body like Enceladus is larger. For
the range of likely shell and ocean thicknesses of Enceladus and Europa,
the thin shell approximation of \citet{Beuthe:2016a} is generally
accurate to less than about 4\%. Explaining Enceladus' endogenic power
radiated from the south polar terrain by ocean tidal heating requires
ocean and shell thicknesses that are significantly smaller than the
values inferred from gravity and topography constraints. The time-averaged
surface distribution of ocean tidal heating is distinct from that
due to dissipation in the solid shell, with higher dissipation near
the equator and poles for eccentricity and obliquity forcing respectively.
This can lead to unique horizontal shell thickness variations if the
shell is conductive. The surface displacement driven by eccentricity
and obliquity forcing can have a phase lag relative to the forcing
tidal potential due to the delayed ocean response. For Europa and
Enceladus, eccentricity forcing generally produces greater tidal amplitudes
due to the large eccentricity values relative to the obliquity values.
Despite the small obliquity values, obliquity forcing generally produces
larger phase lags due to the generation of Rossby-Haurwitz waves.
If Europa's shell and ocean are respectively 10 and 100 km thick,
the tide amplitude and phase lag are 26.5 m and $<1$ degree for eccentricity
forcing, and $<2.5$ m and $<18$ degrees for obliquity forcing. Measurement
of the obliquity phase lag (e.g. by Europa Clipper) would provide
a probe of ocean thickness

\end{abstract}
\end{frontmatter}

\section{Introduction\label{subsec:Introduction}}

A variety of observations suggest the presence of subsurface global
oceans in icy satellites of the outer solar system, significantly
increasing their appeal as habitable worlds. Jupiter's magnetic field
tilt relative to the rotation axis generates a time-varying magnetic
field, and this can in turn generate an induced magnetic field on
the Galilean satellites if they contain a sufficiently conducting
material. Induced magnetic fields have been detected on Europa, Ganymede,
and Callisto, and the conducting material has been interpreted as
a subsurface, salty, ocean \citep{Kivelson:2000,Zimmer:2000,Kivelson:2002}.
Ganymede's internal, permanent, magnetic field \citep{Kivelson:2002}
and the generation of induced magnetic fields by Callisto's ionosphere
\citep{Hartkorn:2017} complicates the interpretation of a subsurface
ocean. Recent Hubble Space Telescope observations of Ganymede's aurora
\citep{Saur:2015}, which is sensitive to induced magnetic fields,
favor the presence of a subsurface ocean. Saturnian satellites do
not experience a strong time-varying magnetic field due to the close
alignment of the magnetic field with the rotation axis, limiting the
generation of induced magnetic fields. Gravity data obtained by the
Cassini mission has provided strong evidence for subsurface oceans
on Titan and Enceladus. These data can be used to quantify the satellite
deformation in response to the time-varying forcing tidal potential
using the degree-two tidal Love number, $k_{2}^{T}$. For Titan, the
large $k_{2}^{T}\sim0.6$ requires a highly deformable interior over
the tidal forcing period, suggesting the presence of a global subsurface
ocean \citep{Iess:2012fj}. The large libration amplitude of Enceladus
requires decoupling the outer shell from the interior, implying the
presence of a global subsurface ocean \citep{Thomas:2016ex}. We refer
the reader to \citet{Nimmo:2016} for a comprehensive review of observational
constraints for subsurface oceans in icy satellites. 

As a long-term energy source, ocean tidal heating can influence the
thermal, rotational, and orbital evolution of icy satellites, and
the sustainability of subsurface oceans. The thermal, rotational,
and orbital evolution are coupled because tidal heating is a main
energy source that depends on the rotational and orbital parameters.
Previous studies have investigated various aspects of this coupled
problem \citep[e.g.][]{Ojakangas:1989a,Ross:1990,Sohl:1995,Showman:1997,Hussmann:2004,Behounkova:2012}.
However, these studies ignore ocean tidal heating and assume that
energy dissipation occurs only in solid regions. This assumption is
not justified on Earth, where most of the tidal heating is generated
in the oceans. As the observational evidence for subsurface oceans
in icy satellites increased, more recent studies have considered ocean
tidal heating \citep{Tyler:2008,Tyler:2009,Tyler:2011,Chen:2011,Tyler:2014gf,Matsuyama:2014,Chen:2014,Hay:2017}.
However, these studies ignore the presence of an overlying solid shell,
with the exception of \citet{Beuthe:2016a} who models this effect
using a thin shell approximation.

In this paper, we present a new theoretical treatment for tidal heating
in thin subsurface oceans that takes into account the effect of an
overlying shell of arbitrary thickness. As shown below with applications
to Enceladus and Europa, this is important because an overlying shell
damps ocean tides, reducing tidal heating. Additionally, ocean tidal
heating can generate distinct horizontal shell thickness variations
and surface displacement phase lags relative to the forcing tidal
potential, which has implications for spacecraft observations.

As discussed above, Enceladus' large libration amplitude requires
the presence a global subsurface ocean, and is consistent with an
ocean $26-31$ km thick and a solid shell $21-26$ km thick \citep{Thomas:2016ex}.
Bayesian inversion using gravity and topography data constrains the
ocean and shell thicknesses to $38\pm4$ km and $23\pm4$ km respectively
\citep{Beuthe:2016}. We assume a spherically symmetric three-layer
interior structure with the parameters in Table \ref{tab:Interior parameters}
and consider a range of shell and ocean thicknesses that are consistent
with these observational constraints.

The ocean and shell thicknesses of Europa are only weakly constrained.
A combined ocean and shell thickness in the range of about 80 to 170
km was estimated from the mean moment of inertia inferred from gravity
data, $I/(MR^{2})=0.346\pm0.05$, where $M$ and $R$ are the satellite
mass and radius \citep{Anderson:1998}. The mean moment of inertia
uncertainty is likely larger due to the implicit assumption of a hydrostatic
ratio $J_{2}/C_{22}=10/3$ for the degree-two gravity coefficients,
and the error associated with the Radau-Darwin approximation used
to obtain the mean moment of inertia from the degree-two gravity coefficients
\citep{Gao:2013}. We assume the simplest possible interior structure
with a subsurface ocean and shell, a spherically symmetric three-layer
interior structure model with the parameters in Table \ref{tab:Interior parameters},
and consider a large range of shell and ocean thicknesses.

\begin{table}[]
\begin{centering}
\resizebox{!}{!}{ 
\begin{tabular}{lllllll}
\toprule 
 & & & & \textbf{Enceladus} &  & \textbf{Europa}
\tabularnewline
\midrule 
Parameter & & Symbol & & Value & & 
\tabularnewline
\cmidrule{1-1} \cmidrule{3-3} \cmidrule{5-5} \cmidrule{7-7} 
Mass & &  $M$  &  & $1.08\times10^{20}$ kg  &  &  $4.8\times10^{22}$ kg
\tabularnewline
Radius & & $R$ &  & 252.1 km &  & 1561 km 
\tabularnewline
Rotation rate & & $\Omega$&  & $5.31\times10^{-5}$ rad s$^{-1}$  &   &  $2.05\times10^{-5}$ rad s$^{-1}$ 
\tabularnewline
Orbital eccentricity &  & $e$ &  & 0.0047 &  & 0.0094 
\tabularnewline
 Predicted obliquity &  & $\theta_{0}$ &  & $4.5\times10^{-4}$ degree &  & 0.1 degree 
\tabularnewline
 Core shear modulus  &  &  $\mu_{c}$  &  & $40\times10^{9}$ Pa  &  & $40\times10^{9}$ Pa 
\tabularnewline
 Ocean density &  & $\rho_{o}$  &  & $10^{3}$ kg m$^{-3}$  &  & $10^{3}$ kg m$^{-3}$ 
\tabularnewline
Ocean thickness &  &  $h_{o}$  &  & 38 km  &  & 100 km 
\tabularnewline
 
Shell density 
 &  
 
 &  
$\rho_{s}$ 
 &  
 
 &  
940 kg m$^{-3}$ 
 &  
 
 &  
940 kg m$^{-3}$ 
\tabularnewline
 
Shell thickness 
 &  
 
 &  
$h_{s}$ 
 &  
 
 &  
23 km 
 &  
 
 &  
10 km 
\tabularnewline
 
Shell shear modulus 
 &  
 
 &  
$\mu_{s}$ 
 &  
 
 &  
$3.5\times10^{9}$ Pa  
 &  
 
 &  
$3.5\times10^{9}$ Pa  
\tabularnewline
 
Linear drag coefficient 
 &  
 
 &  
$\alpha$ 
 &  
 
 &  
$10^{-11}-10^{-5}$ s$^{-1}$ 
 &  
 
 &  
$10^{-11}-10^{-5}$ s$^{-1}$ 
\tabularnewline
\bottomrule
\end{tabular} }
\par\end{centering}
 
\caption{\label{tab:Interior parameters}Model parameters for Enceladus and
Europa. Given the ocean and shell thicknesses, we calculate the core
density self-consistently so as to satisfy the mean density constraint. }
\end{table}
\begin{figure}[h]
\begin{centering}
\includegraphics[width=10cm]{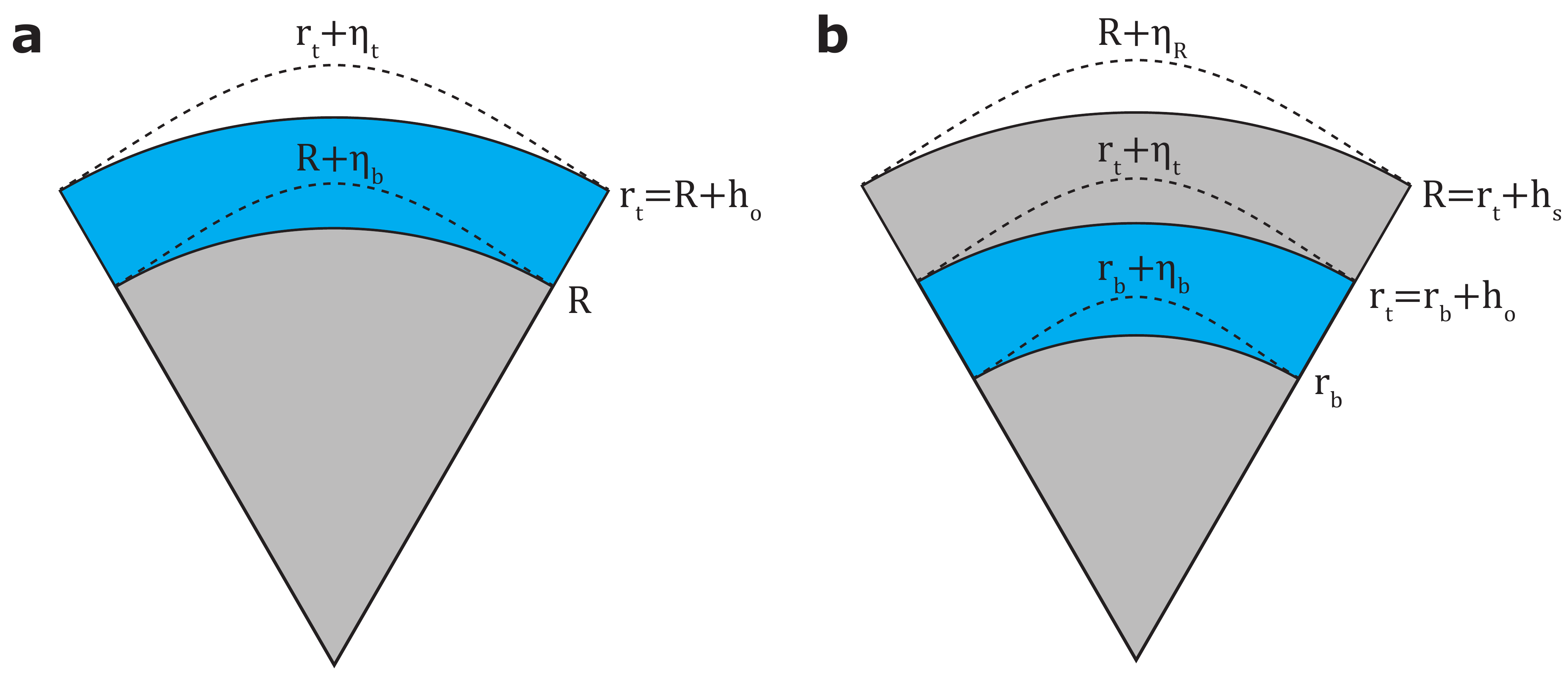}
\par\end{centering}
\caption{\label{fig:illustration}Comparison of radial displacements for (a)
surface and (b) subsurface oceans. In both cases, the ocean thickness
is $h_{o}$, and the radial tide $\eta$ is defined as the difference
between the radial displacement at the top and bottom of the ocean,
$\eta\equiv\eta_{t}-\eta_{b}$. For a surface ocean, the ocean bottom
radius is the surface radius $R$ and the ocean top is $R+h_{o}$.
For a subsurface ocean, the ocean top and bottom radii are related
by $r_{t}=r_{b}+h_{o}$, the surface and ocean top radii are related
by $R=r_{t}+h_{s}$, where $h_{s}$ is the overlying shell thickness,
and the radial tide at the surface is $\eta_{R}$. }
\end{figure}

\section{Theory\label{sec:Theory}}

The Laplace tidal equations (LTE) describing dynamic ocean tides can
be obtained from the momentum and mass conservation equations by assuming
a thin, homogeneous surface ocean on a spherical planet \citep{Lamb:1993}.
For a surface or subsurface ocean, these equations can be written
as
\begin{equation}
\partial_{t}\eta+h_{o}\nabla\cdot\mathbf{u}=0\label{eq:mass cons.}
\end{equation}
and 
\begin{equation}
\partial_{t}\mathbf{u}+2\mathbf{\mathbf{\Omega}}\times\mathbf{u}+\alpha\mathbf{u}+\frac{c_{D}}{h_{o}}|\mathbf{u}|\mathbf{u}+\nu\nabla^{2}\mathbf{u}=-\frac{1}{\rho_{o}}\nabla P+\nabla U;\label{eq:momentum cons.}
\end{equation}
where $h_{o}$ is a reference, uniform ocean thickness; $\mathbf{u}=(u_{\theta},\,u_{\phi})$
is the depth averaged horizontal velocity vector in spherical coordinates
$\theta$ and $\phi$ (colatitude and longitude); $\alpha$, $c_{D}$,
and $\nu$ are linear, bottom, and Navier-Stokes drag coefficients;
and $\nabla$ is a horizontal gradient operator. That is,
\begin{align}
\nabla\cdot\mathbf{u} & =(r\sin\theta)^{-1}\left[\partial_{\theta}(\sin\theta u_{\theta})+\partial_{\phi}u_{\phi}\right]\nonumber \\
\nabla U & =r^{-1}\hat{\mathbf{e}}_{\theta}\partial_{\theta}U+\hat{\mathbf{e}}_{\phi}(r\sin\theta)^{-1}\partial_{\phi}U,\label{eq:nabla horizontal}
\end{align}
where $\mathbf{\hat{\mathbf{e}}_{\theta}}$ and $\mathbf{\hat{\mathbf{e}}_{\phi}}$
are unit vectors in the $\theta$ and $\phi$ directions \citep[e.g., ][Eqs. A.117, A.118 and A.141]{Dahlen:1998}.
In the mass conservation equation (\ref{eq:mass cons.}), the radial
tide $\eta$ is defined as the difference between the radial displacement
at the top and bottom of the ocean (Fig. \ref{fig:illustration}),
\begin{equation}
\eta\equiv\eta_{t}-\eta_{b}\equiv\sum_{n=m}^{\infty}\sum_{m=0}^{2}\eta_{nm}Y_{nm}(\theta,\,\phi)e^{-i\omega t},\label{eq:radial tide}
\end{equation}
which we expand in spherical harmonics $Y_{nm}(\theta,\,\phi)\equiv P_{nm}(\theta)e^{im\phi}$
(\ref{sec:Numerical-solution}). The expansion coefficients
$\eta_{nm}$ are imaginary and $P_{nm}$ is the associated Legendre
function \citep{Arfken:1995}. As described below, the forcing tidal
potential is decomposed into eastward ($\omega=\Omega$) and westward
($\omega=-\Omega$) traveling components. In the momentum conservation
equation (\ref{eq:momentum cons.}), $\mathbf{\Omega}$ is the rotation
vector, $\rho_{o}$ is the ocean density, $P$ is the radial pressure,
and $U$ is the forcing potential. 

The time-dependent part of the tidal forcing potential contains eccentricity
and obliquity contributions, $U^{T}=U_{ecc}^{T}+U_{obliq}^{T}$, where 

\begin{align}
U_{ecc}^{T} & =\Omega^{2}r^{2}e\left\{ -\frac{3}{2}P_{20}(\cos\theta)\cos(\Omega t)+\frac{1}{8}P_{22}(\cos\theta)\left[7\cos(2\phi-\Omega t)-\cos(2\phi+\Omega t)\right]\right\} \nonumber \\
U_{obliq}^{T} & =\frac{1}{2}\Omega^{2}r^{2}\theta_{0}P_{21}(\cos\theta)\left[\cos(\phi-\Omega t)+\cos(\phi+\Omega t)\right]\label{eq:U_ecc_obliq}
\end{align}
to lowest order in eccentricity, $e$, and obliquity, $\theta_{0}$
\citep{Tyler:2011}. Eccentricity forcing causes the total tidal bulge
(static part included) to librate in longitude and to vary in amplitude,
and obliquity forcing causes the tidal bulge to librate in latitude,
producing time-varying ocean tides. We only consider $n=2$ contributions
and ignore higher order terms because the forcing tidal potentials
scale with $(r/a)^{n}$ and $r\ll a$, where $a$ is the semi-major
axis of the satellite. The obliquities of Enceladus and Europa are
not directly constrained by observations. We assume obliquities of
$4.5\times10^{-4}$ deg and 0.1 deg for Enceladus and Europa respectively
under the assumption that tidal energy dissipation has driven the
obliquities to Cassini state values \citep{Bills:2005,Chen:2011,Baland:2016}.

We extend the method of \citet{Longuet-Higgins:1968} to solve the
mass and momentum conservation equations for a thin ocean with an
overlying incompressible elastic shell of arbitrary thickness (\ref{sec:Numerical-solution}). This requires ignoring bottom ($c_{D}=0$)
and Navier-Stokes ($\nu=0$) drag, and in this case the dissipated
energy per unit surface area and time is 

\begin{equation}
F_{diss}=-\rho_{o}h_{o}\alpha\text{\textbf{u}\ensuremath{\cdot\textbf{u}}}.\label{eq:dissipated flux}
\end{equation}
Earth's ocean tidal heating studies consider both linear and bottom
friction drag formalisms. \citet{Egbert:2000jf} and \citet{Egbert:2001}
assume linear drag with $\alpha^{\prime}=\alpha h_{o}\sim0.03$ m
s$^{-1}$, which yields $\alpha\sim10^{-5}$ s$^{-1}$ assuming an
average Earth ocean thickness $h_{o}\sim4$ km. \citet{Webb:1980}
assume $\alpha\sim1/\tau$ with $\tau\sim24-60$ hours, which also
yields $\alpha\sim10^{-5}$ s$^{-1}$. The bottom drag formalism is
based on the assumption that drag arises due to turbulent flow interacting
with a bottom boundary. A nominal bottom drag coefficient $\sim(2-3)\times10^{-3}$
has been assumed to model tidal heating in the Earth \citep[e.g., ][]{Lambeck:1980earth,Jayne:2001,Egbert:2001,Green:2013},
Titan \citep{Sagan:1982}, and Jovian planets \citep{Goldreich:1966}.
We can estimate a corresponding linear drag coefficient by comparing
the energy flux due to linear drag (Eq. (\ref{eq:dissipated flux}))
with the energy flux due to bottom drag, $F_{diss}=\rho_{o}c_{D}(\text{\textbf{u}\ensuremath{\cdot\textbf{u }}})^{3/2}$.
This order-of-magnitude estimate yields
\begin{equation}
c_{D}\sim\alpha^{3/2}\left(\frac{\rho_{o}h_{o}^{3}}{F_{diss}}\right)^{1/2}.\label{eq:cd_alpha_scaling}
\end{equation}
There are no constraints for the linear or bottom drag coefficients
in icy satellites; therefore, we consider a large range of possible
values. Using the tidal heating fluxes in section \ref{sec:General-results}
and the nominal value $c_{D}\sim10^{-3}$, this estimates yields $\alpha\sim10^{-11}$
s$^{-1}$ for eccentricity and obliquity forcing on Enceladus, $\alpha\sim10^{-10}$
s$^{-1}$ for eccentricity forcing on Europa, and $\alpha\sim10^{-9}$
s$^{-1}$ for obliquity forcing on Europa. We adopt these values as
lower limits and the Earth value, $\alpha\sim10^{-5}$ s$^{-1}$,
as an upper limit. 

Tidal heating in the solid regions of a satellite is commonly quantified
by a tidal quality factor defined as $Q\equiv2\pi E_{max}/E_{diss},$
where $E_{max}$ is the maximum energy stored in the tidal deformation
and $E_{diss}$ is the energy dissipated in one cycle. For dissipation
in solid regions, it is possible to calculate the elastic energy stored
due to tidal deformation and the corresponding $Q$. For ocean energy
dissipation, however, tidal deformation does not produce elastic energy.
It is possible to introduce a tidal quality factor $Q\equiv\Omega/(2\alpha)$
for ocean tidal heating by redefining $E_{max}$ as the maximum kinetic
energy of the ocean \citep{Tyler:2011}. This definition has been
used in previous ocean tidal heating studies \citep{Tyler:2011,Matsuyama:2014,Chen:2014,Beuthe:2016a}.
However, it can lead to counter-intuitive results such as decreasing
energy dissipation with decreasing $Q$ (\ref{sec:Tidal-quality-factor})
because the kinetic and dissipated energies are coupled \citep{Matsuyama:2014}.
Although we can introduce an alternative definition of $Q$ that is
physically more intuitive (\ref{sec:Tidal-quality-factor}),
we do not favor its use because $Q$ is not a fundamental quantity,
but a phenomenological factor whose definition depends on the particular
context. This can introduce significant errors even when considering
solid tides. For example, the neglect of self-gravity and hydrostatic
pre-stress in the traditional relationship between $Q$ and the tidal
phase delay can lead to order of magnitude errors \citep{Zschau:1978jp}.
The relevant quantity for computing the effect of tidal heating on
the thermal, rotational, and orbital evolution is the energy dissipation
rate and our thick shell theory provides a method for computing it. 

The mass and momentum conservation equations (\ref{eq:mass cons.})
and (\ref{eq:momentum cons.}), and the energy dissipation equation
(\ref{eq:dissipated flux}) are applicable to both surface and subsurface
oceans; however, the pressure and forcing potential terms in the momentum
conservation equation are different for each case, as described in
sections \ref{subsec:Surface-oceans} and \ref{subsec:Subsurface-oceans}
below.

\subsection{Surface oceans\label{subsec:Surface-oceans}}

For surface oceans, the pressure at a reference radius $r$ in the
ocean is $P=\rho_{o}g(r_{t}+\eta_{t}-r),$ where $r_{t}\equiv R+h_{o}$
is the constant ocean top radius (Fig. \ref{fig:illustration}a) and
we assume a constant gravitational acceleration $g$ in the ocean,
as expected for a thin ocean. Thus, the pressure gradient term in
the momentum conservation equation (\ref{eq:momentum cons.}) is 
\begin{equation}
-\frac{\nabla P}{\rho_{0}}=-g\nabla\eta_{t},\label{surface ocean: pressure gradient}
\end{equation}
where we ignore density and gravitational acceleration variations.
The former is justified by the assumption of an incompressible ocean,
and the latter is justified by the assumption of small amplitude tides
($\eta\ll R$). 

The total forcing tidal potential is given by 
\begin{equation}
U_{nm}=\left[1+k_{n}^{T}(R)\right]U_{nm}^{T}+\left[1+k_{n}^{L}(R)\right]U_{nm}^{L},\label{surface ocean:nabla U}
\end{equation}
where we expand the potential $U$ in spherical harmonics $U_{nm}$
and take into account the effects of ocean self-gravity and deformation
of the solid regions using Love number theory \citep{Hendershott:1972,Matsuyama:2014}.
The tidal Love number $k_{n}^{T}$ describes the response to the forcing
tidal potential $U_{nm}^{T}$, and the load Love number $k_{n}^{L}$
describes the response to the ocean loading potential $U_{nm}^{L}$.
We use the term ocean loading potential to refer to the gravitational
potential arising from self-gravity of the tides (\citealp[p. 305, Eq. (13)]{Lamb:1993};
\citealp{Hendershott:1972}; \citealp{Lambeck:1980earth}; \citealp{Matsuyama:2014}).
The Love numbers must be evaluated at the solid surface ($r=R)$ for
a surface ocean (Fig. \ref{fig:illustration}a) . The ocean loading
potential is related to the radial tide by 
\begin{equation}
U_{nm}^{L}=\frac{3}{(2n+1)}\frac{\rho_{o}}{\bar{\rho}}g\eta_{nm}=g\xi_{n}\eta_{nm},\label{surface ocean:U ocean}
\end{equation}
where $\bar{\rho}$ is the mean density of the satellite and we define
the degree-\emph{n} density ratio
\begin{equation}
\xi_{n}\equiv\frac{3}{(2n+1)}\frac{\rho_{o}}{\bar{\rho}}.\label{surface ocean: xi definition}
\end{equation}

The ocean bottom displacement (Fig. \ref{fig:illustration}a) is given
by
\[
\eta_{nm}^{b}=h_{n}^{T}(R)U_{nm}^{T}/g(R)+h_{n}^{L}(R)U_{nm}^{L}/g(R),
\]
where the displacement Love numbers $h_{n}^{T}$ and $h_{n}^{L}$
must be evaluated at the surface. Combining Eqs. (\ref{eq:radial tide})-(\ref{surface ocean: xi definition})
to eliminate the ocean loading potential yields
\begin{equation}
\partial_{t}\mathbf{u}+2\mathbf{\mathbf{\Omega}}\times\mathbf{u}+\alpha\mathbf{u}=-g(R)\nabla\sum_{n=m}^{\infty}\sum_{m=0}^{2}(1-\xi_{n}\gamma_{n}^{L})\eta_{nm}Y_{nm}(\theta,\,\phi)+\gamma_{2}^{T}\nabla\sum_{m=0}^{2}U_{2m}^{T}Y_{2m}(\theta,\,\phi),\label{surface ocean: momentum}
\end{equation}
where the tilt factors are
\begin{align}
\gamma_{n}^{L} & \equiv1+k_{n}^{L}(R)-h_{n}^{L}(R)\nonumber \\
\gamma_{2}^{T} & \equiv1+k_{2}^{T}(R)-h_{2}^{T}(R).\label{surface ocean: tilt factors}
\end{align}

The Love numbers can be computed using the propagator matrix method
(\ref{sec:Love numbers liquid layers}) or the analytic expressions
for a homogeneous, incompressible body (\ref{sec:Homogeneous interior Love numbers})
for the case of a uniform interior beneath the ocean.

\subsection{Subsurface oceans\label{subsec:Subsurface-oceans}}

There are two main aspects that must be taken into account when considering
ocean tidal heating in subsurface oceans. First, energy dissipation
due to drag can occur at both the top and bottom of the ocean. Second,
the overlying solid shell provides an additional source of pressure
that resists ocean tides. In the absence of information about drag
at the ocean top and bottom, the first effect can be taken into account
by simply assuming a drag coefficient that is twice as large compared
with the value assumed for a surface ocean \citep{Tyler:2014gf}.
The second effect requires a significant extension of the theory,
as described below.

For a thin subsurface ocean, the total forcing potential in the momentum
conservation equation (\ref{eq:momentum cons.}) can be written as
\begin{equation}
U_{nm}=\left[(r/R)^{n}+k_{n}^{T}(r)\right]U_{nm}^{T}(R)+k_{n}^{P}(r)U_{nm}^{P}(r),\label{sub ocean: Unm}
\end{equation}
where $k_{n}^{T}$ and $k_{n}^{P}$ are internal, degree-$n$, tidal
and pressure Love numbers \citep{Hinderer:1986ik}, $U_{nm}^{T}$
are the expansion coefficients of the tidal potential, and $U_{nm}^{P}(r_{t})$
are the expansion coefficients of a pressure potential representing
the dynamic part of ocean forcing on the shell. The first term in
Eq. (\ref{sub ocean: Unm}), $(r/R)^{n}U_{nm}^{T}(R)$, corresponds
to the forcing tidal potential, and the terms $k_{n}^{T}(r)U_{nm}^{T}(R)$
and $k_{n}^{P}(r)U^{P}(r_{t})$ describe the gravitational potentials
arising from the static and dynamic parts of the deformation in response
to the forcing potential. This decomposition into static and dynamic
components was first used to study the effect of fluid core dynamics
on Earth nutations (\citealp{Sasao:1980iv}, Eq. (51); \citealp{Sasao:1981},
Eq. (3.10)), and later adopted to study Earth surface gravity perturbations
due to fluid core oscillations using the terminology of internal pressure
Love numbers (\citealp{Hinderer:1986ik}, Eq. (2); \citealp{Hinderer:1989jx},
Eq. (2.9)). 

The dynamic ocean pressure generates radial stress discontinuities
of magnitude $\rho_{o}U_{nm}^{P}(r_{t})$ and $\rho_{o}U_{nm}^{P}(r_{b})$
at the ocean top and bottom, respectively (\ref{sec:Interior pressure Love numbers liquid},
Eqs. (\ref{app:ocean_pressure_pert_top}) and (\ref{app:ocean_pressure_pert_bot})).
In the limit of a thin ocean, $U_{nm}^{P}(r_{b})=-U_{nm}^{P}(r_{t})$,
which allows us to describe pressure forcing at both the ocean top
and bottom using a single Love number instead of the traditional approach
of describing each forcing with a different Love number, and we choose
$U_{nm}^{P}(r_{t})$ as the reference pressure potential (Appendix
\ref{sec:Interior pressure Love numbers liquid}). Under the same
thin ocean approximation, Eq. (\ref{sub ocean: Unm}) can be evaluated
at the ocean top ($r=r_{t}$) or bottom ($r=r_{b}$), and we choose
$r_{t}$ as the reference ocean radius. 

In our formalism using interior Love numbers, the gravitational potential
arising from the mass redistribution due to the static and dynamic
ocean tide (self-gravity) is taken into account by the $k_{n}^{T}(r)U_{nm}^{T}(R)$
and $k_{n}^{P}(r)U_{nm}^{P}(r_{t})$ terms, respectively, in Eq. (\ref{sub ocean: Unm}).
Therefore, an additional ocean loading term is not required in the
momentum conservation equation. In contrast, the subsurface
ocean treatment of \citet{Beuthe:2016a} using a thin shell approximation
includes an ocean loading term describing ocean self-gravity in the
momentum conservation equation. Despite this apparent difference in
the theoretical treatment, our thick shell solutions converge to the
thin shell solutions as the overlying shell thickness decreases, as
shown below. 

Using the tidal and pressure displacement Love numbers $h_{n}^{T}$
and $h_{n}^{P}$, the radial tides (Fig. \ref{fig:illustration}b)
at the surface ($r=R$), ocean top ($r=r_{t}$), and ocean bottom
($r=r_{b}$) are given by 
\begin{align}
\eta_{nm}^{R} & =h_{n}^{T}(R)U_{nm}^{T}(R)/g(R)+h_{n}^{P}(R)U_{nm}^{P}(r_{t})/g(R)\nonumber \\
\eta_{nm}^{t} & =h_{n}^{T}(r_{t})U_{nm}^{T}(R)/g(R)+h_{n}^{P}(r_{t})U_{nm}^{P}(r_{t})/g(R)\nonumber \\
\eta_{nm}^{b} & =h_{n}^{T}(r_{b})U_{nm}^{T}(R)/g(R)+h_{n}^{P}(r_{b})U_{nm}^{P}(r_{t})/g(R).\label{sub ocean: displacements}
\end{align}
Once again, we describe pressure forcing at the ocean top and bottom
using a single Love number and choose $r=r_{t}$ as the reference
ocean radius. The pressure potential can be written in terms of the
ocean tide $\eta_{nm}\equiv\eta_{nm}^{t}-\eta_{nm}^{b}$ and the forcing
tidal potential using Eq. (\ref{sub ocean: displacements}), 
\begin{equation}
U_{nm}^{P}(r_{t})=\frac{g(R)\eta_{nm}-\delta h_{n}^{T}U_{nm}^{T}(R)}{\delta h_{n}^{P}}\label{sub:Up_displacement}
\end{equation}
where we define $\delta h_{n}^{T}\equiv h_{n}^{T}(r_{t})-h_{n}^{T}(r_{b})$
and $\delta h_{n}^{P}\equiv h_{n}^{P}(r_{t})-h_{n}^{P}(r_{b})$. If
the pressure potential is zero, then the ocean tide is equal to the
equilibrium ocean tide ($\eta_{nm}=\delta h_{n}^{T}U_{nm}^{T}/g(R)$),
as expected. Although we do not use it in this paper, the approximation
$\eta_{nm}\sim\eta_{nm}^{t}$ could be used because the displacement
Love numbers at the ocean bottom are significantly smaller than those
at the ocean top due to mechanical decoupling (\ref{sec:Love-numbers}). 

For a thin subsurface ocean, the radial pressure at a reference radius
$r$ in the ocean can be written as

\begin{equation}
P=\sigma_{rr}^{T}+\sigma_{rr}^{P}+\rho_{o}g(r_{t})\eta_{t}+\rho_{o}g(r_{t})(r_{t}-r),\label{sub ocean:pressure}
\end{equation}
where $\sigma_{rr}^{T}$ and $\sigma_{rr}^{P}$ are the radial stresses
at the shell-ocean boundary due to tidal and dynamic pressure forcing
respectively, and $\rho_{o}$ is the ocean density (Fig. \ref{fig:illustration}b).
To obtain the compact form of the momentum conservation equation below,
it is useful to describe the radial stress due to tidal forcing in
terms of the difference between an ideal fluid equipotential displacement,
$\left[(r/R)^{n}+k_{n}^{T}(r)\right]U_{nm}^{T}(R)/g(r_{t})$, and
the actual displacement, $h_{2}^{T}(r_{t})U_{2m}^{^{T}}/g(R)$ (\citealp{Saito:1974},
Eq. (14); \citealp{Jara-Orue:2011}, Eq. (A.12); \citealp{Beuthe:2015a},
Eq. (25)) , 
\begin{equation}
\sigma_{rr}^{T}=\rho_{o}g(r_{t})\sum_{m=0}^{2}\sum_{n=m}^{\infty}\left[\frac{x^{n}+k_{n}^{T}(r_{t})}{g(r_{t})}-\frac{h_{n}^{T}(r_{t})}{g(R)}\right]U_{nm}^{T}(R)Y_{nm}(\theta,\,\phi).\label{sub:trr_tidal}
\end{equation}
Similarly, for the radial stress due to dynamic pressure
forcing,
\begin{equation}
\sigma_{rr}^{P}=\rho_{o}g(r_{t})\sum_{m=0}^{2}\sum_{n=m}^{\infty}\left[\frac{k_{n}^{P}(r_{t})}{g(r_{t})}-\frac{h_{n}^{P}(r_{t})}{g(R)}\right]U_{nm}^{P}(r_{t})Y_{nm}(\theta,\,\phi)+\rho_{o}\sum_{m=0}^{2}\sum_{n=m}^{\infty}U_{nm}^{P}(r_{t}),\label{sub_trr_press}
\end{equation}
where the last term describes the radial stress discontinuity at the
ocean top due to the dynamic pressure forcing. We adopt a sign convention
that implies that a positive pressure potential leads to a positive
radial displacement (\ref{sec:Interior pressure Love numbers liquid}).
Combining Eqs. (\ref{sub ocean: Unm}) and (\ref{sub ocean:pressure})-(\ref{sub_trr_press}),
the forcing term in the right-hand-side (RHS) of the momentum conservation
equation (\ref{eq:momentum cons.}) reduces to 
\begin{equation}
-\frac{1}{\rho_{o}}\nabla P_{nm}+\nabla U_{nm}=-\nabla U_{nm}^{P}(r_{t})=-\frac{1}{\rho_{o}}\nabla P_{nm}^{dyn}(r_{t}),\label{sub:dynamic_pressure}
\end{equation}
where the dynamic pressure $P_{nm}^{dyn}\equiv\rho_{o}U^{P}$. Finally,
we replace Eq. (\ref{sub:Up_displacement}) in this forcing term to
write the momentum conservation equation (\ref{eq:momentum cons.})
as

\begin{equation}
\partial_{t}\mathbf{u}+2\Omega\times\mathbf{u}+\alpha\mathbf{u}=-g(R)\nabla\sum_{m=0}^{2}\sum_{n=m}^{\infty}\beta_{n}\eta_{nm}Y_{nm}(\theta,\:\phi)+\upsilon_{2}\nabla\sum_{m=0}^{2}U_{2m}^{T}(R)Y_{2m}(\theta,\:\phi),\label{sub ocean: momentum eq.}
\end{equation}
where 
\begin{align}
\beta_{n}\equiv & \frac{1}{\delta h_{n}^{P}}\nonumber \\
\upsilon_{n}\equiv & \frac{\delta h_{n}^{T}}{\delta h_{n}^{P}}.\label{sub ocean: nu beta}
\end{align}
Comparison of Eqs. (\ref{sub ocean: momentum eq.}) and (\ref{surface ocean: momentum})
shows that $\beta_{n}$ and $\upsilon_{2}$ are the equivalent of
$1-\xi_{n}\gamma_{n}^{L}$ and $\gamma_{2}^{T}$ for a surface ocean. 

The dynamic deformation can also be described using Love numbers that
take into account dynamic terms in the momentum conservation equation.
This approach has been used to study global oscillations and the tides
of the Earth \citep[e.g.][Eq. (82)]{Takeuchi:1972kh}, and tidal resonances
in icy satellites with subsurface oceans \citep{Kamata:2015,Beuthe:2015a}.
However, this approach does not capture the complete ocean dynamics
because the Coriolis term in the momentum equation (\ref{sub ocean: momentum eq.})
breaks the assumption of spherical symmetry. As described above, we
use tidal and pressure Love numbers to describe the static and dynamic
parts of the deformation in response to tidal forcing. This allows
us to couple the non-spherically symmetric LTE with a thick shell
and mantle that are spherically symmetric. 

Before considering dynamic tides, it is useful to consider equilibrium
tides, $\eta^{eq}$, without dynamic pressure forcing ($\mathbf{u}=U_{nm}^{P}=0$).
The equilibrium tide can be written as 

\begin{equation}
\eta_{2m}^{eq}=Z_{2}\frac{U_{2m}^{T}}{g(R)},\label{surface oean:equil. tide}
\end{equation}
where $Z_{2}$ is a non-dimensional admittance. For surface oceans,
ignoring the LHS of Eq. (\ref{surface ocean: momentum}) associated
with the horizontal velocity, $Z_{2}\equiv\gamma_{2}^{T}/(1-\xi_{2}\gamma_{2}^{L}),$
where $\gamma_{2}^{T}$ and $\gamma_{2}^{L}$ are given by Eq. (\ref{surface ocean: tilt factors}).
Similarly, for subsurface oceans, ignoring the LHS of Eq. (\ref{sub ocean: momentum eq.}),
$Z_{2}=\upsilon_{2}/\beta_{2}$, where $\upsilon_{2}$ and $\beta_{2}$
are given by Eq. (\ref{sub ocean: nu beta}), and
\begin{equation}
\eta_{2m}^{eq}=\frac{\upsilon_{2}}{\beta_{2}}\frac{U_{2m}^{T}}{g(R)}=\delta h_{2}^{T}\frac{U_{2m}^{T}}{g(R)},\label{sub:eq_tide}
\end{equation}
as expected from the definition of the displacement Love numbers.
Fig. \ref{fig:beuthe_comparison} illustrates that the admittance
and corresponding equilibrium tide decrease with increasing shell
thickness, as expected. 

We use a propagator matrix method to compute the tidal and pressure
Love numbers (\ref{sec:Interior pressure Love numbers liquid}).
In this paper, we assume a conductive shell that is entirely elastic.
For a convecting shell, only the near-surface part is likely to be
elastic and this can be taken into account in the Love numbers calculation.
We solve the mass and momentum conservation equations (Eqs. (\ref{eq:mass cons.})
and (\ref{sub ocean: momentum eq.})) by extending the method of \citet{Longuet-Higgins:1968},
as described in \ref{sec:Numerical-solution}. 

As dynamic effects vanish in the limit of equilibrium tide, the LTE
must be coupled to deviations from the equilibrium tide. This is indeed
the case: deviations from the equilibrium tide are proportional to
the dynamic pressure potential $U_{nm}^{P}(r_{t})$ generating the
forcing (Eq. (\ref{sub:dynamic_pressure})), the transfer function
being the pressure Love number differential $\delta h_{n}^{P}$ (Eq.
(\ref{sub:Up_displacement})) normalized by $g(R)$.

\subsection{Comparison with thin shell approximation solutions\label{subsec:Comparison-with-the}}

\begin{figure}[h]
\begin{centering}
\includegraphics[width=14cm]{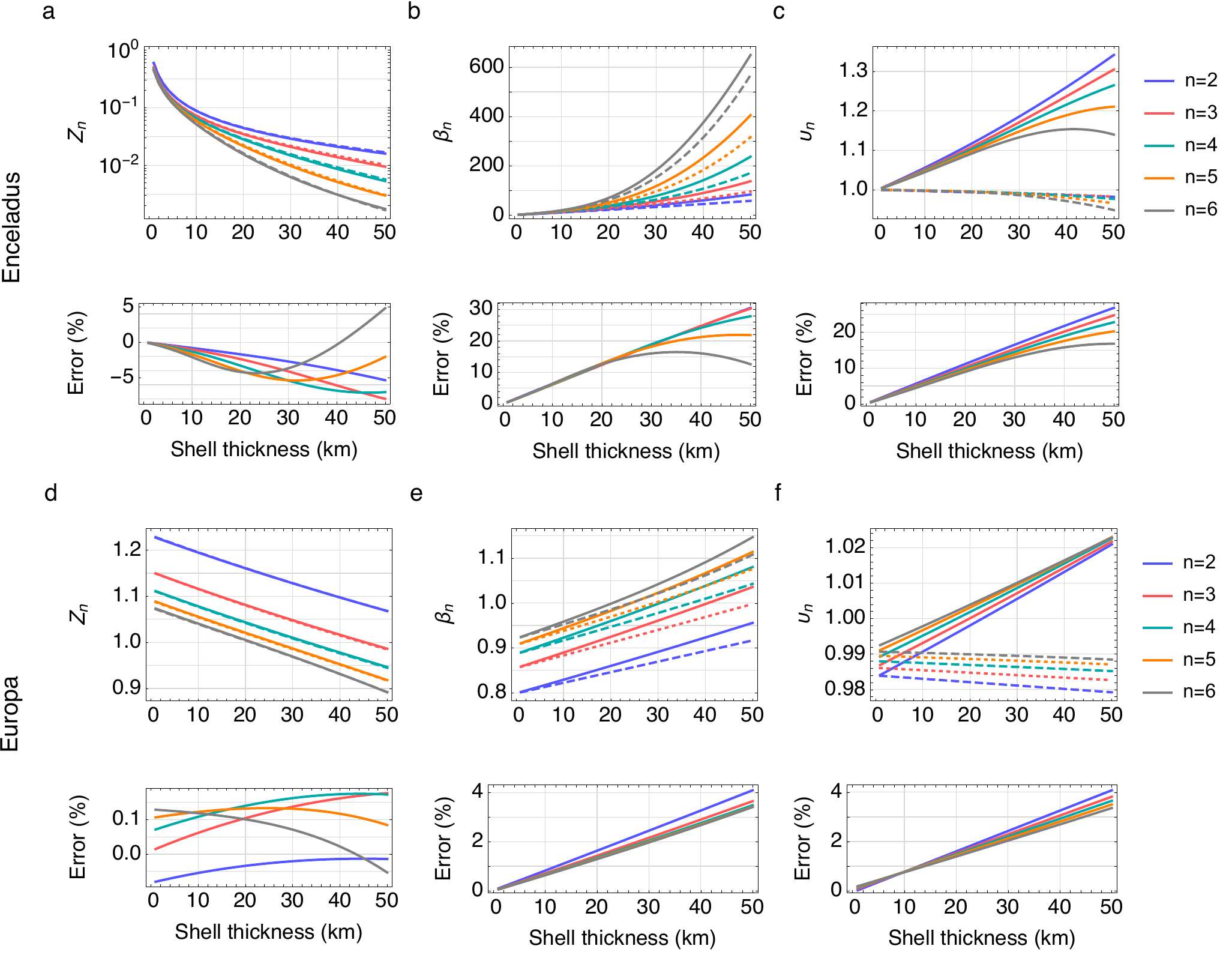}
\par\end{centering}
\caption{\label{fig:beuthe_comparison}Admittance, $Z_{n}$, and momentum conservation
equation terms $\beta_{n}$, and $\upsilon_{n}$ (Eq. (\ref{sub ocean: nu beta}))
as a function of shell thickness for different spherical harmonic
degrees, $n$. Solid lines are thick shell solutions, dashed and dotted
lines are thin shell approximation solutions \citep{Beuthe:2016a},
and bottom panels show the difference between the two solutions. We
assume the interior structure parameters in Table \ref{tab:Interior parameters}
and a 1 km thick ocean for both Enceladus and Europa. }
\end{figure}

\citet{Beuthe:2016a} models the effect of an overlying shell on the
LTE using a thin shell approximation. In this case, the terms $\beta_{n}$
and $\upsilon_{n}$ in the momentum conservation equation are given
by Eq. (27) of \citet{Beuthe:2016a}. Fig. \ref{fig:beuthe_comparison}
compares the thick and thin shell approximation solutions of these
terms and the corresponding admittance $Z_{n}$. We assume a three-layer
interior structure with the parameters in Table \ref{tab:Interior parameters},
a 1 km thick ocean for both Enceladus and Europa, and a uniform core
beneath the ocean. In this case, we can compute the thin shell approximation
solutions using the analytic expressions for the Love numbers of a
homogeneous body (\ref{sec:Homogeneous interior Love numbers})
with the uniform core values. For a differentiated interior structure
beneath the ocean, the Love numbers at the ocean bottom can be computed
using the propagator matrix method (\ref{sec:Love numbers liquid layers}).
The thick shell solutions converge to the thin shell solutions, illustrating
that the momentum conservation equation using the thick shell theory
converges to that of the thin shell approximation as the shell thickness
decreases, as expected. 

\begin{figure}[h]
\begin{centering}
\includegraphics[width=13cm]{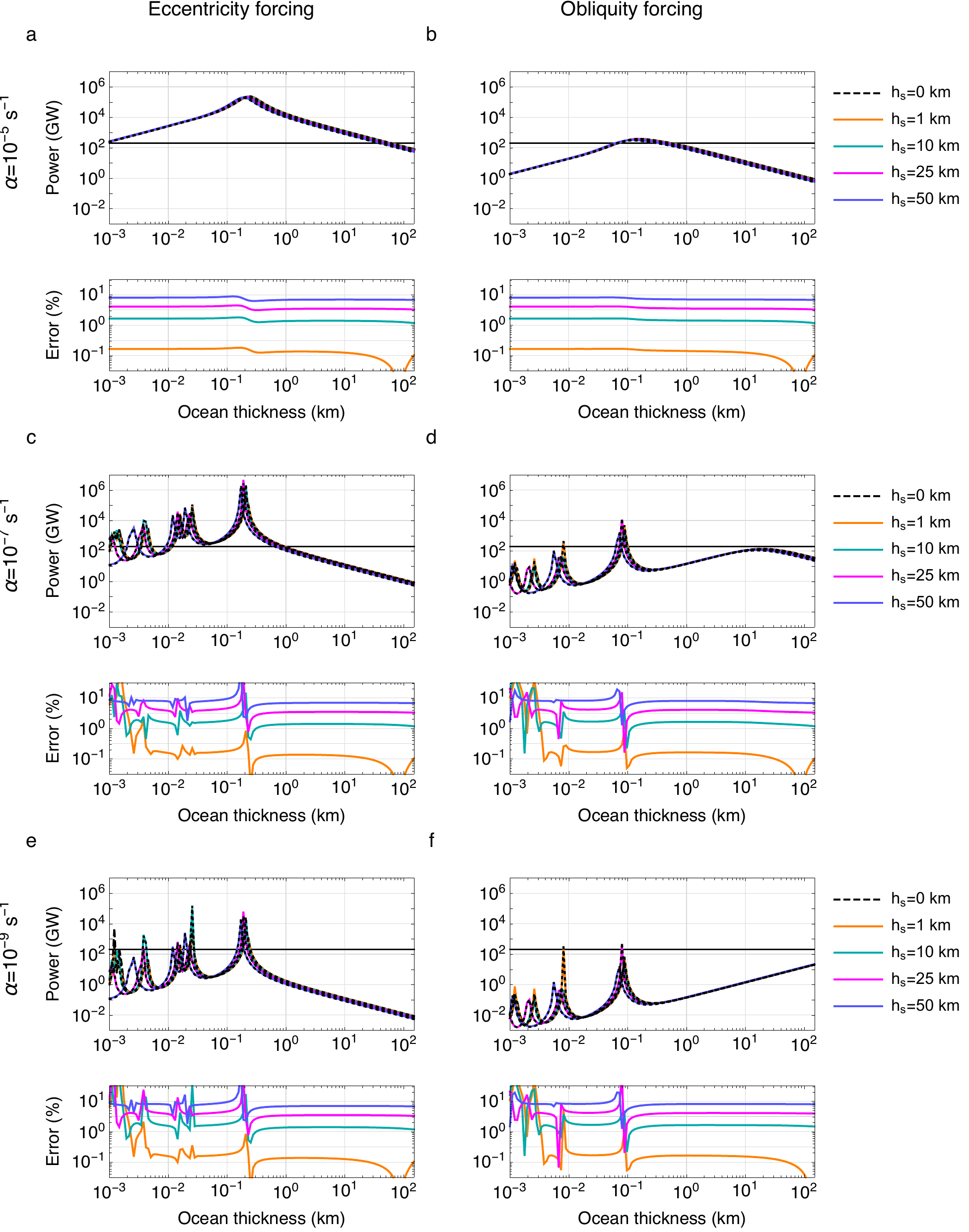}
\par\end{centering}
\caption{\label{fig:europa_power_ocean}Ocean tidal heating power in Europa
due to eccentricity and obliquity forcing as a function of ocean thickness
for different shell thicknesses, $h_{s}$, and linear drag coefficients,
$\alpha$. Solid lines are thick shell solutions, dotted lines are
thin shell approximation solutions \citep{Beuthe:2016a}, and bottom
panels show the difference between the two solutions. Dashed lines
are surface ocean solutions without an overlying solid shell ($h_{s}=0$).
The solid horizontal black line is the estimated radiogenic heating
power (200 GW). We assume the interior structure parameters in Table
\ref{tab:Interior parameters}. }
\end{figure}

\begin{figure}[h]
\begin{centering}
\includegraphics[width=13cm]{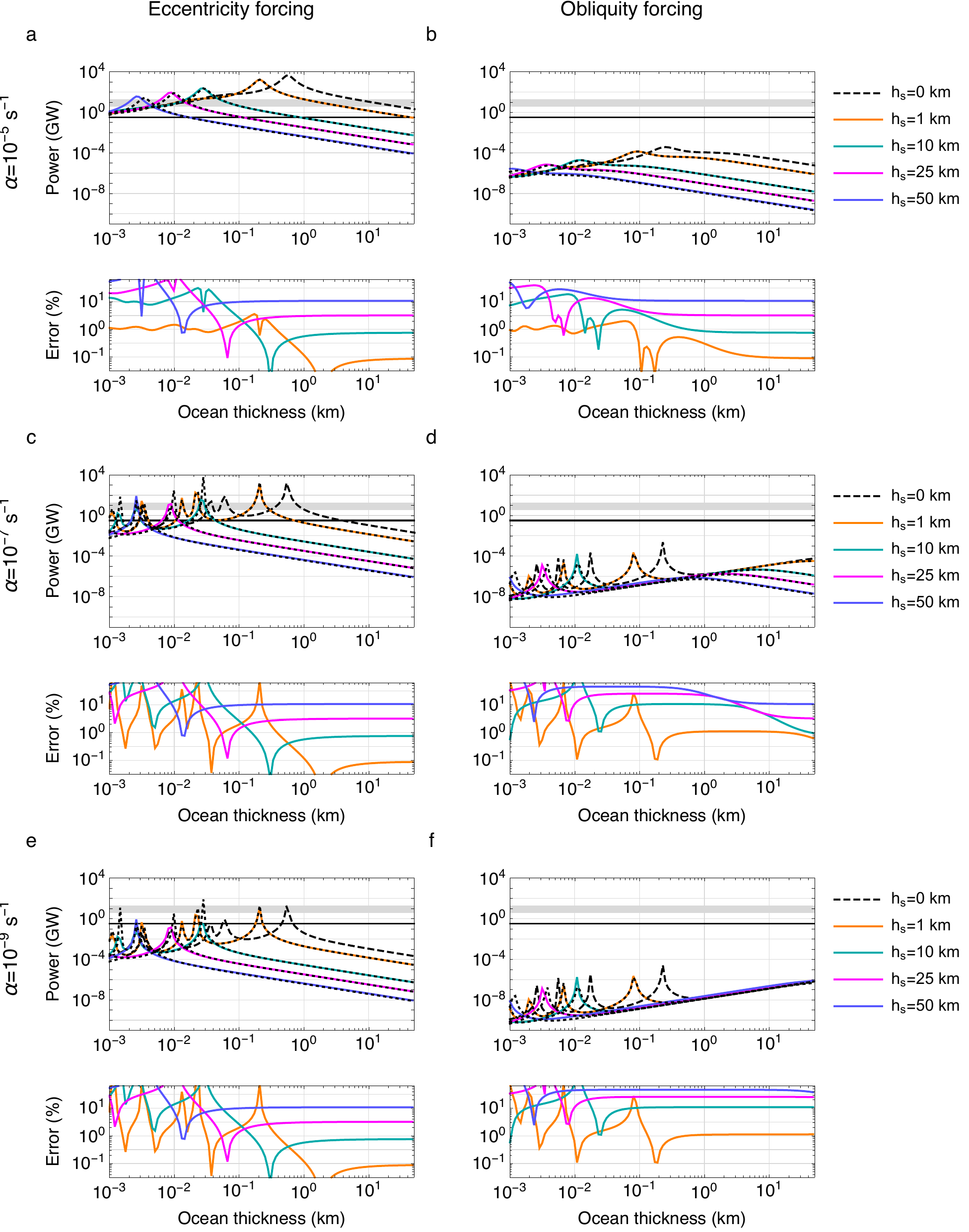}
\par\end{centering}
\caption{\label{fig:enceladus_power_ocean}Ocean tidal heating power in Enceladus
due to eccentricity and obliquity forcing as a function of ocean thickness
for different shell thicknesses, $h_{s}$, and linear drag coefficients,
$\alpha$. Solid lines are thick shell solutions, dotted lines are
thin shell approximation solutions \citep{Beuthe:2016a}, and bottom
panels show the difference between the two solutions. Dashed lines
are surface ocean solutions without an overlying solid shell ($h_{s}=0$).
The shaded gray region corresponds to the observational constraint
of $3.9-18.9$ GW, and the solid horizontal black line is the estimated
radiogenic heating power (0.3 GW). We assume the interior structure
parameters in Table \ref{tab:Interior parameters}. }
\end{figure}

The convergence of the momentum conservation equation implies that
ocean tidal heating solutions using the thick shell theory must also
converge to the thin shell approximation solutions. We verify this
by comparing the time- and surface-averaged tidal heating (Appendix
\ref{sec:Numerical-solution}, Eqs. (\ref{numsol:power1}) and (\ref{numsol:power2}))
on Enceladus and Europa in Figs. \ref{fig:europa_power_ocean} and
\ref{fig:enceladus_power_ocean}. The difference between the thin
and thick shell solutions can be large near resonant ocean thicknesses
(discussed below). However, these resonances occur for oceans thinner
than about 1 km, which is significantly smaller than the likely ocean
thicknesses. For ocean thicknesses larger than 1 km, the thick shell
solutions converge to the thin shell solutions as the shell thickness
decreases, as expected. Assuming a 1 km thick shell for both Enceladus
and Europa, the difference between the two solutions is less than
1\% (Figs. \ref{fig:europa_power_ocean} and \ref{fig:enceladus_power_ocean}). 

Assuming a likely range of shell and ocean thicknesses for Europa
and Enceladus, $h_{s}<25$ km and $h_{o}>10$ km, the accuracy of
the thin shell approximation solutions is 3.5\% and 4.1\% for Europa
(Fig. \ref{fig:europa_power_ocean}) and 3.2\% and 26\% for Enceladus
(Fig. \ref{fig:enceladus_power_ocean}) for eccentricity tides and
obliquity tides, respectively. The larger difference for obliquity
forcing on Enceladus only occurs for small linear drag coefficients
($\alpha\lesssim10^{-9}$ s$^{-1}$, Fig. \ref{fig:enceladus_power_ocean}f)
and becomes similar to the difference for eccentricity forcing (3.2\%)
for larger linear drag coefficients (Fig. \ref{fig:enceladus_power_ocean}b,
d).


\begin{figure}[h]
\begin{centering}
\includegraphics[width=13cm]{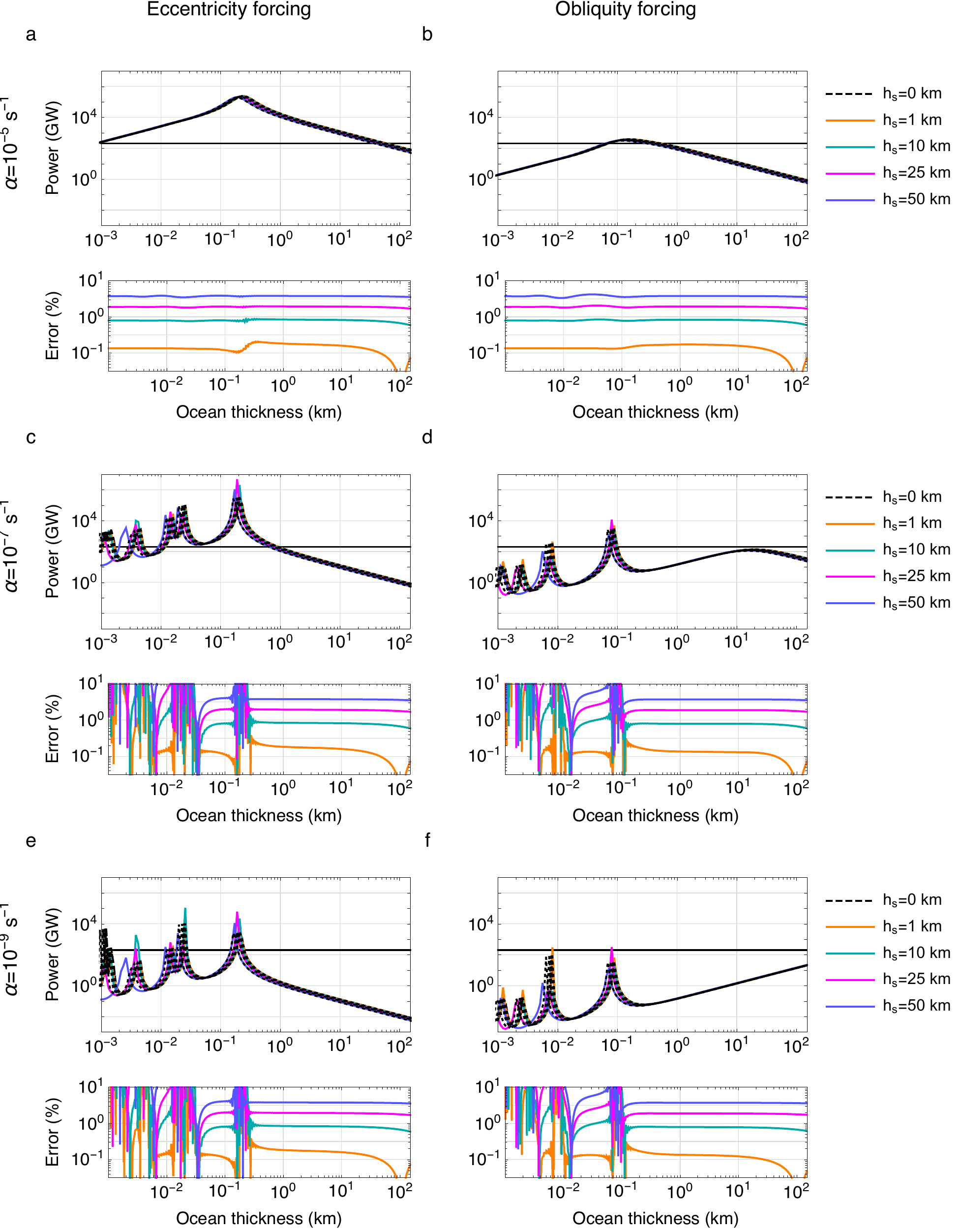}
\par\end{centering}
\caption{\label{fig:rescale_europa}Europa's ocean tidal heating power due
to the obliquity and eccentricity forcing as a function of ocean thickness
for different shell thicknesses, $h_{s}$, and linear drag coefficients,
$\alpha$. Solid lines are thick shell solutions, dotted lines are
rescaled surface ocean solutions (Eq. (\ref{eq:rescale_factor2})),
and bottom panels show the difference between the two solutions. Dashed
lines are surface ocean solutions without an overlying solid shell
($h_{s}=0$). The solid horizontal black line is the estimated radiogenic
heating power (200 GW). We assume the interior structure parameters
in Table \ref{tab:Interior parameters}.}
\end{figure}
\begin{figure}[h]
\begin{centering}
\includegraphics[width=13cm]{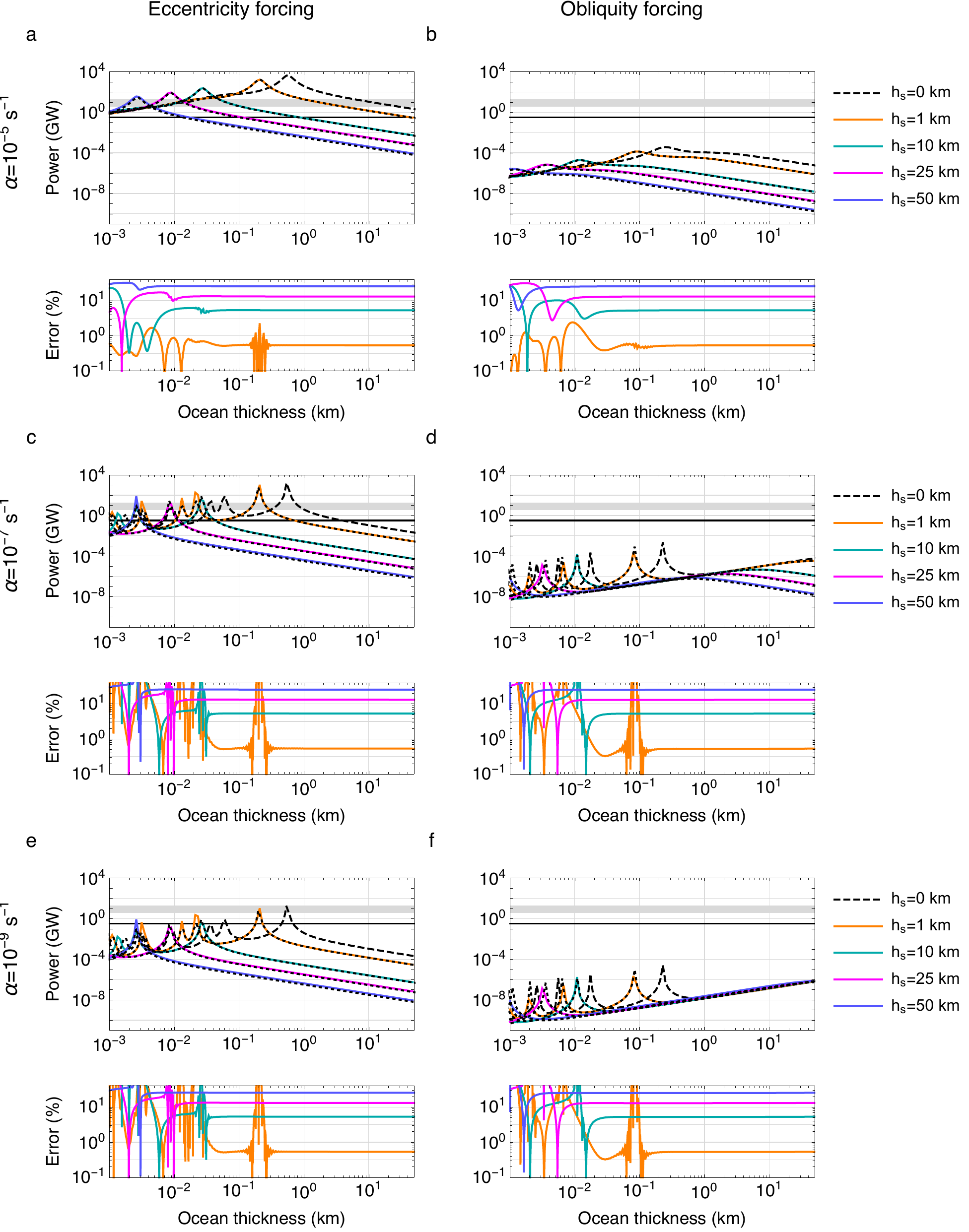}
\par\end{centering}
\caption{\label{fig:rescale_enceladus}Enceladus' ocean tidal heating power
due to the obliquity and eccentricity forcing as a function of ocean
thickness for different shell thicknesses, $h_{s}$, and linear drag
coefficients, $\alpha$. Solid lines are thick shell solutions, dotted
lines are rescaled surface ocean solutions (Eq. (\ref{eq:rescale_factor2})),
and bottom panels show the difference between the two solutions. Dashed
lines are surface ocean solutions without an overlying solid shell
($h_{s}=0$). The shaded gray region corresponds to the observational
constraint of $3.9-18.9$ GW, and the solid horizontal black line
is the estimated radiogenic heating power (0.3 GW). We assume the
interior structure parameters in Table \ref{tab:Interior parameters}.}
\end{figure}

\subsection{Rescaling surface ocean solutions to subsurface ocean solutions\label{sec:Rescaling-surface-ocean}}

Assuming that the solutions to the LTE are dominated by spherical
harmonic degree-two, it is possible to rescale surface ocean solutions
to obtain subsurface ocean solutions \citep{Beuthe:2016a}. We can
rescale the surface ocean solutions by a factor
\begin{equation}
\left(\frac{r_{t}}{R}\right)^{2}\left[\frac{(1-\xi_{2}\gamma_{2}^{L})}{\beta_{2}}\right]\label{eq:rescale_factor1}
\end{equation}
in ocean thickness, and by a factor 
\begin{equation}
\left(\frac{r_{t}}{R}\right)^{2}\left[\frac{(1-\xi_{2}\gamma_{2}^{L})}{\beta_{2}}\frac{\upsilon_{2}}{\gamma_{2}^{T}}\right]\label{eq:rescale_factor2}
\end{equation}
in energy dissipation (\ref{sec:Numerical-solution}, Eqs.
(\ref{numsol:power1}) and (\ref{numsol:power2})), where $\gamma_{2}^{L}$
and $\gamma_{2}^{T}$ are given by Eq. (\ref{surface ocean: tilt factors})
and $\upsilon_{2}$ and $\beta_{2}$ are given by Eq. (\ref{sub ocean: nu beta}).
The first term in Eqs. (\ref{eq:rescale_factor1}) and (\ref{eq:rescale_factor2})
accounts for the change in Lamb's parameter from $\text{\ensuremath{\epsilon}=}4\Omega^{2}R^{2}/(g(R)h_{o})$
for a surface ocean to $\epsilon=4\Omega^{2}r_{t}^{2}/(g(R)h_{o})$
for a subsurface ocean in our solutions based on the method of \citet{Longuet-Higgins:1968}
(\ref{sec:Numerical-solution}). The second term accounts
for the change in the momentum conservation equation terms from $1-\xi_{2}\gamma_{n}^{L}$
and $\gamma_{2}^{T}$ for a surface ocean (Eq. (\ref{surface ocean: momentum}))
to $\upsilon_{n}$ and $\beta_{n}$ for a subsurface ocean (Eq. (\ref{sub ocean: momentum eq.})).
The resonant ocean thicknesses are independent of the forcing; therefore
the scaling factor (\ref{eq:rescale_factor1}) for the ocean thickness
does not include the scaling between the forcing terms ($\upsilon_{2}/\gamma_{2}^{T}$).
Because the Love numbers in Eq. (\ref{surface ocean: tilt factors})
for $\gamma_{2}^{L}$ and $\gamma_{2}^{T}$ are generally significantly
smaller than unity for typical rigidities, these terms can be approximated
as $\sim1$. 

Figs. \ref{fig:rescale_europa} and \ref{fig:rescale_enceladus} compare
subsurface ocean solutions with rescaled surface ocean solutions for
energy dissipation (\ref{sec:Numerical-solution}). The difference
between the two solutions decreases with decreasing shell thickness,
as expected. As in the thin shell approximation, the difference between
the two solutions can be large near resonant ocean thicknesses (discussed
below) but is small for the likely ocean thicknesses larger than about
1 km. Assuming the same likely range of shell and ocean thicknesses
for Europa and Enceladus as above ($h_{s}<25$ km and $h_{o}>10$
km), the accuracy of the rescaled solutions is 2\% for Europa (Fig.
\ref{fig:rescale_europa}) and 13\% for Enceladus (Fig. \ref{fig:rescale_enceladus}).

\section{Time and surface-averaged tidal heating\label{sec:General-results}}

Having illustrated the stabilizing effect of an overlying shell on
equilibrium tides, we consider dynamic tides and the corresponding
time- and surface-averaged ocean tidal heating in Figs. \ref{fig:europa_power_ocean}-\ref{fig:rescale_enceladus}.
We assume the three-layer interior structures for Enceladus and Europa
described in Table \ref{tab:Interior parameters} and consider a range
of shell and ocean thicknesses. Given a velocity solution of the Laplace
tidal equation, the dissipated energy per unit time and surface area
can be found with Eq. (\ref{eq:dissipated flux}), and the time- and
surface-averaged energy dissipation can be found by integrating this
equation (\ref{sec:Numerical-solution}, Eq. (\ref{numsol:power1}).
Energy conservation requires that the time- and surface-averaged dissipated
energy and work done by the tide must be equal, which provides an
alternative expression (\ref{sec:Numerical-solution}, Eq.
(\ref{numsol:power2}).

Tidal heating decreases with shell thickness, as expected due to the
stabilizing effect of an overlying shell. This effect is more pronounced
on Enceladus than on Europa (e.g. compare Figs \ref{fig:europa_power_ocean}
and \ref{fig:enceladus_power_ocean}) because the effective rigidity
on a small body like Enceladus is larger (\ref{subsec:Homogeneous-thick-shell}).
This difference can also be seen in the Love numbers. The magnitude
of the Love numbers decreases rapidly with shell thickness for Enceladus
while it remains relatively constant for Europa (\ref{sec:Love-numbers}).
A 1 km thick shell can reduce tidal heating by about an order of magnitude
for Enceladus, and by tens of percent for Europa, compared with tidal
heating solutions without an overlying shell.

As previously shown for surface oceans \citep{Tyler:2008,Tyler:2009,Tyler:2011,Matsuyama:2014},
tidal heating can increase sharply for ocean thicknesses for which
the tidal flow is resonantly enhanced (the sharp peaks in Figs. \ref{fig:europa_power_ocean}
and \ref{fig:enceladus_power_ocean}). Eccentricity and obliquity
forcing can generate gravity wave resonances for oceans thinner than
about 1 km, limiting their impact because the oceans in Enceladus
and Europa are likely significantly thicker. Obliquity forcing can
also generate Rossby-Haurwitz waves \citep{Tyler:2008}, and this
produces a tidal heating increase with ocean thickness for small linear
drag coefficients (Figs. \ref{fig:europa_power_ocean}f and \ref{fig:enceladus_power_ocean}f).
In addition to reducing the magnitude of tidal heating, increasing
the shell thickness also decreases the resonant ocean thicknesses,
as illustrated by \citet{Beuthe:2016a} using a thin shell approximation.

Enceladus' total endogenic power radiated from the south polar terrain
(SPT) has been estimated to be $15.8\pm3.1$ GW based on Cassini infrared
emission observations \citep{Howett:2011}. The actual uncertainty
is likely larger due to the difficulty of estimating and subtracting
the thermal emission associated with absorbed solar radiation \citep{Spencer:2013}.
This may explain the difference from the previous estimate of $5.8\pm1.9$
GW \citep{Spencer:2006}. We adopt a range of $3.9-18.9$ GW as a
conservative constraint that includes all available estimates. For
the range of likely shell and ocean thicknesses inferred from gravity
and topography constraints, $h_{s}=23\pm4$ km and $h_{o}=38\pm4$
km \citep{Beuthe:2016}, the total power is lower than the observed
value (Fig. \ref{fig:enceladus_power_ocean}). Ocean tidal heating
driven by eccentricity forcing can match the observed value; however,
this requires ocean and shell thicknesses that are significantly smaller
than the values inferred from gravity and topography constraints (Fig.
\ref{fig:enceladus_power_ocean}). Additionally, a more direct comparison
with the observed value requires computing the power dissipated beneath
SPT alone, which would worsen the discrepancy. 

The radiogenic heating power is about  200 GW and 0.3 GW for Europa
and Enceladus respectively assuming the fiducial values in Table \ref{tab:Interior parameters}
and a present-day, chondritic radiogenic heating rate of $4.5\times10^{-12}$
W kg$^{-1}$ for the core \citep{Spohn:2003iw}. For Europa, ocean
tidal heating is comparable to radiogenic heating for eccentricity
forcing, $\alpha\gtrsim10^{-5}$ s$^{-1}$, and ocean thicknesses
$\lesssim50$ km (Fig. \ref{fig:europa_power_ocean}a); or obliquity
forcing, $\alpha\lesssim10^{-7}$ s$^{-1}$, and ocean thicknesses
$\sim25$ km (Fig. \ref{fig:europa_power_ocean}d, f). For Enceladus,
ocean tidal heating is comparable to radiogenic heating for eccentricity
forcing and $\alpha\gtrsim10^{-5}$ s$^{-1}$ (Fig. \ref{fig:enceladus_power_ocean}a);
however, this requires ocean and shell thicknesses that are significantly
smaller than the values inferred from gravity and topography constraints;
and ocean tidal heating due to obliquity forcing is generally smaller
than that due to eccentricity forcing due to Enceladus' small obliquity
(compare the left and right panels in Fig. \ref{fig:enceladus_power_ocean}).
It is worth noting that ocean tidal heating may be generally weaker
than solid-body tidal heating \citep{Chen:2014,Beuthe:2016a}; however,
the effect of ocean dynamics remains important because ocean tides
can increase solid-body tidal heating. 

\section{Temporal and surface distribution of tidal heating and tides\label{sec:Temporal-and-surface}}

Thus far we have focused on the time- and surface-averaged ocean tidal
heating. While this is important for the global average tidal heating
and has implications for the long-term thermal, rotational, and orbital
evolution, the temporal and surface distribution of ocean tidal heating
contains unique features that may be observable, as shown in sections
\ref{subsec:lateral_shell_variations} and \ref{subsec:phase_lag}. 

\subsection{Surface distribution of ocean tidal heating\label{subsec:lateral_shell_variations}}

\begin{figure}[h]
\begin{centering}
\includegraphics[width=15cm]{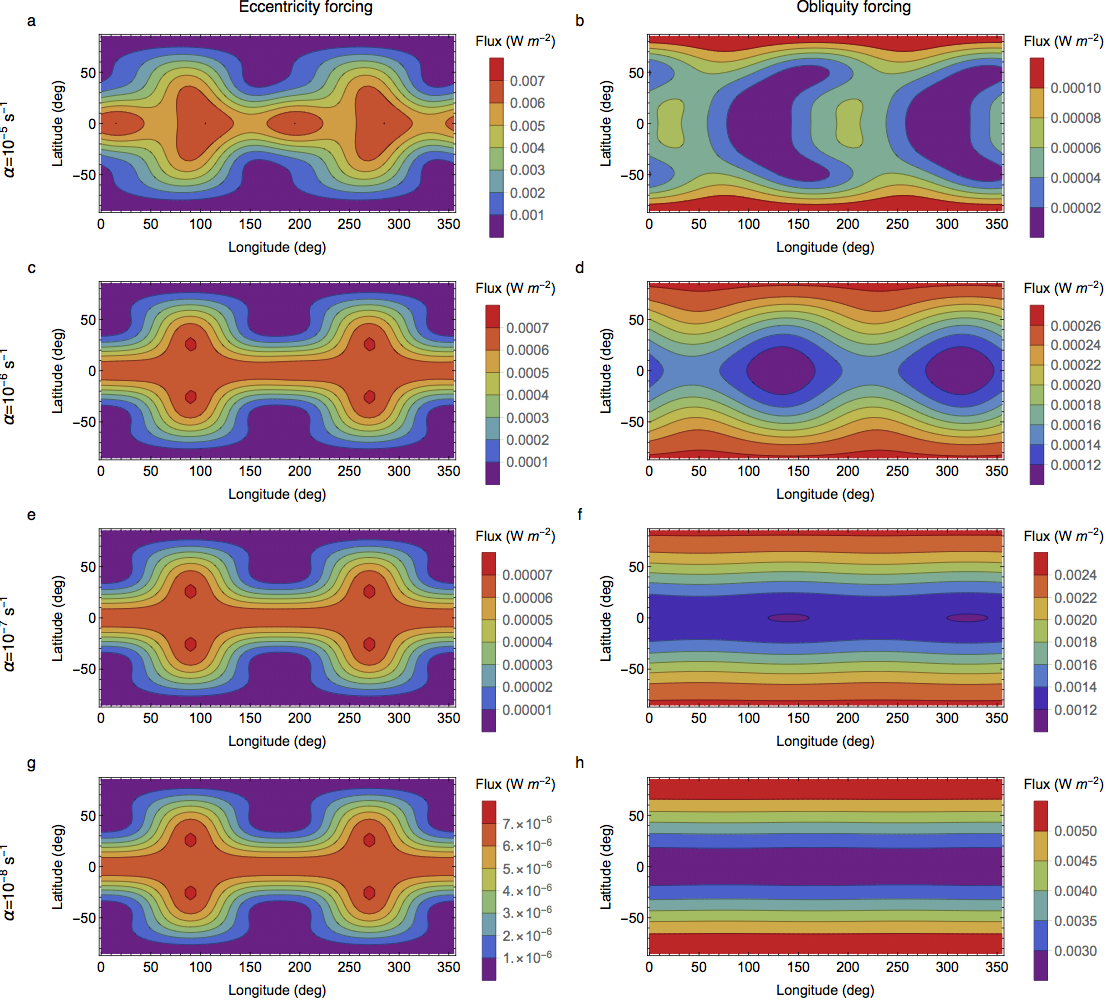}
\par\end{centering}
\caption{\label{fig:eur_surface_dist}Europa's surface distribution of ocean
tidal heating due to eccentricity and obliquity forcing for different
linear drag coefficients, $\alpha$. Contours show the energy flux
averaged over the tidal forcing period. We assume the parameters in
Table \ref{tab:Interior parameters}.}
\end{figure}

\begin{figure}[h]
\begin{centering}
\includegraphics[width=15cm]{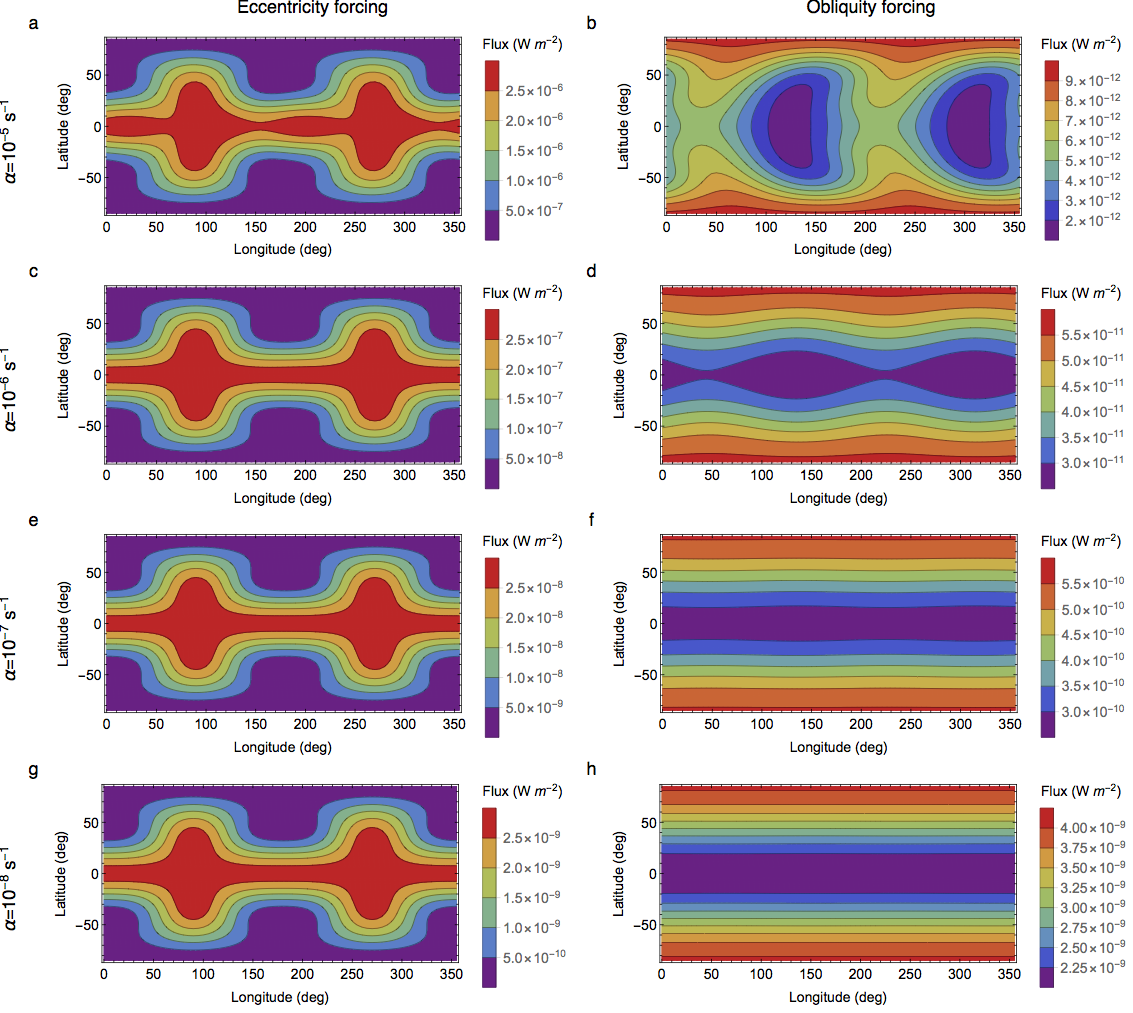}
\par\end{centering}
\caption{\label{fig:enc_surface_dist}Enceladus' surface distribution of ocean
tidal heating due to eccentricity and obliquity forcing for different
linear drag coefficients, $\alpha$. Contours show the energy flux
averaged over the tidal forcing period. We assume the parameters in
Table \ref{tab:Interior parameters}.}
\end{figure}

Figs. \ref{fig:eur_surface_dist} and \ref{fig:enc_surface_dist}
show the surface distribution of the tidal heating flux (Eq. (\ref{eq:dissipated flux}))
averaged over the tidal forcing period. We assume the Enceladus and
Europa model parameters in Table \ref{tab:Interior parameters}.

For Enceladus, we assume the likely shell and ocean thicknesses inferred
from gravity and topography constraints for Enceladus, $h_{s}=23$
km and $h_{o}=38$ km \citep{Beuthe:2016}. For Europa, we assume
values that are consistent with the constraint of a combined ocean
and shell thickness between 80 and 170 km \citep{Anderson:1998},
$h_{s}=10$ km and $h_{o}=100$ km. The surface distribution of the
time-averaged tidal heating flux is similar to that of surface oceans
(compare Figs. \ref{fig:eur_surface_dist} and \ref{fig:enc_surface_dist}
with Fig. 7 of \citealp{Chen:2014}) for small linear drag coefficients
($\alpha\lesssim10^{-7}$ s$^{-1}$). 

Tidal heating in a subsurface ocean can generate horizontal shell
thickness variations if the shell is not convective \citep{Ojakangas:1986,Ojakangas:1989,Nimmo:2007,Nimmo:2010}.
The surface distribution of the time-averaged tidal heating flux is
distinct from that due solid dissipation in the shell \citep{Ojakangas:1989,Tobie:2005a,Beuthe:2013},
with higher dissipation near the equator and poles for eccentricity
and obliquity forcing respectively (Figs. \ref{fig:eur_surface_dist}
and \ref{fig:enc_surface_dist}). Therefore, observations of these
variations can be used to constrain ocean tidal heating. The tidal
heating flux is orders of magnitude larger for Europa than for Enceladus
(compare Figs. \ref{fig:eur_surface_dist} and \ref{fig:enc_surface_dist})
due to Enceladus' larger effective rigidity (\ref{subsec:Homogeneous-thick-shell})
and small obliquity, which has implications for observable horizontal
shell thickness variations.

The surface heat flux due to radiogenic heating is about $7\times10^{-3}$
W m$^{-2}$ and $4\times10^{-4}$ W m$^{-2}$ for Europa and Enceladus
respectively assuming the fiducial values described above and a present-day,
chondritic radiogenic heating rate of $4.5\times10^{-12}$ W kg$^{-1}$
for the core \citep{Spohn:2003iw}. As discussed above for the time-
and surface-averaged tidal heating results, ocean tidal heating in
Europa is comparable to radiogenic heating for eccentricity forcing
and $\alpha\gtrsim10^{-5}$ s$^{-1}$ (Fig. \ref{fig:eur_surface_dist}a)
or obliquity forcing and $\alpha\lesssim10^{-7}$ s$^{-1}$ (Fig.
\ref{fig:eur_surface_dist}f, h). Ocean tidal heating in Enceladus
is significantly weaker than radiogenic heating (Fig. \ref{fig:enc_surface_dist}).


\subsection{Surface displacement phase lag and amplitude\label{subsec:phase_lag}}

The surface displacements driven by eccentricity and obliquity forcing
can have phase lags relative to the forcing tidal potential due to
the delayed ocean response. To compute this phase lag, we compare
the dynamic and equilibrium surface displacements $\eta^{R}$ (Eq.
(\ref{sub ocean: displacements})). The former is the solution of
the LTE coupled to a thick shell, while the latter assumes
that the satellite deforms instantaneously in response to the forcing
tidal potential and can be computed by ignoring the pressure potential
term, $U_{nm}^{P}$, in Eq. (\ref{sub ocean: displacements}). 

\begin{figure}[h]
\begin{centering}
\includegraphics[width=15cm]{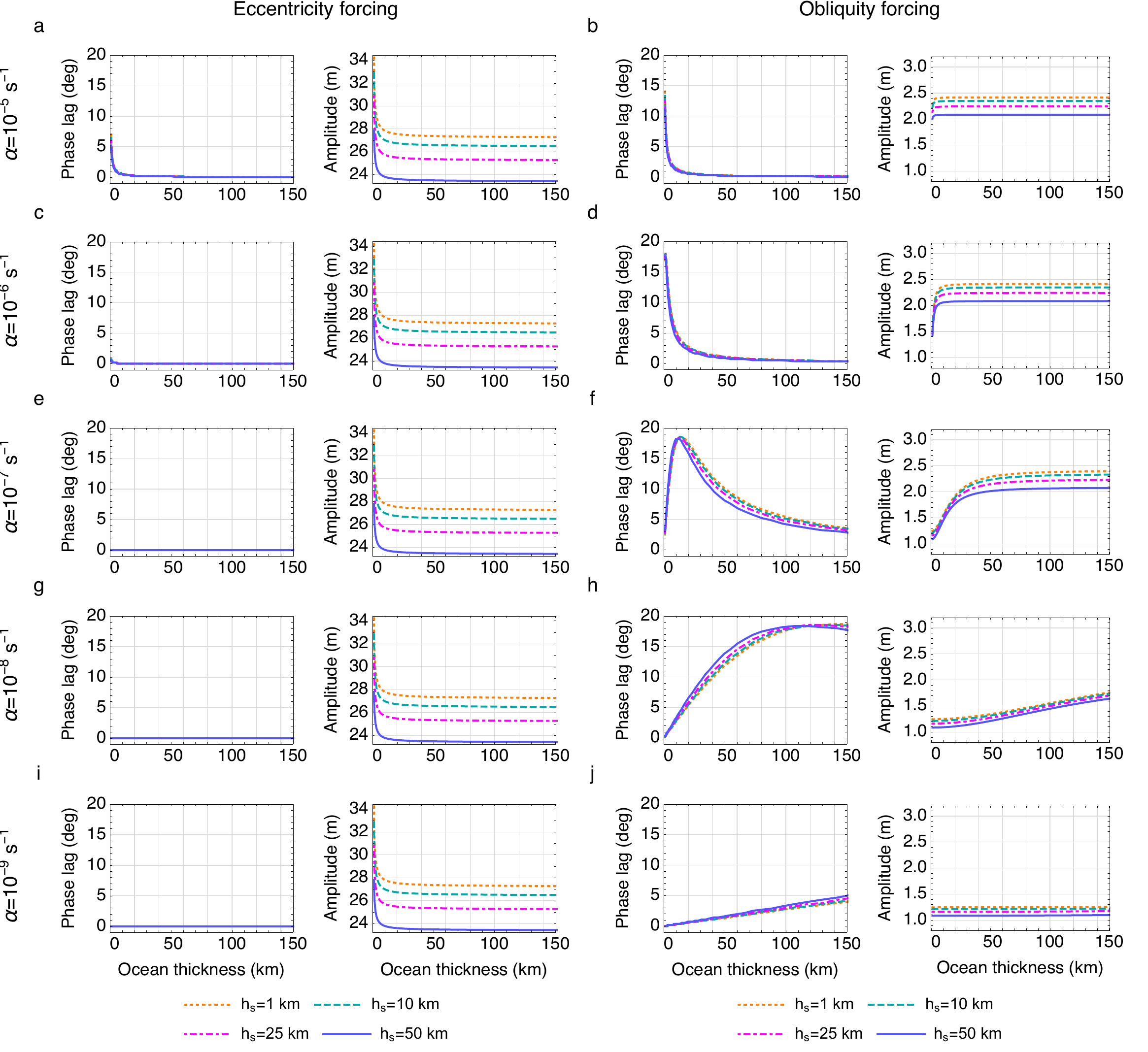}
\par\end{centering}
\caption{\label{fig:europa phase lag}Surface displacement phase lag and amplitude
due to eccentricity and obliquity forcing on Europa as a function
of ocean thickness for different shell thicknesses and linear drag
coefficients. The phase lag and amplitude are computed at 0$^{\circ}$
and $45^{\circ}$ latitudes for eccentricity and obliquity forcings
respectively, where the amplitude is maximum. The phase lag is given
by $\delta\phi=\Omega\delta t$, where $\Omega$ is the rotation rate
and $\delta t$ is the time lag. We assume the interior structure
parameters in Table \ref{tab:Interior parameters}. }
\end{figure}

\begin{figure}[h]
\begin{centering}
\includegraphics[width=15cm]{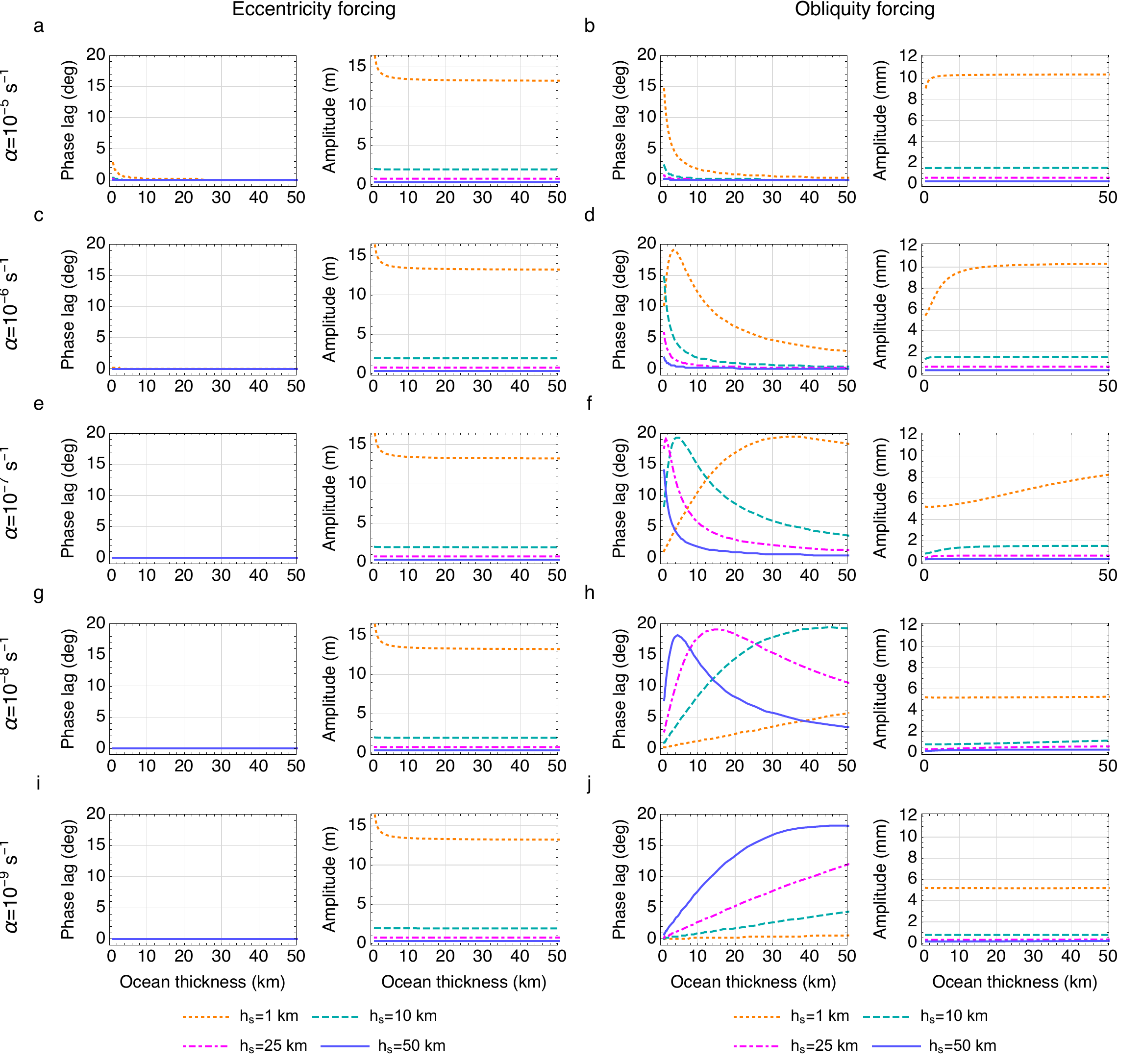}
\par\end{centering}
\caption{\label{fig:enc phase lag}Surface displacement phase lag and amplitude
due to eccentricity and obliquity forcing on Enceladus as a function
of ocean thickness for different shell thicknesses and linear drag
coefficients. The phase lag and amplitude are computed at 0$^{\circ}$
and $45^{\circ}$ latitudes for eccentricity and obliquity forcings
respectively, where the amplitude is maximum. The phase lag is given
by $\delta\phi=\Omega\delta t$, where $\Omega$ is the rotation rate
and $\delta t$ is the time lag. We assume the interior structure
parameters in Table \ref{tab:Interior parameters}. }
\end{figure}

Figs. \ref{fig:europa phase lag} and \ref{fig:enc phase lag} and
show the surface displacement phase lag and amplitude for Europa and
Enceladus. We compute the surface displacement time lag $\delta t$
and corresponding phase lag $\delta\phi=\Omega\delta t$ by finding
the wave maximum using Eq. (\ref{sub ocean: displacements}).
We assume the model parameters in Table \ref{tab:Interior parameters}
and consider a wide range of ocean and shell thicknesses and linear
drag coefficients. Eccentricity forcing generally produces greater
tidal amplitudes due to the large eccentricity values relative to
the obliquity values. Despite the small obliquity values, obliquity
forcing generally produces larger phase lags due to the generation
of Rossby-Haurwitz waves.

The amplitudes of the equilibrium and dynamic surface displacement
are comparable, and the latter scales linearly with the forcing tidal
potential, which in turn scales linearly with eccentricity, $e$,
and obliquity, $\theta_{0}$ (Eq. \ref{eq:U_ecc_obliq}). Assuming
the minimum energy Cassini state values in Table \ref{tab:Interior parameters},
$e/\theta_{0}\sim10$ and $e/\theta_{0}\sim10^{3}$ for Europa and
Enceladus respectively (note that the obliquities in the forcing potential
must be in radians), which is consistent with the differences in the
obliquity and eccentricity surface displacement amplitudes in Figs.
\ref{fig:europa phase lag} and \ref{fig:enc phase lag}. The surface
displacement amplitude increases with decreasing shell thickness,
as expected due to the weaker resistance to deformation. 


In the absence of Rossby-Haurwitz waves, the surface displacement
phase lag decreases with increasing ocean thickness (Figs. \ref{fig:europa phase lag}
and \ref{fig:enc phase lag}, eccentricity tide results). In this
case, dynamic effects decrease as the ocean thickness increases away
from the resonant thicknesses, decreasing the phase lag because the
tide response becomes more static. We also verify this for obliquity
forcing by removing the westward component of the obliquity forcing
tidal potential (Eq. \ref{eq:U_ecc_obliq}), which prevents the generation
of Rossby-Haurwitz waves. The generation of Rossby-Haurwitz waves
by obliquity forcing produces phase lags that are larger than those
produced by eccentricity forcing, and introduces a complex dependence
of the phase lag on ocean thickness (Figs. \ref{fig:europa phase lag}
and \ref{fig:enc phase lag}). For Enceladus, the phase lag is also
sensitive to the overlying shell thickness (Fig. \ref{fig:enc phase lag}).

Fig. \ref{fig:europa phase lag} shows the amplitude and phase lag
of the surface displacement on Europa. Assuming the fiducial shell
and ocean thicknesses ($h_{s}=10$ km and $h_{o}=100$ km) and linear
drag coefficients $\alpha<10^{-5}$ s$^{-1}$, the amplitude and phase
lag are 26.5 m and $<1$ degree for eccentricity forcing, and $<2.5$
m and $<18$ degrees for obliquity forcing. The larger obliquity phase
lags correspond to linear drag coefficients $\lesssim10^{-7}$ s$^{-1}$
and smaller amplitudes (e.g. $5$ degrees and 2.5 m for $\alpha=10^{-7}$
s$^{-1}$, and $18$ degrees and 1.6 m for $\alpha=10^{-8}$ s$^{-1}$),
and measurement of this lag (e.g. by Europa Clipper) would provide
a probe of ocean thickness. Ignoring the dynamic ocean response, there
can be a phase lag due to the viscoelastic behavior of the shell.
This phase lag is smaller than 2 degrees for shell thicknesses smaller
than 100 km \citep{Moore:2000}; therefore, the phase lag due to the
delayed ocean response can be significantly larger for this range
of shell thicknesses. 

Fig. \ref{fig:enc phase lag} shows the amplitude and phase lag of
the surface displacement on Enceladus. Assuming the fiducial shell
and ocean thicknesses ($h_{s}=23$ km and $h_{o}=38$ km) and linear
drag coefficients $\alpha<10^{-5}s^{-1}$, the amplitude and phase
lag are 0.8 m and $<1$ degrees for eccentricity forcing, and $<0.6$
mm and $<13$ degrees for obliquity forcing. The amplitude increases
rapidly with decreasing shell thickness (e.g. 13.2 m and $<11$ mm
for eccentricity and obliquity forcing respectively assuming a shell
thickness of 1 km and $\alpha=10^{-5}$ s$^{-1}$) due to Enceladus'
small size and large effective rigidity \textcolor{black}{(Appendix
\ref{subsec:Homogeneous-thick-shell})}. As discussed above, the phase
lag dependence on the ocean and shell thicknesses is complex due to
the generation of Rossby-Haurwitz waves. The large phase lags driven
by obliquity forcing may have implications for the observed phase
lag of about 50 degrees (time delay of about 5 hours) in plume activity
with respect to predictions based on tidal stress models \citep{Hedman:2013,Nimmo:2014}.
However, obliquity forcing can only produce surface displacement amplitudes
of the order of mm, generating stresses that are negligible. Furthermore,
all modeling so far has assumed that eccentricity forcing is the source
of tidal modulation \citep{Hedman:2013,Nimmo:2014}, and the predicted
phase lag due to eccentricity forcing is smaller than 2.7 degrees
for shells thicknesses larger than 1 km.

\section{Discussion}

We consider tidal heating in the subsurface oceans of Enceladus and
Europa using a new theoretical treatment that is applicable to icy
satellites with thin oceans overlaid by incompressible elastic shells
of arbitrary thickness. 

The shell's resistance to dynamic ocean tides reduces ocean tidal
heating, and this effect is larger on Enceladus than on Europa due
to the Enceladus' small size and larger effective rigidity \textcolor{black}{(Appendix
\ref{subsec:Homogeneous-thick-shell})}. Tidal heating driven by eccentricity
forcing is generally dominant over that driven by obliquity forcing,
with the exception of Europa models with ocean thicknesses $\gtrsim10$
km and linear drag coefficient $\alpha\lesssim10^{-7}$ s$^{-1}$
(Fig. \ref{fig:europa_power_ocean}), and Enceladus models with ocean
and shell thicknesses $\gtrsim10$ km and linear drag coefficients
$\alpha\lesssim10^{-9}$ s$^{-1}$ (Fig. \ref{fig:enceladus_power_ocean}).

We assume a conductive shell that is entirely elastic. For a convecting
shell, only the near-surface part is likely to be rigid, reducing
the rigid shell thickness and its resistance to tides. This increases
energy dissipation in the shell and ocean. \textcolor{black}{Applying
the thin shell approximation of \citet{Beuthe:2016a} to Enceladus
and assuming the likely range of shell and ocean thicknesses, convection
in the shell increases energy dissipation in the shell and ocean by
about an order of magnitude and tens of percent, respectively \citep[Figs. 13 and 14]{Beuthe:2016a}.
The larger effect on shell energy dissipation arises because a convective
shell is more dissipative than a conductive one \citep[Eqs. (101) and (102)]{Beuthe:2016a}.}
Our assumed shear modulus of 3.5 GPa is uncertain and may be lower
due to fractures or periodic tidal forcing \citep{Wahr:2006}. This
has not been observed in laboratory experiments; however, only a limited
range of temperatures and forcing frequencies that are significantly
higher than those in icy satellites is accessible \citep{Hammond:2018}.
The deformation of the shell is sensitive to both it thickness and
shear modulus, and decreases with the product of these two quantities
\citep{Wahr:2006,Beuthe:2015,Beuthe:2015a}. Thus, reducing the shear
modulus has a similar effect as reducing the thickness, and the wide
range of shell thicknesses considered in this paper can also be interpreted
as a wide range of possible shear moduli.

The time-averaged surface distribution of ocean tidal heating (Figs.
\ref{fig:eur_surface_dist} and \ref{fig:enc_surface_dist}) is distinct
from that due to dissipation in the solid shell, with higher dissipation
near the equator and poles for eccentricity and obliquity forcing
respectively (Figs. \ref{fig:eur_surface_dist} and \ref{fig:enc_surface_dist}).
This can lead to unique horizontal shell thickness variations if the
shell is conductive, providing constraints on ocean tidal heating.
Characterizing the expected horizontal shell thickness variations
requires solving the coupled thermal-orbital evolution, and previous
studies have investigated various aspects of this coupled problem
\citep[e.g.][]{Ojakangas:1986,Ross:1990,Fischer:1990,Showman:1997,Hussmann:2004,Behounkova:2012}.
However, these studies ignore ocean tidal heating and assume that
energy dissipation occurs only in solid regions. \citet{Chen:2016}
investigated the role of tidal heating in a surface magma ocean on
the early evolution of the Earth-Moon system. Our thick shell theory
can be coupled to thermal and orbital evolution models to study the
effect of tidal heating in subsurface oceans.

The surface displacement driven by eccentricity and obliquity forcing
can have a phase lag relative to the forcing tidal potential due to
the delayed ocean response. For Europa and Enceladus, eccentricity
forcing generally produces greater tidal amplitudes due to the large
eccentricity values relative to the obliquity values. Despite the
small obliquity values, obliquity forcing generally produces larger
phase lags due to the generation of Rossby-Haurwitz waves. Figs. \ref{fig:europa phase lag}
and \ref{fig:enc phase lag} summarize the surface displacement results
for Europa and Enceladus. The phase lag due to the delayed ocean response
can be significantly larger than the phase lag due to the viscoelastic
behavior of the shell \citep[e.g. ][]{Moore:2000,Kamata:2016}. Ignoring
the horizontal ocean dynamics, a small phase lag (less than about
10 degrees) can be used as a proxy for the presence of a subsurface
ocean in Ganymede \citep{Kamata:2016}. Given our result that large
phase lags can be generated by obliquity forcing on Europa and the
similar Cassini state obliquities of Ganymede and Europa \citep{Bills:2005},
future Ganymede studies should consider the horizontal ocean dynamics
captured by the LTE. Although eccentricity tides likely dominate obliquity
tides at Europa, the expected obliquity tide amplitude (about 2.5
m) and phase lag (up to 18 degrees) are potentially detectable by
a future spacecraft mission such as Europa Clipper. Measuring this
effect would help determine the ocean thickness and effective drag
coefficient.

For the range of likely shell and ocean thicknesses, the thin shell
approximation of \citet{Beuthe:2016a} is generally accurate to less
than about 4\% for Enceladus and Europa (Figs. \ref{fig:europa_power_ocean}
and \ref{fig:enceladus_power_ocean}), with the exception of obliquity
forcing on Enceladus and small linear drag coefficients ($\alpha\lesssim10^{-9}$
s$^{-1}$) for which the accuracy is less than 26\% (Fig. \ref{fig:enceladus_power_ocean}f).
It is worth noting that the thick shell theory described in this paper
is general enough to be applicable to any interior rheology. The effect
of compressibility can be taken into account by using compressible
Love numbers instead of incompressible ones \citep[e.g. ][]{Tobie:2005a,Wahr:2009,Kamata:2015}
and is likely small. For example, the assumption of an incompressible
shell introduces errors of about 8\% for the radial tidal displacements
on Enceladus \citep[Fig. 1]{Beuthe:2018}. The viscoelastic response
can be derived from the elastic response using the correspondence
principle \citep{Peltier:1974ki} in the Laplace \citep[e.g.][]{Jara-Orue:2011}
or Fourier \citep[e.g. ][]{Moore:2000,Tobie:2005a,Wahr:2009,Kamata:2015}
domains. The effect of taking into account viscoelasticity is also
likely small. For example, the radial tidal displacement on Europa
increases by about 5\% when viscoelasticity is taken into account
\citep[Fig. 1, solid line]{Wahr:2009}. The elastic limit results
of this paper provide an upper estimate of the shell effect on ocean
tidal heating. The thin shell approximation of \citet{Beuthe:2016a}
provides an explicit treatment of viscoelasticity and compressibility
of the crust in terms of the effective shear modulus and Poisson's
ratio \citep[Eq. (B.4)]{Beuthe:2016a}. More importantly,
the dynamic ocean tides are described by a modified Laplace tidal
equation that assumes a thin homogeneous ocean. This approximation
is likely reasonable for Europa if the combined ocean and shell thickness
is smaller than about  170 km \citep{Anderson:1998}. However, it
may introduce significant errors for Enceladus if the ocean thickness
is $38\pm4$ km \citep{Beuthe:2016}. The assumption of a uniform
thickness for the shell and ocean may also introduce errors given
the likely variations in these parameters \citep{McKinnon:2015,Thomas:2016ex,Beuthe:2016,Beuthe:2018}.

We assume the simplest linear drag formalism to model dissipation.
Earth's ocean tidal heating studies commonly assume a quadratic bottom
drag formalism based on the assumption that drag arises due to turbulent
flow interacting with a bottom boundary. We use an order-of-magnitude
analysis (Eq. \ref{eq:cd_alpha_scaling}) to estimate a linear drag
coefficient from the nominal bottom drag coefficient $c_{D}\sim(2-3)\times10^{-3}$
assumed in Earth \citep{Lambeck:1980earth,Jayne:2001,Egbert:2001,Green:2013},
Titan \citep{Sagan:1982}, and Jovian planets \citep{Goldreich:1966}
studies. The numerical method of \citet{Hay:2017} can solve the LTE
with the bottom drag formalism, and our thick shell theory can be
used to extend this method to include the effect of an overlying solid
shell.

\section*{Acknowledgements}

All data are publicly available, as described in the text. This material
is based upon work supported by the National Aeronautics and Space
Administration (NASA) under Grant No. NNX15AQ88G issued through the NASA
Habitable Worlds program. M. B. is financially supported by the Belgian
PRODEX program managed by the European Space Agency in collaboration
with the Belgian Federal Science Policy Office. H. C. F. C. H. was
financially supported by the NASA Earth and Space Science Fellowship.
S. K. acknowledges support from the Japan Society for the Promotion
of Science KAKENHI grant No. 16K17787.

\appendix

\part*{Appendix}

\section{Love numbers\label{sec:Love-numbers}}

\subsection{Propagator matrix method\label{sec:prop matrix method}}

We compute the tidal, load, and pressure Love numbers using the propagator
matrix method \citep[e.g. ][]{Sabadini:2004a}. Following the notation
of \citet{Sabadini:2004a}, the spheroidal solution vector containing
the spherical harmonic expansion coefficients is 
\begin{equation}
\mathbf{y}_{nm}=(U_{\ell m},\:V_{\ell m},\,R_{\ell m},\,S_{\ell m},\,-\Phi_{nm},\:Q_{nm}),\label{App:y vector}
\end{equation}
where $U_{nm}$ and $V_{nm}$ are the radial and tangential displacements,
$R_{\ell m}$ and $S_{\ell m}$ are the radial and tangential stress,
$\Phi_{nm}$ is the gravitational potential, and $Q_{nm}$ is the
potential stress. Hereafter, we drop the ``n'' and ``m'' subscripts
to simplify the notation. 

The surface boundary conditions for the radial, tangential, and potential
stresses are \citep{Sabadini:2004a,Beuthe:2016a}:
\begin{equation}
\mathbf{b}=(y_{3}(R),\,y_{4}(R),\,y_{6}(R))=\begin{cases}
\left(0,\,0,\,-\frac{2n+1}{R}\right)U^{T}(R) & \mbox{tidal}\\
\left(-\frac{2n+1}{3}\bar{\rho},\,0,\,-\frac{2n+1}{R}\right)U^{L}(R) & \mbox{surface loading}\\
\left(\bar{\rho},\,0,\,0\right)U^{P}(R) & \mbox{surface pressure,}
\end{cases}\label{app: surface bcs}
\end{equation}
where $U^{T}(R)$, $U^{L}(R)$, and $U^{P}(R)$ are the tidal, surface
loading, and surface pressure loading potentials. In this equation,
$R$ and $\bar{\rho}$ are the mean radius and density respectively,
and the subscript in the solution vector now refers to the third,
fourth, and sixth components of the solution vector. Eq. (\ref{app: surface bcs})
is equivalent to Eqs. (C.5), (C.6), and (E.3) of \citet{Beuthe:2016a}
if we take into account the following differences. First, we use the
convention of \citet{Sabadini:2004a} for the solution vector (Eq.
(\ref{App:y vector})), while \citet{Beuthe:2016a} uses the convention
of \citet{Takeuchi:1972kh}.\textcolor{magenta}{{} }The two conventions
differ by exchanging the definitions of $y_{2}$ and $y_{3}$ and
by changing the sign of both $y_{5}$ and $y_{6}$.\textcolor{magenta}{{}
}Second, we normalize the pressure potential as $y_{3}(R)=\bar{\rho}U^{P}(R)$,
whereas \citet{Beuthe:2016a} normalizes it as the pressure part of
a load potential, $y_{3}(R)=-(2n+1)\bar{\rho}U^{P}(R)/3$. Our sign
convention implies that a positive pressure potential leads to a positive
radial displacement, whereas the sign convention of \citet{Beuthe:2016a}
implies that a positive pressure potential leads to a negative radial
displacement (as in the case of a mass load). 

\begin{figure}[h]
\begin{centering}
\includegraphics[width=5cm]{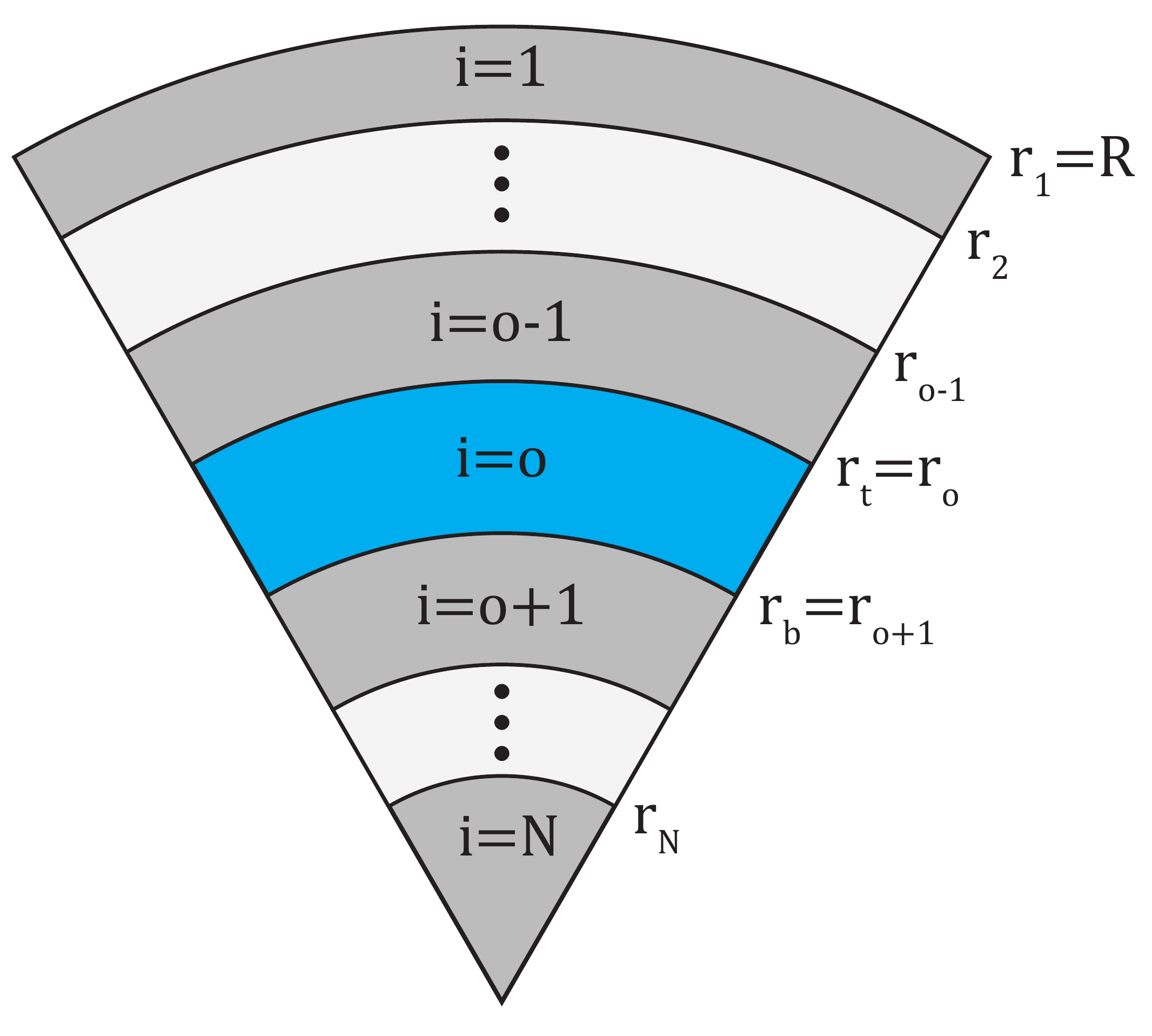}
\par\end{centering}
\caption{\label{fig:interior layers}Definition of the nomenclature used to
describe the internal layers of a satellite. Layers with $1\leq i<o-1$
are the shell overlying the ocean, the layer with $i=0$ is he ocean,
layers with $o<i\leq N$ are the layers below the ocean, and the layer
with $i=N$ is the core. }
\end{figure}

The solution vector at the core is 
\begin{equation}
\mathbf{y}^{(N)}(r_{N})=\mathbf{I}_{c}\text{\textbf{C}}_{c},\label{app: y core}
\end{equation}
where $r_{N}$ is the core radius (Fig. \ref{fig:interior layers}),
the $(N)$ superscript indicates the solution vector corresponds to
that of the core layer, $\mathbf{I}_{c}$ is given by Eq. (1.103)
of \citet{Sabadini:2004a} for a liquid core, and $\mathbf{C}_{c}$
is a constant vector. For an incompressible solid core, $\mathbf{I_{c}}$
is given by the first three columns of the matrix in Eq. (1.74) of
\citet{Sabadini:2004a}. Propagating the solution vector from the
core to the surface, the solution vector at the surface is 
\begin{equation}
\mathbf{y}^{(1)}(R)=\left(\prod_{i=1}^{N-1}\mathbf{Y}^{(i)}(r_{i})\mathbf{Y}^{(i)^{-1}}(r_{i+1})\right)\mathbf{I}_{c}\text{\textbf{C}}_{c},\label{app: yR prop}
\end{equation}
where the fundamental matrices $\mathbf{Y}^{(i)}$ and $\mathbf{Y}^{(i)^{-1}}$
are given by Eqs. (1.74) and (1.75) of \citet{Sabadini:2004a} assuming
incompressible layers. Once again, we drop the spherical harmonic
subscript (``n'' in this work, ``$\ell$'' in \citet{Sabadini:2004a})
to simplify the notation. The superscript ``$i$'' indicates that
the properties (e.g., density, gravity, shear modulus) of the layer
$i$ must be used (Fig. \ref{fig:interior layers}). There is a typographical
error in the sign of third component in Eq. (1.76) of \citet{Sabadini:2004a},
this equation should read
\[
\text{diag}(\mathbf{D}(r))=\frac{1}{2n+1}\left((n+1)r^{-n-1},\,\frac{n(n+1)}{2(2n-1)}r^{-n+1},\,r^{-n+1},\,nr^{n},\,\frac{n(n+1)}{2(2n+3)}r^{n+2},\,-r^{n+1}\right).
\]
Using the boundary conditions at the surface, 

\begin{equation}
\mathbf{P}_{1}\left(\prod_{i=1}^{N-1}\mathbf{Y}^{(i)}(r_{i})\mathbf{Y}^{(i)^{-1}}(r_{i+1})\right)\mathbf{I}_{c}\text{\textbf{C}}_{c}=\mathbf{b},\label{app: surf prop2}
\end{equation}
where 
\begin{equation}
\mathbf{P_{1}}\equiv\left(\begin{array}{cccccc}
0 & 0 & 1 & 0 & 0 & 0\\
0 & 0 & 0 & 1 & 0 & 0\\
0 & 0 & 0 & 0 & 0 & 1
\end{array}\right)\label{app: P1 matrix}
\end{equation}
is a projection matrix for the third, fourth, and sixth components
of the solution vector. Thus, 
\begin{equation}
\mathbf{C}_{c}=\left[\mathbf{P}_{1}\left(\prod_{i=1}^{N-1}\mathbf{Y}^{(i)}(r_{i})\mathbf{Y}^{(i)^{-1}}(r_{i+1})\right)\mathbf{I}_{c}\right]^{-1}\mathbf{b},\label{app: Cc vector}
\end{equation}
and the solution vector at the surface can be written as 
\begin{equation}
\mathbf{y}^{(1)}(R)=\left(\prod_{i=1}^{N-1}\mathbf{Y}^{(i)}(r_{i})\mathbf{Y}^{(i)^{-1}}(r_{i+1})\right)\mathbf{I}_{c}\left[\mathbf{P}_{1}\left(\prod_{i=1}^{N-1}\mathbf{Y}^{(i)}(r_{i})\mathbf{Y}^{(i)^{-1}}(r_{i+1})\right)\mathbf{I}_{c}\right]^{-1}\mathbf{b}.\label{app: surf yR solution}
\end{equation}

Using the boundary conditions in Eq. (\ref{app: surface bcs}), the
tidal and loading Love numbers at the satellite surface are 

\begin{align}
k(R) & =-y_{5}^{(1)}(R)-1\nonumber \\
h(R) & =g(R)y_{1}^{(1)}(R)\nonumber \\
\ell(R) & =g(R)y_{2}^{(1)}(R).\label{app:surface love numbers}
\end{align}
The pressure Love numbers are also defined by this equation, except
$k^{P}(R)=-y_{5}^{(1)}(R)$ because pressure forcing can only produce
an induced gravitational potential.

\subsection{Propagator matrix method with internal liquid layers\label{sec:Love numbers liquid layers}}

The presence of an internal liquid layer requires a special treatment
because this causes the layers above and below the liquid layer to
be mechanically decoupled while remaining gravitationally coupled.
We use the method of \citet{Jara-Orue:2011} to extend the propagator
matrix method to take into account the mechanical decoupling. In this
case, 
\begin{equation}
\left(\begin{array}{c}
y_{1}^{(1)}(R)\\
y_{2}^{(1)}(R)\\
y_{5}^{(1)}(R)
\end{array}\right)\equiv\left(\begin{array}{c}
U(R)\\
V(R)\\
-\Phi(R)
\end{array}\right)=\mathbf{P_{35}}\mathbf{W}_{2}(\mathbf{W}_{1})^{-1}\mathbf{b},\label{app: yR liq layer}
\end{equation}
where $\mathbf{P}_{35}$, $\mathbf{W}_{2}$, $\mathbf{W}_{1}$, and
$\mathbf{b}$ are defined by Eqs. (A.18), (A.19), (A.21)-(A.23), and
(B.20) of \citet{Jara-Orue:2011}. The solution vector at the ocean
top $(r_{o})$ can be found by propagating the surface solution to
the ocean top, 
\begin{equation}
\mathbf{y}^{(o-1)}(r_{o})=\left[\prod_{i=1}^{o-1}\mathbf{Y}^{(i)}(r_{i})\mathbf{Y}^{(i)^{-1}}(r_{i+1})\right]^{-1}\mathbf{y}^{(1)}(R),\label{app: surf y ocean top}
\end{equation}
where 
\begin{equation}
\mathbf{y}^{(1)}(R)=\begin{cases}
\left(U(R),\,V(R),\,0,\,0,\,-\Phi(R),\,-\frac{2n+1}{R}\right)U^{T}(R) & \mbox{tidal}\\
\left(U(R),\,V(R),\,-\frac{2n+1}{3}\bar{\rho},\,0,\,-\Phi(R),\,-\frac{2n+1}{R}\right)U^{L}(R) & \mbox{surface loading}\\
\left(U(R),\,V(R),\,\bar{\rho},\,0,\,-\Phi(R),\,0\right)U^{P}(R) & \mbox{surface pressure,}
\end{cases}\label{app: surf yR full solutions}
\end{equation}
where $U^{T}$, $U^{L}$, and $U^{P}$ are the tidal, surface loading,
and surface pressure loading potentials. The sign convention for pressure
forcing implies that a positive pressure potential leads to a positive
radial displacement. As described above for the case without internal
liquid layers, we drop the spherical harmonic subscript (``n'' in
this work, ``$\ell$'' in \citet{Jara-Orue:2011}) to simplify the
notation.

The solution vector at the ocean bottom ($r=r_{b}$, Fig. \ref{fig:interior layers})
can be found using the $\mathbf{C}_{icy}$ vector containing constant
terms \citep[Eq. A.20]{Jara-Orue:2011} to obtain the 3-component
constant vector $\mathbf{C}_{c}$ for the solution vector at the core:
\begin{equation}
\mathbf{C}_{icy}\equiv(K_{4},\,K_{5},\,K_{1},\,K_{2},\,K_{3})=(\mathbf{W}_{1})^{-1}\mathbf{b},\label{app: surf C icy}
\end{equation}
 
\begin{equation}
\mathbf{C}_{c}\equiv(K_{1},\,K_{2},\,K_{3}),\label{app: surf liq Ccore}
\end{equation}
and 
\begin{equation}
\mathbf{y}^{(o+1)}(r_{b})=\left(\prod_{i=o+1}^{N-1}\mathbf{Y}^{(i)}(r_{i})\mathbf{Y}^{(i)^{-1}}(r_{i+1})\right)\mathbf{I_{c}C_{c}}.\label{app: y ocean bottom}
\end{equation}
For a single core layer beneath the ocean, $\mathbf{y}^{(o+1)}(r_{b})=\mathbf{I_{c}C_{c}}$. 

Using the boundary conditions in Eq. (\ref{app: surface bcs}), the
tidal and loading Love numbers are 

\begin{align}
k(r) & =-y_{5}^{(i)}(r)-(r/R)^{n}\nonumber \\
h(r) & =g(R)y_{1}^{(i)}(r)\nonumber \\
\ell(r) & =g(R)y_{2}^{(i)}(r),\label{app:interior love numbers}
\end{align}
where $i=1$ and $r=R$ for the surface, $i=o-1$ and $r=r_{t}$ for
the ocean top, and $i=o+1$ and $r=r_{b}$ for the ocean bottom (Fig.
\ref{fig:interior layers}). Once again, the pressure Love numbers
are also defined by this equation, except $k(r)=-y_{5}^{(i)}(r)$
because pressure forcing can only produce an induced gravitational
potential.

\subsection{Interior pressure forcing in a subsurface ocean\label{sec:Interior pressure Love numbers liquid}}

As discussed in the main text, we introduce pressure Love numbers
to describe the dynamic pressure forcing in the ocean and couple the
LTE with the Love number equations (the mass and momentum conservation
equations and Poisson's equation). Computing the Love numbers due
to interior pressure forcing in a subsurface ocean requires further
modifications to the method of \citet{Jara-Orue:2011}. 

Dynamic pressure forcing in the ocean changes the boundary conditions
at the ocean top and bottom. For pressure forcing at the ocean top
($r=r_{t}$), the third component of Eqs (A.12) in \citet{Jara-Orue:2011}
becomes
\begin{equation}
\sigma_{rr}^{(o-1)}(r_{t})=\rho_{o}g(r_{t})K_{4}-\rho_{o}U^{P}(r_{t}),\label{app:ocean_pressure_pert_top}
\end{equation}
where $i=o-1$ corresponds to the layer overlying the ocean (Fig.
\ref{fig:interior layers}). For pressure forcing at the ocean bottom
($r=r_{b}$), the third component of Eqs (A.14) in \citet{Jara-Orue:2011}
becomes 
\begin{equation}
\sigma_{rr}^{(o+1)}(r_{b})=-\rho_{o}g(r_{t})K_{6}+\rho_{o}U^{P}(r_{b}),\label{app:ocean_pressure_pert_bot}
\end{equation}
where $i=o+1$ corresponds to the layer below the ocean (Fig. \ref{fig:interior layers}).
The sign convention for the pressure load $U^{P}$ implies that a
positive pressure potential leads to a positive radial displacement.
In the thin ocean limit, 
\begin{equation}
U^{P}(r_{b})=-U^{P}(r_{t}),\label{app:ocean_pressure_top_bot}
\end{equation}
and Eqs. (\ref{app:ocean_pressure_pert_top}) and (\ref{app:ocean_pressure_pert_bot})
can be written as a function of a single forcing term evaluated at
the ocean top or bottom. This allow us to describe ocean pressure
forcing in terms of a single Love number evaluated at the ocean top
or bottom instead of the traditional approach describing each forcing
with a different Love number, as described below.

\subsubsection{Interior pressure forcing at the ocean top}

Dynamic pressure forcing at the ocean top can be taken into account
by modifying the boundary condition in Eq. (A.20) of \citet{Jara-Orue:2011},
\begin{equation}
\mathbf{b}+\mathbf{b}^{t}=W_{1}C_{icy},\label{app:ocean forcing top1}
\end{equation}
where $\mathbf{b}=(y_{3}(R),\,y_{4}(R),\,y_{6}(R))=(0,\,0,\,0)$ for
interior pressure forcing and 
\begin{equation}
\mathbf{b^{t}=}\rho_{o}U^{P}(r_{t})\left(0,\,0,\,B_{33}^{si},\,B_{43}^{si},\,B_{63}^{si}\right).\label{app:ocean forcing top2}
\end{equation}
With this modified boundary condition, the constants vector $C_{icy}=(K_{4},\,K_{5},\,K_{1},\,K_{2},\,K_{3})=W_{1}^{-1}(\mathbf{b}+\mathbf{b}^{t})$,
the core constants vector $C_{c}\equiv(K_{1},\,K_{2},\,K_{3})$.

The unconstrained solution at the surface (Eq. (B.19) of \citet{Jara-Orue:2011})
becomes 
\begin{equation}
\mathbf{X}=P_{35}W_{2}C_{icy}-\rho_{o}U^{P}(r_{t})\left(\begin{array}{c}
B_{13}^{si}\\
B_{23}^{si}\\
B_{53}^{si}
\end{array}\right),\label{app:ocean forcing top X}
\end{equation}
where the ice shell propagation matrix is 
\begin{equation}
\mathbf{B}^{si}=\prod_{i=1}^{o-1}\mathbf{Y}^{(i)}(r_{i})\mathbf{Y}^{(i)^{-1}}(r_{i+1}).\label{app:ocean forcing top Bsi}
\end{equation}
The solution vector at the surface is given by $\mathbf{y}^{(1)}(R)=(X_{1},\,X_{2},\,0,\,0,\,X_{3},\,0)$,
the solution vector at the ocean top, $\mathbf{y}^{(o-1)}(r_{t})$,
can be obtained by propagating the surface solution through the shell
(Eq. (\ref{app: surf y ocean top})), and the solution vector at the
ocean bottom is given by Eq. (\ref{app: y ocean bottom}) or $\mathbf{y}^{(o+1)}(r_{o+1})=\mathbf{I_{c}C_{c}}$
for a single core layer beneath the ocean. 

Given the solution vector at a radius $r$, the corresponding Love
numbers are given by Eq. (\ref{app:interior love numbers}), except
$k(r)=-y_{5}^{(i)}(r)$ because pressure forcing can only produce
an induced gravitational potential.

\subsubsection{Interior pressure forcing at the ocean bottom}

Dynamic pressure forcing at the ocean top can be taken into account
by modifying the boundary condition in Eq. (A.20) of \citet{Jara-Orue:2011},
\begin{equation}
\mathbf{b}+\mathbf{b}^{b}=W_{1}C_{icy},\label{app:ocean_forcing bottom1}
\end{equation}
where $\mathbf{b}=(y_{3}(R),\,y_{4}(R),\,y_{6}(R))=(0,\,0,\,0)$ for
interior pressure forcing, 

\begin{equation}
\mathbf{b^{b}=}\rho_{o}U^{P}(r_{b})\left(-\frac{1}{\rho_{o}g(r_{b})},\,0,\,A_{1},\,A_{2},\,A_{3}\right),\label{app:pressure forcing bottom2}
\end{equation}
 we define
\begin{equation}
A_{i}\equiv\frac{4\pi G}{g(r_{b})}\left\{ B_{12}^{f}\left[\frac{B_{i1}^{R1}}{g(r_{t})}-B_{i2}^{R1}-B_{i3}^{R1}\left(\frac{n+1}{r_{t}}-\frac{4\pi G\rho_{o}}{g(r_{t})}\right)\right]-B_{22}^{f}B_{i3}^{R1}\right\} ,\label{app:ocean_forcing_Ai bottom}
\end{equation}
and $\mathbf{B}^{f}$ and $\mathbf{B}^{R1}$ are given by Eqs. (A.11)
and (B.5) of \citet{Jara-Orue:2011}. The first component in Eq. (\ref{app:pressure forcing bottom2})
arises from the change of Eq. (B.15) of \citet{Jara-Orue:2011} to
\begin{equation}
-\frac{U^{P}(r_{b})}{g(r_{b})}=L_{i}C_{c,\,i},\label{app:pressure forcing bottom3}
\end{equation}
where we adopt the Einstein summation convention for the subscript
$i=1,2,\,3$ and $L_{i}$ is given by Eq. (B.16) of \citet{Jara-Orue:2011}.
To obtain this modified equation, we use $\sigma_{rr}^{(o+1)}(r_{b})=B_{3i}^{sm}C_{c,\,i}$,
$\Phi^{(o+1)}(r_{b})=B_{5i}^{sm}C_{c,\,i}$, and $U^{(o+1)}(r_{b})=B_{1i}^{sm}C_{c,\,i}$,
where the propagator matrix
\begin{equation}
\mathbf{B}^{sm}\equiv\prod_{i=o+1}^{N-1}\mathbf{Y}^{(i)}(r_{i})\mathbf{Y}^{(i)^{-1}}(r_{i+1})\mathbf{I_{c}}.\label{app:pressure forcing bottom Bsm}
\end{equation}
For a single core layer beneath the ocean, $\mathbf{B}^{sm}=\mathbf{I}_{c}$,
and is given by the first three columns of the matrix in Eq. (1.74)
of \citet{Sabadini:2004a}. Using the modified boundary condition,
the constants vector $C_{icy}=(K_{4},\,K_{5},\,K_{1},\,K_{2},\,K_{3})=W_{1}^{-1}(\mathbf{b}+\mathbf{b}^{b})$,
the core constants vector $C_{c}\equiv(K_{1},\,K_{2},\,K_{3})$. 

The unconstrained solution at the surface (Eq. (B.19) of \citet{Jara-Orue:2011})
becomes 
\begin{equation}
\mathbf{X}=P_{35}W_{2}C_{icy}-\rho_{o}U^{P}(r_{o+1})\left(\begin{array}{c}
A_{1}^{\prime}\\
A_{2}^{\prime}\\
A_{3}^{\prime}
\end{array}\right),\label{app:ocean forcing bottom5}
\end{equation}
where we define 
\begin{equation}
A_{i}^{\prime}\equiv\frac{4\pi G}{g(r_{b})}\left\{ B_{12}^{f}\left[\frac{B_{i1}^{R2}}{g(r_{t})}-B_{i2}^{R2}-B_{i3}^{R2}\left(\frac{n+1}{r_{t}}-\frac{4\pi G\rho_{o}}{g(r_{t})}\right)\right]-B_{22}^{f}B_{i3}^{R2}\right\} \label{app:ocean forcing bottom Aiprime}
\end{equation}
and $\mathbf{B}^{R2}$ is given by Eq. (B.22) of \citet{Jara-Orue:2011}.
The solution vector at the surface is given by $\mathbf{y}^{(1)}(R)=(X_{1},\,X_{2},\,0,\,0,\,X_{3},\,0)$,
the solution vector at the ocean top, $\mathbf{y}^{(o-1)}(r_{t})$,
can be obtained by propagating the surface solution through the shell
(Eq. (\ref{app: surf y ocean top})), and the solution vector at the
ocean bottom is given by Eq. (\ref{app: y ocean bottom}) or $\mathbf{y}^{(o+1)}(r_{o+1})=\mathbf{I_{c}C_{c}}$
for a single core layer beneath the ocean. 

\subsubsection{Interior pressure forcing at the ocean top and bottom in a thin ocean}

In the thin ocean limit, $U^{P}(r_{b})=-U^{P}(r_{t})$, which allows
us to describing dynamic pressure forcing at both the ocean top and
bottom using a single Love number instead of the traditional approach
of describing each forcing with a separate Love number, and we choose
$U^{P}(r_{t})$ as the reference pressure potential. 

Taking into account pressure forcing at both the ocean top and bottom
in the boundary conditions, Eq. (A.20) of \citet{Jara-Orue:2011}
becomes

\begin{equation}
\mathbf{b}+\mathbf{b}'=W_{1}C_{icy},\label{app:ocean_forcing1}
\end{equation}
where $\mathbf{b}=(y_{3}(R),\,y_{4}(R),\,y_{6}(R))=(0,\,0,\,0)$ for
interior pressure forcing, 

\begin{equation}
\mathbf{b^{\prime}\equiv}\mathbf{b}^{t}+\mathbf{b}^{b}=\rho_{o}U^{P}(r_{o})\left(\begin{array}{c}
1/(\rho_{o}g(r_{b}))\\
0\\
B_{33}^{si}-A_{1}\\
B_{43}^{si}-A_{2}\\
B_{63}^{si}-A_{3}
\end{array}\right),\label{app:ocean_forcing_b}
\end{equation}
and $\mathbf{B}^{si}$ and $A_{i}$ is given by Eqs. (\ref{app:ocean forcing top Bsi})
and (\ref{app:ocean_forcing_Ai bottom}). Using the modified boundary
condition, the constants vector
\begin{equation}
C_{icy}=(K_{4},\,K_{5},\,K_{1},\,K_{2},\,K_{3})=W_{1}^{-1}(\mathbf{b}+\mathbf{b}^{\prime}),\label{app:ocean_forcing_Cicy}
\end{equation}
the core constants vector 
\begin{equation}
C_{c}\equiv(K_{1},\,K_{2},\,K_{3}).\label{app:ocean forcing Ccore}
\end{equation}

The unconstrained solution at the surface (Eq. (B.19) of \citet{Jara-Orue:2011})
becomes 
\[
\mathbf{X}=P_{35}W_{2}C_{icy}+\rho_{o}U^{P}(r_{o})\left(\begin{array}{c}
A_{1}^{\prime}-B_{13}^{si}\\
A_{2}^{\prime}-B_{23}^{si}\\
A_{3}^{\prime}-B_{53}^{si}
\end{array}\right),
\]
where $A_{i}^{\prime}$ is given by Eq. (\ref{app:ocean forcing bottom Aiprime}).
The solution vector at the surface is
\[
\mathbf{y}^{(1)}(R)=(X_{1},\,X_{2},\,0,\,0,\,X_{3},\,0),
\]
the solution vector at the ocean top, $\mathbf{y}^{(o-1)}(r_{t})$,
can be obtained by propagating the surface solution through the shell
(Eq. (\ref{app: surf y ocean top})), and the solution vector at the
ocean bottom is given by Eq. (\ref{app: y ocean bottom}) or $\mathbf{y}^{(o+1)}(r_{b})=\mathbf{I_{c}C_{c}}$
for a single core layer beneath the ocean. The pressure Love numbers
describing pressure forcing at both the ocean top and bottom in a
thin ocean are

\begin{align}
k(r) & =-y_{5}^{(i)}(r)\nonumber \\
h(r) & =g(R)y_{1}^{(i)}(r)\nonumber \\
\ell(r) & =g(R)y_{2}^{(i)}(r),\label{app:interior love numbers pressure}
\end{align}
where $i=1$ and $r=R$ for the surface, $i=o-1$ and $r=r_{t}$ for
the ocean top, and $i=o+1$ and $r=r_{b}$ for the ocean bottom (Fig.
\ref{fig:interior layers}).

\begin{figure}[h]
\begin{centering}
\includegraphics[width=11cm]{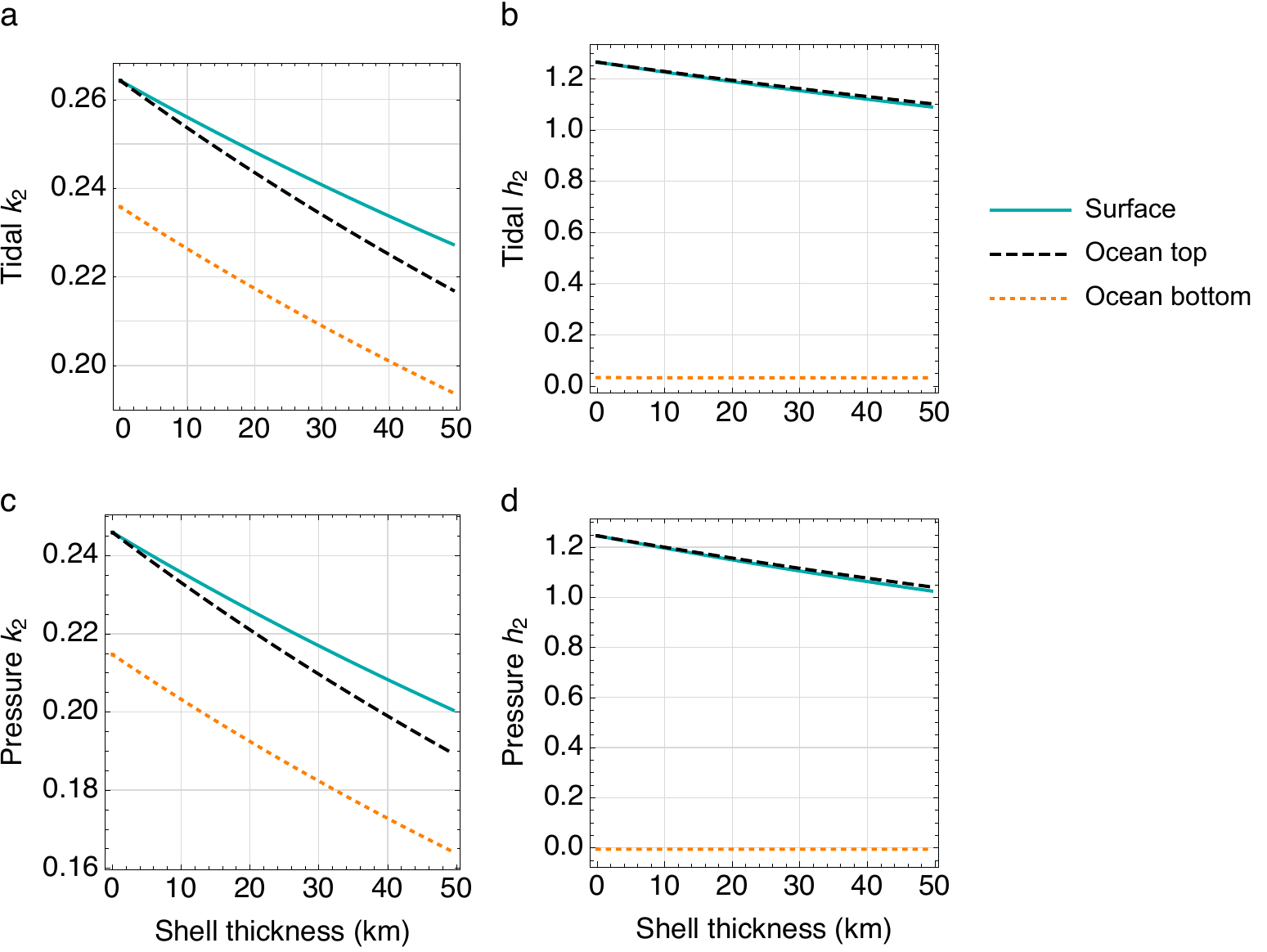}
\par\end{centering}
\caption{\label{fig:Europa Love numbers}Europa's tidal and pressure $k_{2}$
and $h_{2}$ Love numbers as a function of shell thickness. We assume
a three-layer interior structure with the parameters in Table \ref{tab:Interior parameters}.
The core density is calculated self-consistently so as to satisfy
the mean density constraint ($\bar{\rho}=3.013$ g cm$^{-3}$). }
\end{figure}

\begin{figure}[h]
\begin{centering}
\includegraphics[width=11cm]{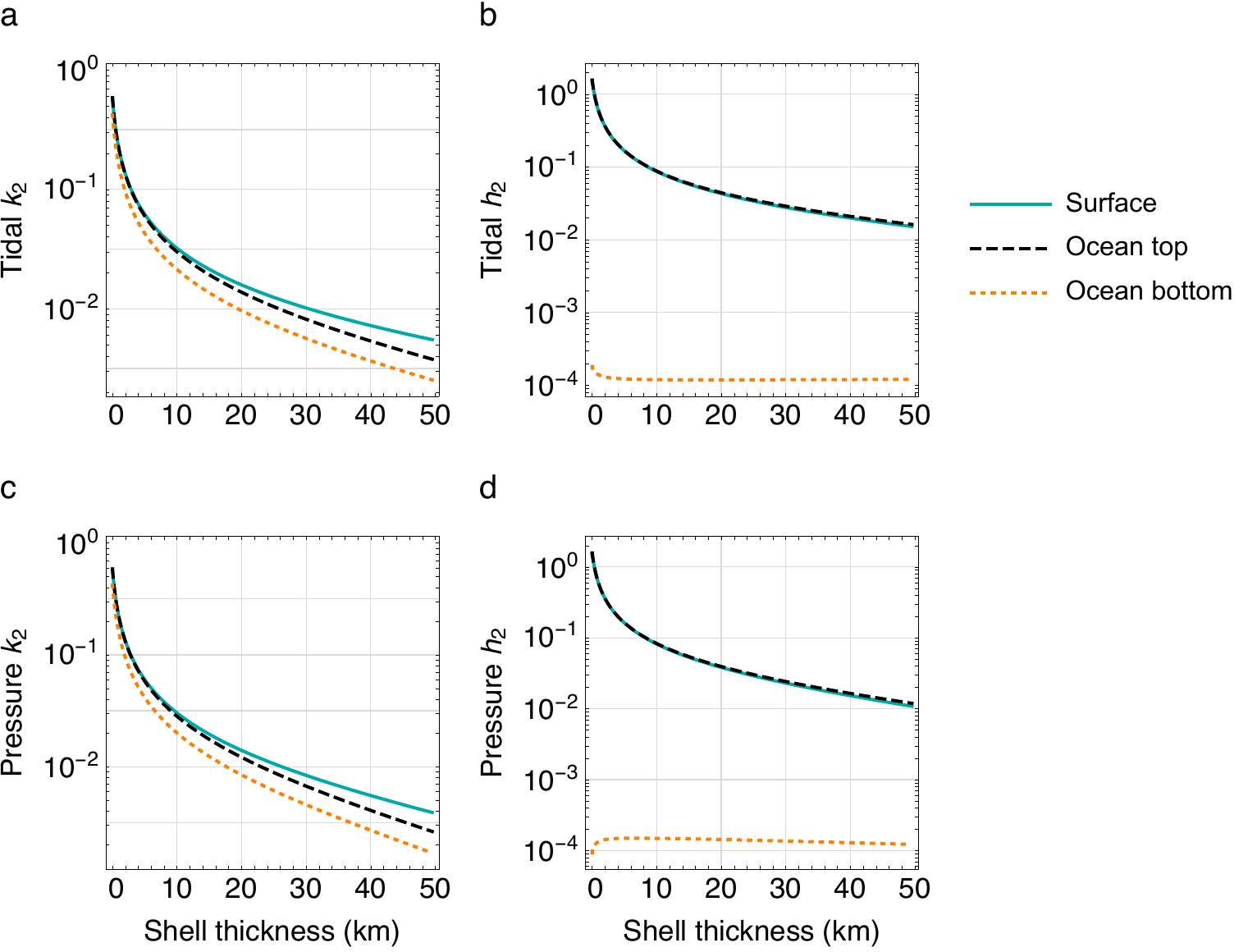}
\par\end{centering}
\caption{\label{fig:Enceladus Love numbers}Enceladus' tidal and pressure $k_{2}$
and $h_{2}$ Love numbers as a function of shell thickness. We assume
a three-layer interior structure with the parameters in Table \ref{tab:Interior parameters}.
The core density is calculated self-consistently so as to satisfy
the mean density constraint ($\bar{\rho}=1.609$ g cm$^{-3}$). In
panel d, the displacement pressure Love number at the ocean bottom,
$h_{2}^{P}(r_{b})$, is negative because a positive pressure potential
at the ocean bottom corresponds to a negative radial displacement,
and the dotted orange line corresponds to $|h_{2}^{P}(r_{b})|$. }
\end{figure}

Figs. \ref{fig:Europa Love numbers} and \ref{fig:Enceladus Love numbers}
illustrate the effect of mechanical decoupling due to a subsurface
ocean. The ocean top and bottom remain gravitationally coupled, and
therefore the $k_{2}$ Love numbers describing gravitational perturbations
are nearly identical at these radii for thin shells and remain of
the same order of magnitude for thick shells (Figs. \ref{fig:Europa Love numbers}a,
c; Figs. \ref{fig:Enceladus Love numbers}a, c). On the other hand,
the ocean top and bottom are mechanically decoupled, and thus the
$h_{2}$ Love numbers describing radial displacements differ by orders
of magnitude at these radii (Figs. \ref{fig:Enceladus Love numbers}b,
d; Figs. \ref{fig:Europa Love numbers}b, d). The Love numbers are
not sensitive to the assumed ocean thickness.


\section{Analytic Love number solutions}

The propagator matrix method can be used to calculate the Love numbers
of bodies with arbitrary interior structures. Although simplified
models are less realistic, the analytic solutions for these models
provide valuable physical insight. 

\subsection{Homogeneous interior\label{sec:Homogeneous interior Love numbers}}

Assuming a uniform, incompressible, interior with shear modulus $\mu$
and density $\rho$, the solution vector at the surface is $\mathbf{y}(R)=\mathbf{I}_{c}\mathbf{C}_{c}$,
where $\mathbf{I_{c}}$ is given by the first three columns of the
matrix in Eq. (1.74) of \citet{Sabadini:2004a}, and the boundary
conditions at the surface yield $\mathbf{C}_{c}=(\mathbf{P}_{1}\mathbf{I}_{c})^{-1}\mathbf{b}$.
Thus, the solution vector at the surface can be written as 
\begin{equation}
\mathbf{y}(R)=\mathbf{I}_{c}(\mathbf{P}_{1}\mathbf{I}_{c})^{-1}\mathbf{b}.\label{app: yR unifrom interior}
\end{equation}
Using the boundary conditions (\ref{app: surface bcs}) and the Love
number definitions (\ref{app:surface love numbers}), 
\begin{align}
(k_{n}^{T},\,h_{n}^{T},\,\ell_{n}^{T}) & =\frac{1}{1+\hat{\mu}}\left\{ \frac{3}{2(n-1)},\,\frac{2n+1}{2(n-1)},\,\frac{3}{2n(n-1)}\right\} \nonumber \\
(k_{n}^{L},\,h_{n}^{L},\,\ell_{n}^{L}) & =-\frac{1}{1+\hat{\mu}}\left\{ 1,\,\frac{2n+1}{3},\,\frac{1}{n}\right\} \nonumber \\
(k_{n}^{P},\,h_{n}^{P},\,\ell_{n}^{P}) & =(k_{n}^{T},\,h_{n}^{T},\,\ell_{n}^{T})
\end{align}
where 
\begin{equation}
\hat{\mu}\equiv\frac{2n^{2}+4n+3}{n}\frac{\mu}{\rho gR}\label{T1:mu_effective}
\end{equation}
is a dimensionless effective rigidity. Note that the pressure Love
numbers here describe the deformation in response to pressure forcing
at the surface of the satellite. The expressions for the tidal and
load Love numbers are consistent with those given by Eqs. (2.1.9a)-(2.1.9c)
of \citet{Lambeck:1980earth} for $n=2$. We note that Eqs. (5.6.1)
and (5.7.1) in \citet{Munk:1960} contain typographical errors for
the $\ell$ Love numbers.

\subsection{Homogeneous thick shell\label{subsec:Homogeneous-thick-shell}}

An incompressible 3-layer body consisting of a core, ocean, and shell
is a good toy model for icy satellites with subsurface oceans. If
the ocean and shell have the same density, gravitational effects that
change with shell thickness can be avoided, providing analytic Love
number solutions for arbitrary shell thicknesses. Assuming static
tides and a core with infinity rigidity (i.e. non-deformable) \citep[Eqs. (62) and  (L.2)]{Beuthe:2018},
\begin{align}
h_{n}^{T} & =\left(\frac{1}{1-\xi_{n}}\right)\left[1+\left(\frac{1}{1-\xi_{n}}\right)z_{h}\frac{\mu}{\rho_{o}gR}\right]^{-1}\label{hom_thick_shell:ht}
\end{align}
where the degree-\emph{n} density ratio $\xi_{n}$ is given by Eq
(\ref{surface ocean: xi definition}), $g$ is the surface gravity,
and $\mu$ is the shear modulus of the shell. The dimensionless factors
$z_{h}$ and $z_{\ell}$ depend on $d/R$, where $d$ is the shell
thickness \citep[Eq. (L.5)]{Beuthe:2018}. For a fluid shell, $\mu=0$
and $h_{n}^{T}=1/(1-\xi_{n})$. At degree 2, $z_{h}$ decreases with
shell thickness from $19/5$ for a homogeneous interior to $0$ for
the membrane limit of the shell, with an asymptotic behavior $\sim\frac{24}{11}\frac{d}{R}$
\citep[Fig. 13]{Beuthe:2015}. Thus, the concept of the effective
rigidity can be generalized from a homogenous body to a body in which
a global subsurface ocean decouples the shell from the core if Eq.
(\ref{T1:mu_effective}) is modified to 
\begin{equation}
\hat{\mu}\equiv\left(\frac{1}{1-\xi_{n}}\right)z_{h}\frac{\mu}{\rho_{o}gR}.\label{hom_thick_shell:effective mu}
\end{equation}
In the homogeneous body limit, $\frac{1}{1-\xi_{n}}=\frac{2n+1}{2(n-1)}$,
$z_{h}=\frac{2(n-1)}{2n+1}\frac{2n^{2}+4n+3}{n}$ \citep[Eq. (L.6)]{Beuthe:2018},
and Eq. (\ref{hom_thick_shell:effective mu}) reduces to Eq. (\ref{T1:mu_effective}),
as expected. If the shell is thin ($d/R\lesssim0.2$), the effective
rigidity decreases nearly linearly with shell thickness: $z_{h}\sim\frac{6(n-1)(n+2)}{2(n-1)(n+2)+3}\frac{d}{R}$.
At degree 2, this approximation underestimates $z_{h}$ by (1, 6,
17)\% if the shell thickness is (2, 10, 20)\% of the surface radius,
respectively. With this definition of the effective rigidity, it is
clear that Europa behaves as a `soft shell' body ($\hat{\mu}\lesssim1$)
whereas Enceladus behaves as a `hard shell' body ($\hat{\mu}\gg1$)
\citep[Section 4.3.2]{Beuthe:2018}.

\section{Numerical solution of the Laplace tidal equations\label{sec:Numerical-solution}}

We extend the method of \citet{Matsuyama:2014}, based on the method
of \citet{Longuet-Higgins:1968}, to solve the mass and momentum conservation
equations (Eqs. (\ref{eq:mass cons.}) and (\ref{sub ocean: momentum eq.}))
for ocean tides with an overlying solid shell and no bottom or Navier-Stokes
drag ($c_{D}=\nu=0$).

The velocity is specified using a Helmholtz decomposition \citep[Section 1.16]{Arfken:1995},
\begin{equation}
\mathbf{u}=\nabla\Phi+\nabla\times(\Psi\mathbf{\hat{\mathbf{e}}_{r}})=\hat{\mathbf{e}}_{\theta}\left[r^{-1}\partial_{\theta}\Phi+(r\sin\theta)^{-1}\partial_{\phi}\Psi\right]+\hat{\mathbf{e}}_{\phi}\left[(r\sin\theta)^{-1}\partial_{\phi}\Phi-r^{-1}\partial_{\theta}\Psi\right],\label{eq:Helmoltz}
\end{equation}
where $\Phi$ has the properties of a potential and $\Psi$ has the
properties of a stream function. We expand the forcing potential $U^{T},$
radial tide $\eta$, $\Phi$, and $\Psi$ in spherical harmonics as
\begin{align}
\eta(r,\,\theta,\,\phi) & =\frac{1}{2}\sum_{m=0}^{2}\sum_{n=m}^{\infty}\eta_{nm}(r)Y_{nm}(\theta,\,\phi)e^{-i\omega t}+c.c.\nonumber \\
U^{T}(R,\,\theta,\,\phi) & =\frac{1}{2}\sum_{m=0}^{2}\sum_{n=m}^{\infty}U_{nm}^{T}(R)Y_{nm}(\theta,\,\phi)e^{-i\omega t}+c.c.\nonumber \\
\Phi(r,\,\theta,\,\phi) & =\frac{1}{2}\sum_{m=0}^{2}\sum_{n=m}^{\infty}\Phi_{nm}(r)Y_{nm}(\theta,\,\phi)e^{-i\omega t}+c.c.\nonumber \\
\Psi(r,\,\theta,\,\phi) & =\frac{1}{2}\sum_{m=0}^{2}\sum_{n=m}^{\infty}\Psi_{nm}(r)Y_{nm}(\theta,\,\phi)e^{-i\omega t}+c.c.,\label{eq:harmonics_expansions}
\end{align}
where $\omega=\Omega$ and $\omega=-\Omega$ for the eastward and
westward components respectively, 
\begin{equation}
Y_{nm}(\theta,\,\phi)\equiv P_{nm}(\cos\theta)e^{im\phi},\label{eq:Ylm}
\end{equation}
$P_{nm}$ are the unnormalized associated Legendre functions define
by the Legendre polynomials equation \citep[Eq. (12.81)]{Arfken:1995}

\begin{equation}
P_{nm}(x)=\left(1-x^{2}\right)^{m/2}\frac{d^{m}}{dx^{m}}P_{n}(x),\label{eq:Pnm}
\end{equation}
and $c.c.$ denotes the complex conjugate. Note that this definition
of does not include the Condon-Shortley phase factor of $(-1)^{m}$.
The nonzero forcing potential coefficients are listed on Table \ref{tab:forcing_coef}. 

\begin{table}[h]
 
\begin{centering}
\resizebox{!}{!}{ 
\begin{tabular}{llccccccc}
\toprule

 &  
 
 & \multicolumn{3}{c}{ 
\textbf{Obliquity tide} 
} &  
 
 & \multicolumn{3}{c}{ 
\textbf{Eccentricity tide} 
}\tabularnewline
\midrule 
 
\emph{m} 
 &  
 
 &  
Eastward 
 &  
 
 &  
Westward 
 &  
 
 &  
Eastward 
 &  
 
 &  
Westward 
\tabularnewline
\cmidrule{1-1} \cmidrule{3-3} \cmidrule{5-5} \cmidrule{7-7} \cmidrule{9-9} 
 
0 
 &  
 
 &  
- 
 &  
 
 &  
- 
 &  
 
 &  
$-(3/2)\Omega^{2}r^{2}e$ 
 &  
 
 &  
- 
\tabularnewline
 
1 
 &  
 
 &  
$(1/2)\Omega^{2}r^{2}\theta_{0}$ 
 &  
 
 &  
$(1/2)\Omega^{2}r^{2}\theta_{0}$ 
 &  
 
 &  
- 
 &  
 
 &  
- 
\tabularnewline
 
2 
 &  
 
 &  
- 
 &  
 
 &  
- 
 &  
 
 &  
$(7/8)\Omega^{2}r^{2}e$ 
 &  
 
 &  
$-(1/8)\Omega^{2}r^{2}e$ 
\tabularnewline
\bottomrule
\end{tabular} }
\par\end{centering}
 
\caption{\label{tab:forcing_coef}Tidal potential expansion coefficients for
obliquity ($\theta_{0}$) and eccentricity ($e$) tides (compare Eqs.
(\ref{eq:U_ecc_obliq}) and (\ref{eq:harmonics_expansions})). Note
that the decomposition into eastward and westward components is irrelevant
for $m=0$ and only one coefficient should be used in this case.}
\end{table}

Replacing Eqs. (\ref{eq:Helmoltz}) and (\ref{eq:harmonics_expansions})
in the LTE and following the procedure outlined in \citep{Longuet-Higgins:1968}
yields
\begin{align}
\frac{\upsilon_{n}}{2\Omega}U_{nm}^{T} & =K_{n}\Im(\Phi_{nm})+b\Re(\Phi_{nm})+p_{\ell+1}\Re(\Psi_{n+1,\,m})+q_{n-1}\Re(\Psi_{n-1,\,m})\nonumber \\
0 & =K_{n}\Re(\Phi_{nm})-b\Im(\Phi_{nm})-p_{\ell+1}\Im(\Psi_{n+1,\,m})-q_{n-1}\Im(\Psi_{n-1,\,m})\nonumber \\
0 & =L_{n}\Im(\Psi_{nm})+b\Re(\Psi_{nm})-p_{\ell+1}\Re(\Phi_{n+1,\,m})-q_{n-1}\Re(\Phi_{n-1,\,m})\nonumber \\
0 & =L_{n}\Re(\Psi_{nm})-b\Im(\Psi_{nm})+p_{\ell+1}\Im(\Phi_{n+1,\,m})+q_{n-1}\Im(\Phi_{n-1,\,m}),\label{numsol:matrices1}
\end{align}
where $\Re$ and $\Im$ denote the real and imaginary parts, 
\begin{align}
b & \equiv\frac{\alpha}{2\Omega}\nonumber \\
K_{n} & \equiv\lambda+\frac{m}{n(n+1)}-\beta{}_{n}\frac{n(n+1)}{\epsilon\lambda}\nonumber \\
L_{n} & \equiv\lambda+\frac{m}{n(n+1)}\nonumber \\
p_{n} & \equiv\frac{(n+1)(n+m)}{n(2n+1)}\nonumber \\
q_{n} & \equiv\frac{n(n+1-m)}{(n+1)(2n+1)},\label{musol:matrices_definitions}
\end{align}
$\lambda\equiv\omega/(2\alpha)$, and 
\begin{equation}
\epsilon(r)\equiv\frac{4\Omega^{2}r^{2}}{g(R)h_{o}}\label{eq:lamb_parameters}
\end{equation}
is Lamb's parameter. In Eqs. (\ref{numsol:matrices1}) and (\ref{musol:matrices_definitions}),
$\upsilon_{n}$ and $\beta_{n}$ are given by Eq. (\ref{sub ocean: nu beta}).
The surface ocean solutions can be obtained by replacing $\upsilon_{n}$
and $\beta_{n}$ by $\gamma_{n}^{T}$ and $1-\xi_{2}\gamma_{n}^{L}$
(Eq. (\ref{surface ocean: tilt factors})), respectively, and evaluating
Lamb's parameter at the surface ($r=R$). Eq. (\ref{numsol:matrices1})
can be written in the more compact form

\begin{align}
\frac{\upsilon_{n}}{2\Omega}U_{nm}^{T}(R)= & -iK_{n}\Phi_{nm}(r)+b\Phi_{nm}(r)+p_{\ell+1}\Psi_{n+1,\,m}(r)+q_{n-1}\Psi_{n-1,\,m}(r)\nonumber \\
0= & -iL_{\ell}\Psi_{nm}(r)+b\Psi_{nm}(r)-p_{\ell+1}\Phi_{n+1,\,m}(r)-q_{n-1}\Phi_{n-1,\,m}(r),\label{numsol:matrices}
\end{align}
which is slightly different from Eq. (A.5) of \citet{Matsuyama:2014}
due to typographical errors in that paper. 

The resonant ocean modes are given by the eigenvalues of the matrices
describing Eq. (\ref{numsol:matrices}). Given the solutions for $\Phi$
and $\Psi$ by solving Eq. (\ref{numsol:matrices}), the velocity
is given by the Helmholtz decomposition (\ref{eq:Helmoltz}), where
we can use \citep[p. 725, Eq. (12.87)]{Arfken:1995}
\begin{equation}
\partial_{\theta}P_{nm}=\frac{1}{2}(n+m)(n+1-m)P_{n,\,m-1}-\frac{1}{2}P_{n,\,m+1}\label{eq:Pnm_derivative}
\end{equation}
to evaluate the $\partial_{\theta}$ derivatives. The radial tide
can be found using the mass mass conservation Eq. (\ref{eq:mass cons.})
and the Helmholtz decomposition (\ref{eq:Helmoltz}), $i\omega\eta_{nm}=(n(n+1)/r^{2})(h/\omega)\Phi_{nm},$
or 
\begin{align}
\Re(\eta_{nm}(r)) & =-\frac{n(n+1)}{r^{2}}\frac{h_{o}}{\omega}\Im(\Phi_{nm}(r))\nonumber \\
\Im(\eta_{nm}(r)) & =\frac{n(n+1)}{r^{2}}\frac{h_{o}}{\omega}\Re(\Phi_{nm}(r)).\label{nusol:tide}
\end{align}

Given a velocity solution of the Laplace tidal equation, the dissipated
energy per unit time and surface area can be found with Eq. (\ref{eq:dissipated flux}).
Integrating this equation over the tidal forcing period and the satellite
surface assuming a thin ocean yields the time- and surface-averaged
power, 
\begin{equation}
<\dot{E}_{diss}>\equiv r^{2}\int_{0}^{T}dt\int_{0}^{\pi}\int_{0}^{2\pi}d\phi\,d\theta\sin\theta\,F_{diss}=-2\pi\rho_{o}h_{o}\alpha\sum_{m=0}^{2}\sum_{n=m}^{\infty}N_{nm}\left(|\Phi_{nm}(r)|^{2}+|\Psi_{nm}(r)|^{2}\right),\label{numsol:power1}
\end{equation}
where 
\begin{equation}
N_{nm}\equiv\frac{n(n+1)}{2n+1}\frac{(n+m)!}{(n-m)!}\label{numsol:harmonics_normalization}
\end{equation}
is a normalization constant and the radius can be evaluated at the
ocean top ($r=r_{t}$) or bottom ($r=r_{b}$) in the thin ocean limit.
Energy conservation requires that the time- and surface-averaged dissipated
energy and work done by the tide to be equal \citep{Tyler:2011,Chen:2014},
which allows us to derive the alternative expression
\begin{equation}
<\dot{E}_{diss}>=-2\pi\rho_{o}h_{o}\upsilon_{2}\sum_{m=0}^{2}N_{2m}U_{2m}^{T}(R)\Re(\Phi_{2m}(r)).\label{numsol:power2}
\end{equation}
Thus, we can verify that our solutions satisfy energy conservation
by comparing energy dissipation results using Eqs. (\ref{numsol:power1})
and (\ref{numsol:power2}), or Eqs. (\ref{eq:dissipated flux}) and
(\ref{numsol:power2}).

Eq (\ref{numsol:power2}) is equivalent to Eq. (82) of \citet{Beuthe:2016a}
for thin shells if we take into account the following differences.
First, $\upsilon_{2}=\gamma_{2}^{T}+\delta\gamma_{2}^{T}$ in the
thin shell limit (\citealt[Eq. (27)]{Beuthe:2016a}; Figure \ref{fig:beuthe_comparison}).
Second, \citet{Beuthe:2016a} uses normalized spherical harmonics,
whereas we use unnormalized spherical harmonics. Third, \citet{Beuthe:2016a}
sums over eastward and westward directions for all $m$, whereas we
only include one coefficient for $m=0$ (Table \ref{tab:forcing_coef}).
Last, $\Re(\Phi_{2m})=\Im(\tilde{\Phi}_{2m})$ using the notation
$\tilde{\Phi}_{nm}=i\Phi_{nm}$ \citep[Eq. (38)]{Beuthe:2016a}. 

Similarly, we can recover Eqs. (26) and (28) of \citet{Chen:2014}
for a surface ocean using Eqs. (\ref{numsol:power1}) and (\ref{numsol:power2})
taking into account the following differences. First, \citet{Chen:2014}
use normalized spherical harmonics, whereas we use unnormalized spherical
harmonics. Second, our Fourier expansion coefficients in Eq. (\ref{eq:harmonics_expansions})
are twice as large because \citet{Chen:2014} define the Fourier expansion
without the factor of $1/2$. Third, \citet{Chen:2014} ignore self-gravity
and the satellite deformation in response to tidal forcing and surface
ocean loading ($\upsilon_{2}=\beta_{n}=1$).

\section{Tidal quality factor\label{sec:Tidal-quality-factor}}

We consider an alternative definition of the tidal quality factor, 

\begin{equation}
Q\equiv2\pi\frac{E_{max}}{E_{diss}}\label{eq:tidalQ}
\end{equation}
where $E_{max}$ is the maximum kinetic energy of the ocean in the
absence of energy dissipation \citep{Hay:2017}. In this case, it
is no longer possible to obtain a simple expression relating the tidal
quality factor and the linear drag coefficient. Instead, the tidal
quality factor must be calculated after solving the LTE with and without
dissipation. Our new definition does not allow for a simple relation
between $Q$ and $\alpha$. However, the relevant quantity for computing
the effect of tidal heating on the thermal, rotational, and orbital
evolution is the energy dissipation rate and our thick shell theory
provides a method for computing it.

\begin{figure}[h]
\begin{centering}
\includegraphics[width=15cm]{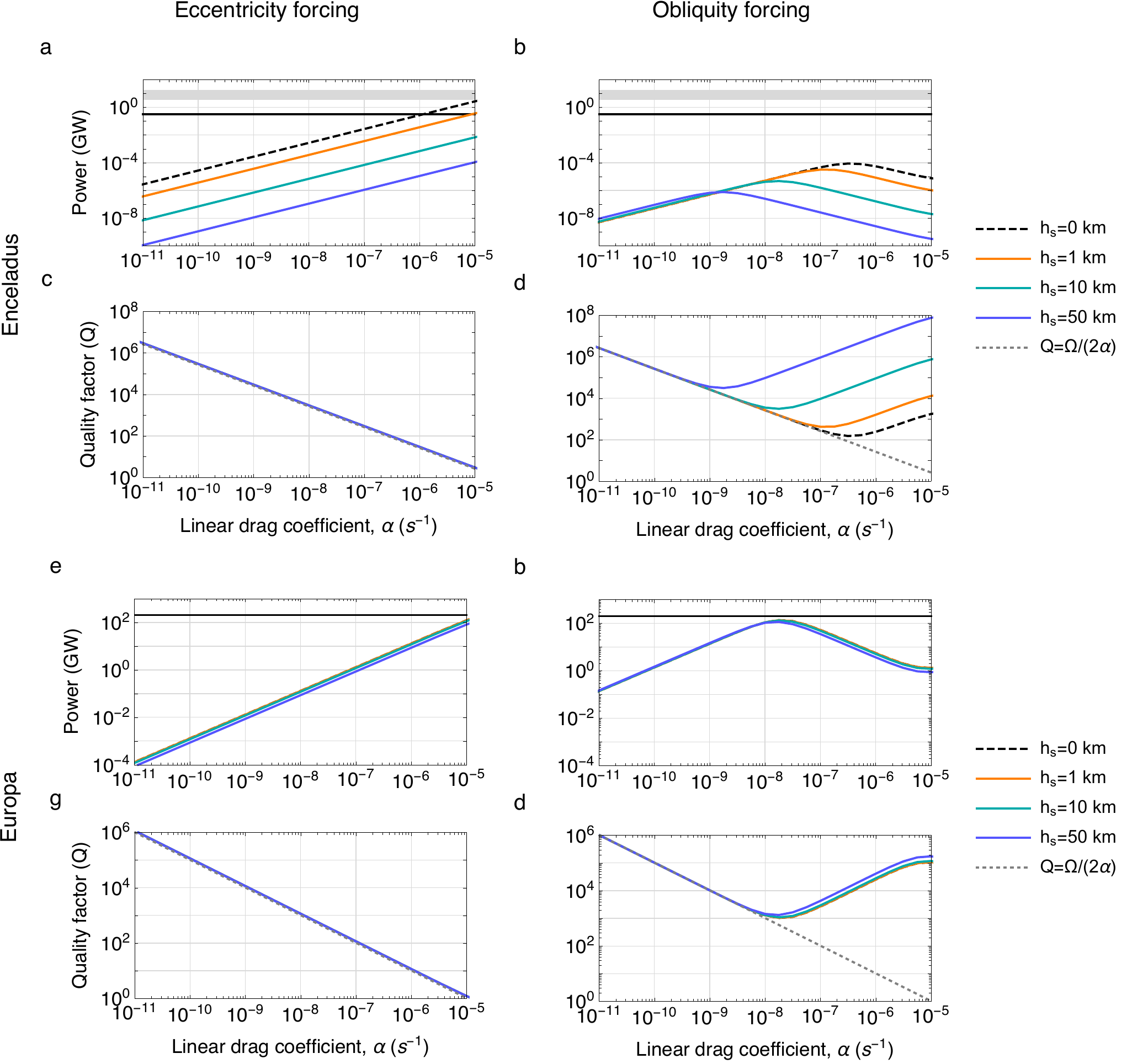}
\par\end{centering}
\caption{\label{fig:tidal_q}Ocean tidal heating power (energy dissipated in
one tidal cycle divided by the tidal forcing period) and tidal quality
factor, $Q$, for eccentricity and obliquity forcing as a function
of the linear drag coefficient, $\alpha$, for different shell thicknesses,
$h_{s}$. Black dashed lines are surface ocean solutions without an
overlying solid shell ($h_{s}=0$), and gray dotted lines are the
prior definition of the tidal quality factor ($Q=\Omega/(2\alpha)$).
The shaded gray region corresponds to the observational constraint
of $3.9-18.9$ GW for Enceladus, and the solid horizontal black line
is the estimated radiogenic heating power (0.3 GW for Enceladus and
200 GW for Europa). We assume the interior structure parameters in
Table \ref{tab:Interior parameters}. }
\end{figure}

Using the new definition of $Q$, higher energy dissipation corresponds
to smaller $Q$ values (Fig. \ref{fig:tidal_q}), as expected. $Q$
converges to the prior definition, $Q=\Omega/(2\alpha)$, as the linear
drag coefficient decreases because this reduces the effect of dissipation
on the kinetic energy. The prior definition results in a tidal quality
factor that decreases indefinitely as the linear drag increases. In
contrast, our definition introduces natural lower limits. Fig. \ref{fig:tidal_q}
shows these natural lower limits for obliquity forcing. The natural
lower limits for $Q$ also emerge for eccentricity forcing if we consider
larger linear drag coefficients. 

The effect of varying the shell thickness on $Q$ is complex because
both the numerator and denominator in Eq. (\ref{eq:tidalQ}) decrease
as the shell thickness increases. For the range of linear drag coefficients
considered, $Q$ is not sensitive to the shell thickness for eccentricity
forcing. For obliquity forcing, the dissipated energy decreases faster
than the maximum kinetic energy, resulting in an increase in $Q$
with increasing shell thickness (Figs. \ref{fig:tidal_q}d and h).

The tidal quality factor for eccentricity forcing reaches small values
that imply energy dissipation in one forcing cycle larger than the
maximum kinetic energy without dissipation (Figs. \ref{fig:tidal_q}c
and g), which may seem unphysical. However, energy conservation requires
the work done on the ocean by the tide-raising potential to be equal
to the dissipated energy, and therefore the dissipated energy can
be larger than the maximum kinetic energy as long as it is balanced
by the work done on the ocean.

Tidal heating increases with the linear drag coefficient, as expected,
with the exception of obliquity forcing with high linear drag coefficients
(Fig. \ref{fig:tidal_q}). Eccentricity forcing generates gravity
waves, while obliquity forcing generates both gravity and Rossby-Haurwitz
waves \citep{Tyler:2008,Tyler:2009,Tyler:2011}. The tidal heating
decrease with increasing linear drag coefficient for the obliquity
forcing is associated with Rossby-Haurwitz waves. We verify this by
removing the westward traveling component of the obliquity forcing
tidal potential (Eq. \ref{eq:U_ecc_obliq}), which prevents the generation
of Rossby-Haurwitz waves. In this case, tidal heating increases with
the linear drag coefficient. 

\section*{References}


\begin{thebibliography}{76}
\expandafter\ifx\csname natexlab\endcsname\relax\def\natexlab#1{#1}\fi
\providecommand{\url}[1]{\texttt{#1}}
\providecommand{\href}[2]{#2}
\providecommand{\path}[1]{#1}
\providecommand{\DOIprefix}{doi:}
\providecommand{\ArXivprefix}{arXiv:}
\providecommand{\URLprefix}{URL: }
\providecommand{\Pubmedprefix}{pmid:}
\providecommand{\doi}[1]{\href{http://dx.doi.org/#1}{\path{#1}}}
\providecommand{\Pubmed}[1]{\href{pmid:#1}{\path{#1}}}
\providecommand{\bibinfo}[2]{#2}
\ifx\xfnm\relax \def\xfnm[#1]{\unskip,\space#1}\fi
\bibitem[{Anderson et~al.(1998)Anderson, Schubert, Jacobson, Lau, Moore and
  Sjogren}]{Anderson:1998}
\bibinfo{author}{Anderson, J.D.}, \bibinfo{author}{Schubert, G.},
  \bibinfo{author}{Jacobson, R.A.}, \bibinfo{author}{Lau, E.L.},
  \bibinfo{author}{Moore, W.B.}, \bibinfo{author}{Sjogren, W.L.},
  \bibinfo{year}{1998}.
\newblock \bibinfo{title}{{Europa's Differentiated Internal Structure:
  Inferences from Four Galileo Encounters}}.
\newblock \bibinfo{journal}{Science} \bibinfo{volume}{281},
  \bibinfo{pages}{2019--2022}.
\newblock \DOIprefix\doi{10.1126/science.281.5385.2019}.
\bibitem[{Arfken and Weber(1995)}]{Arfken:1995}
\bibinfo{author}{Arfken, G.}, \bibinfo{author}{Weber, H.},
  \bibinfo{year}{1995}.
\newblock \bibinfo{title}{Mathematical methods for physicists}.
\newblock \bibinfo{edition}{Fourth} ed., \bibinfo{publisher}{Academic Press}.
\bibitem[{Baland et~al.(2016)Baland, Yseboodt and Van~Hoolst}]{Baland:2016}
\bibinfo{author}{Baland, R.M.}, \bibinfo{author}{Yseboodt, M.},
  \bibinfo{author}{Van~Hoolst, T.}, \bibinfo{year}{2016}.
\newblock \bibinfo{title}{{The obliquity of Enceladus}}.
\newblock \bibinfo{journal}{Icarus} \bibinfo{volume}{268},
  \bibinfo{pages}{12--31}.
\newblock \DOIprefix\doi{10.1016/j.icarus.2015.11.039}.
\bibitem[{B{\v e}hounkov{\'a} et~al.(2012)B{\v e}hounkov{\'a}, Tobie, Choblet
  and {\v C}adek}]{Behounkova:2012}
\bibinfo{author}{B{\v e}hounkov{\'a}, M.}, \bibinfo{author}{Tobie, G.},
  \bibinfo{author}{Choblet, G.}, \bibinfo{author}{{\v C}adek, O.},
  \bibinfo{year}{2012}.
\newblock \bibinfo{title}{{Tidally-induced melting events as the origin of
  south-pole activity on Enceladus}}.
\newblock \bibinfo{journal}{Icarus} \bibinfo{volume}{219},
  \bibinfo{pages}{655--664}.
\newblock \DOIprefix\doi{10.1016/j.icarus.2012.03.024}.
\bibitem[{Beuthe(2013)}]{Beuthe:2013}
\bibinfo{author}{Beuthe, M.}, \bibinfo{year}{2013}.
\newblock \bibinfo{title}{{Spatial patterns of tidal heating}}.
\newblock \bibinfo{journal}{Icarus} \bibinfo{volume}{223},
  \bibinfo{pages}{308--329}.
\newblock \DOIprefix\doi{10.1016/j.icarus.2012.11.020}.
\bibitem[{Beuthe(2015a)}]{Beuthe:2015a}
\bibinfo{author}{Beuthe, M.}, \bibinfo{year}{2015}a.
\newblock \bibinfo{title}{{Tidal Love numbers of membrane worlds: Europa,
  Titan, and Co.}}
\newblock \bibinfo{journal}{Icarus} \bibinfo{volume}{258},
  \bibinfo{pages}{239--266}.
\newblock \DOIprefix\doi{10.1016/j.icarus.2015.06.008}.
\bibitem[{Beuthe(2015b)}]{Beuthe:2015}
\bibinfo{author}{Beuthe, M.}, \bibinfo{year}{2015}b.
\newblock \bibinfo{title}{{Tides on Europa: The membrane paradigm}}.
\newblock \bibinfo{journal}{Icarus} \bibinfo{volume}{248},
  \bibinfo{pages}{109--134}.
\newblock \DOIprefix\doi{10.1016/j.icarus.2014.10.027}.
\bibitem[{Beuthe(2016)}]{Beuthe:2016a}
\bibinfo{author}{Beuthe, M.}, \bibinfo{year}{2016}.
\newblock \bibinfo{title}{{Crustal control of dissipative ocean tides in
  Enceladus and other icy moons}}.
\newblock \bibinfo{journal}{Icarus} \bibinfo{volume}{280},
  \bibinfo{pages}{278--299}.
\newblock \DOIprefix\doi{10.1016/j.icarus.2016.08.009}.
\bibitem[{Beuthe(2018)}]{Beuthe:2018}
\bibinfo{author}{Beuthe, M.}, \bibinfo{year}{2018}.
\newblock \bibinfo{title}{{Enceladus's crust as a non-uniform thin shell: I
  tidal deformations}}.
\newblock \bibinfo{journal}{Icarus} \bibinfo{volume}{302},
  \bibinfo{pages}{145--174}.
\newblock \DOIprefix\doi{10.1016/j.icarus.2017.11.009}.
\bibitem[{Beuthe et~al.(2016)Beuthe, Rivoldini and Trinh}]{Beuthe:2016}
\bibinfo{author}{Beuthe, M.}, \bibinfo{author}{Rivoldini, A.},
  \bibinfo{author}{Trinh, A.}, \bibinfo{year}{2016}.
\newblock \bibinfo{title}{{Enceladus's and Dione's floating ice shells
  supported by minimum stress isostasy}}.
\newblock \bibinfo{journal}{Geophys. Res. Lett.} \bibinfo{volume}{43},
  \bibinfo{pages}{10088--10096}.
\newblock \DOIprefix\doi{10.1002/2016GL070650}.
\bibitem[{Bills(2005)}]{Bills:2005}
\bibinfo{author}{Bills, B.G.}, \bibinfo{year}{2005}.
\newblock \bibinfo{title}{{Free and forced obliquities of the Galilean
  satellites of Jupiter}}.
\newblock \bibinfo{journal}{Icarus} \bibinfo{volume}{175},
  \bibinfo{pages}{233--247}.
\newblock \DOIprefix\doi{10.1016/j.icarus.2004.10.028}.
\bibitem[{Chen and Nimmo(2011)}]{Chen:2011}
\bibinfo{author}{Chen, E.M.A.}, \bibinfo{author}{Nimmo, F.},
  \bibinfo{year}{2011}.
\newblock \bibinfo{title}{{Obliquity tides do not significantly heat
  Enceladus}}.
\newblock \bibinfo{journal}{Icarus} \bibinfo{volume}{214},
  \bibinfo{pages}{779--781}.
\newblock \DOIprefix\doi{10.1016/j.icarus.2011.06.007}.
\bibitem[{Chen and Nimmo(2016)}]{Chen:2016}
\bibinfo{author}{Chen, E.M.A.}, \bibinfo{author}{Nimmo, F.},
  \bibinfo{year}{2016}.
\newblock \bibinfo{title}{{Tidal dissipation in the lunar magma ocean and its
  effect on the early evolution of the Earth-Moon system}}.
\newblock \bibinfo{journal}{Icarus} \bibinfo{volume}{275},
  \bibinfo{pages}{132--142}.
\newblock \DOIprefix\doi{10.1016/j.icarus.2016.04.012}.
\bibitem[{Chen et~al.(2014)Chen, Nimmo and Glatzmaier}]{Chen:2014}
\bibinfo{author}{Chen, E.M.A.}, \bibinfo{author}{Nimmo, F.},
  \bibinfo{author}{Glatzmaier, G.A.}, \bibinfo{year}{2014}.
\newblock \bibinfo{title}{{Tidal heating in icy satellite oceans}}.
\newblock \bibinfo{journal}{Icarus} \bibinfo{volume}{229},
  \bibinfo{pages}{11--30}.
\newblock \DOIprefix\doi{10.1016/j.icarus.2013.10.024}.
\bibitem[{Dahlen and Tromp(1998)}]{Dahlen:1998}
\bibinfo{author}{Dahlen, F.}, \bibinfo{author}{Tromp, J.},
  \bibinfo{year}{1998}.
\newblock \bibinfo{title}{Theoretical Global Seismology}.
\newblock \bibinfo{publisher}{Princeton University Press}.
\bibitem[{Egbert and Ray(2000)}]{Egbert:2000jf}
\bibinfo{author}{Egbert, G.D.}, \bibinfo{author}{Ray, R.D.},
  \bibinfo{year}{2000}.
\newblock \bibinfo{title}{{Significant dissipation of tidal energy in the deep
  ocean inferred from satellite altimeter data}}.
\newblock \bibinfo{journal}{Nature} \bibinfo{volume}{405},
  \bibinfo{pages}{775--778}.
\newblock \DOIprefix\doi{10.1038/35015531}.
\bibitem[{Egbert and Ray(2001)}]{Egbert:2001}
\bibinfo{author}{Egbert, G.D.}, \bibinfo{author}{Ray, R.D.},
  \bibinfo{year}{2001}.
\newblock \bibinfo{title}{{Estimates of M2 tidal energy dissipation from
  TOPEX/Poseidon altimeter data}}.
\newblock \bibinfo{journal}{Journal of Geophysical Research: Oceans}
  \bibinfo{volume}{106}, \bibinfo{pages}{22475--22502}.
\newblock \DOIprefix\doi{10.1029/2000JC000699}.
\bibitem[{Fischer and Spohn(1990)}]{Fischer:1990}
\bibinfo{author}{Fischer, H.J.}, \bibinfo{author}{Spohn, T.},
  \bibinfo{year}{1990}.
\newblock \bibinfo{title}{{Thermal-orbital histories of viscoelastic models of
  Io (J1)}}.
\newblock \bibinfo{journal}{Icarus} \bibinfo{volume}{83},
  \bibinfo{pages}{39--65}.
\newblock \DOIprefix\doi{10.1016/0019-1035(90)90005-T}.
\bibitem[{Gao and Stevenson(2013)}]{Gao:2013}
\bibinfo{author}{Gao, P.}, \bibinfo{author}{Stevenson, D.J.},
  \bibinfo{year}{2013}.
\newblock \bibinfo{title}{{Nonhydrostatic effects and the determination of icy
  satellites{\textquoteright} moment of inertia}}.
\newblock \bibinfo{journal}{Icarus} \bibinfo{volume}{226},
  \bibinfo{pages}{1185--1191}.
\newblock \DOIprefix\doi{10.1016/j.icarus.2013.07.034}.
\bibitem[{Goldreich and Soter(1966)}]{Goldreich:1966}
\bibinfo{author}{Goldreich, P.}, \bibinfo{author}{Soter, S.},
  \bibinfo{year}{1966}.
\newblock \bibinfo{title}{{Q in the Solar System}}.
\newblock \bibinfo{journal}{Icarus} \bibinfo{volume}{5},
  \bibinfo{pages}{375--389}.
\bibitem[{Green and Nycander(2013)}]{Green:2013}
\bibinfo{author}{Green, J.A.M.}, \bibinfo{author}{Nycander, J.},
  \bibinfo{year}{2013}.
\newblock \bibinfo{title}{{A Comparison of Tidal Conversion Parameterizations
  for Tidal Models}}.
\newblock \bibinfo{journal}{Journal of Physical Oceanography}
  \bibinfo{volume}{43}, \bibinfo{pages}{104--119}.
\newblock \DOIprefix\doi{10.1175/JPO-D-12-023.1}.
\bibitem[{Hammond et~al.(2018)Hammond, Barr, Cooper, Caswell and
  Hirth}]{Hammond:2018}
\bibinfo{author}{Hammond, N.P.}, \bibinfo{author}{Barr, A.C.},
  \bibinfo{author}{Cooper, R.F.}, \bibinfo{author}{Caswell, T.E.},
  \bibinfo{author}{Hirth, G.}, \bibinfo{year}{2018}.
\newblock \bibinfo{title}{{Experimental Constraints on the Fatigue of Icy
  Satellite Lithospheres by Tidal Forces}}.
\newblock \bibinfo{journal}{J. Geophys. Res.} \bibinfo{volume}{123},
  \bibinfo{pages}{390--404}.
\newblock \DOIprefix\doi{10.1002/2017JE005464}.
\bibitem[{Hartkorn and Saur(2017)}]{Hartkorn:2017}
\bibinfo{author}{Hartkorn, O.}, \bibinfo{author}{Saur, J.},
  \bibinfo{year}{2017}.
\newblock \bibinfo{title}{{Induction signals from Callisto's ionosphere and
  their implications on a possible subsurface ocean}}.
\newblock \bibinfo{journal}{Journal of Geophysical Research: Space Physics}
  \bibinfo{volume}{122}, \bibinfo{pages}{11677--11697}.
\newblock \DOIprefix\doi{10.1002/2017JA024269}.
\bibitem[{Hay and Matsuyama(2017)}]{Hay:2017}
\bibinfo{author}{Hay, H.C.F.C.}, \bibinfo{author}{Matsuyama, I.},
  \bibinfo{year}{2017}.
\newblock \bibinfo{title}{{Numerically modelling tidal dissipation with bottom
  drag in the oceans of Titan and Enceladus}}.
\newblock \bibinfo{journal}{Icarus} \bibinfo{volume}{281},
  \bibinfo{pages}{342--356}.
\newblock \DOIprefix\doi{10.1016/j.icarus.2016.09.022}.
\bibitem[{Hedman et~al.(2013)Hedman, Gosmeyer, Nicholson, Sotin, Brown, Clark,
  Baines, Buratti and Showalter}]{Hedman:2013}
\bibinfo{author}{Hedman, M.M.}, \bibinfo{author}{Gosmeyer, C.M.},
  \bibinfo{author}{Nicholson, P.D.}, \bibinfo{author}{Sotin, C.},
  \bibinfo{author}{Brown, R.H.}, \bibinfo{author}{Clark, R.N.},
  \bibinfo{author}{Baines, K.H.}, \bibinfo{author}{Buratti, B.J.},
  \bibinfo{author}{Showalter, M.R.}, \bibinfo{year}{2013}.
\newblock \bibinfo{title}{{An observed correlation between plume activity and
  tidal stresses on Enceladus}}.
\newblock \bibinfo{journal}{Nature} \bibinfo{volume}{500},
  \bibinfo{pages}{182--184}.
\newblock \DOIprefix\doi{10.1038/nature12371}.
\bibitem[{Hendershott(1972)}]{Hendershott:1972}
\bibinfo{author}{Hendershott, M.C.}, \bibinfo{year}{1972}.
\newblock \bibinfo{title}{{The Effects of Solid Earth Deformation on Global
  Ocean Tides}}.
\newblock \bibinfo{journal}{Geophysical Journal} \bibinfo{volume}{29},
  \bibinfo{pages}{389--402}.
\newblock \DOIprefix\doi{10.1111/j.1365-246X.1972.tb06167.x}.
\bibitem[{Hinderer(1986)}]{Hinderer:1986ik}
\bibinfo{author}{Hinderer, J.}, \bibinfo{year}{1986}.
\newblock \bibinfo{title}{{Resonance Effects of the Earth{\textquoteright}s
  Fluid Core}}, in: \bibinfo{booktitle}{Earth Rotation: Solved and Unsolved
  Problems}. \bibinfo{publisher}{Springer, Dordrecht},
  \bibinfo{address}{Dordrecht}, pp. \bibinfo{pages}{277--296}.
\newblock \DOIprefix\doi{10.1007/978-94-009-4750-4_20}.
\bibitem[{Hinderer and Legros(1989)}]{Hinderer:1989jx}
\bibinfo{author}{Hinderer, J.}, \bibinfo{author}{Legros, H.},
  \bibinfo{year}{1989}.
\newblock \bibinfo{title}{{Elasto-Gravitational Deformation, Relative Gravity
  Changes and Earth dynamics}}.
\newblock \bibinfo{journal}{Geophysical Journal International}
  \bibinfo{volume}{97}, \bibinfo{pages}{481--495}.
\newblock \DOIprefix\doi{10.1111/j.1365-246X.1989.tb00518.x}.
\bibitem[{Howett et~al.(2011)Howett, Spencer, Pearl and Segura}]{Howett:2011}
\bibinfo{author}{Howett, C.J.A.}, \bibinfo{author}{Spencer, J.R.},
  \bibinfo{author}{Pearl, J.}, \bibinfo{author}{Segura, M.},
  \bibinfo{year}{2011}.
\newblock \bibinfo{title}{{High heat flow from Enceladus' south polar region
  measured using 10-600 cm-1 Cassini/CIRS data}}.
\newblock \bibinfo{journal}{J. Geophys. Res.} \bibinfo{volume}{116},
  \bibinfo{pages}{E03003}.
\newblock \DOIprefix\doi{10.1029/2010JE003718}.
\bibitem[{Hussmann and Spohn(2004)}]{Hussmann:2004}
\bibinfo{author}{Hussmann, H.}, \bibinfo{author}{Spohn, T.},
  \bibinfo{year}{2004}.
\newblock \bibinfo{title}{{Thermal-orbital evolution of Io and Europa}}.
\newblock \bibinfo{journal}{Icarus} \bibinfo{volume}{171},
  \bibinfo{pages}{391--410}.
\newblock \DOIprefix\doi{10.1016/j.icarus.2004.05.020}.
\bibitem[{Iess et~al.(2012)Iess, Jacobson, Ducci, Stevenson, Lunine, Armstrong,
  Asmar, Racioppa, Rappaport and Tortora}]{Iess:2012fj}
\bibinfo{author}{Iess, L.}, \bibinfo{author}{Jacobson, R.A.},
  \bibinfo{author}{Ducci, M.}, \bibinfo{author}{Stevenson, D.J.},
  \bibinfo{author}{Lunine, J.I.}, \bibinfo{author}{Armstrong, J.W.},
  \bibinfo{author}{Asmar, S.W.}, \bibinfo{author}{Racioppa, P.},
  \bibinfo{author}{Rappaport, N.J.}, \bibinfo{author}{Tortora, P.},
  \bibinfo{year}{2012}.
\newblock \bibinfo{title}{{The Tides of Titan}}.
\newblock \bibinfo{journal}{Science} \bibinfo{volume}{337},
  \bibinfo{pages}{457--459}.
\newblock \DOIprefix\doi{10.1126/science.1219631}.
\bibitem[{Jara-Oru{\'e} and Vermeersen(2011)}]{Jara-Orue:2011}
\bibinfo{author}{Jara-Oru{\'e}, H.M.}, \bibinfo{author}{Vermeersen, B.L.A.},
  \bibinfo{year}{2011}.
\newblock \bibinfo{title}{{Effects of low-viscous layers and a non-zero
  obliquity on surface stresses induced by diurnal tides and non-synchronous
  rotation: The case of Europa}}.
\newblock \bibinfo{journal}{Icarus} \bibinfo{volume}{215},
  \bibinfo{pages}{417--438}.
\newblock \DOIprefix\doi{10.1016/j.icarus.2011.05.034}.
\bibitem[{Jayne and St~Laurent(2001)}]{Jayne:2001}
\bibinfo{author}{Jayne, S.R.}, \bibinfo{author}{St~Laurent, L.C.},
  \bibinfo{year}{2001}.
\newblock \bibinfo{title}{{Parameterizing tidal dissipation over rough
  topography}}.
\newblock \bibinfo{journal}{Geophys. Res. Lett.} \bibinfo{volume}{28},
  \bibinfo{pages}{811--814}.
\newblock \DOIprefix\doi{10.1029/2000GL012044}.
\bibitem[{Kamata et~al.(2016)Kamata, Kimura, Matsumoto, Nimmo, Kuramoto and
  Namiki}]{Kamata:2016}
\bibinfo{author}{Kamata, S.}, \bibinfo{author}{Kimura, J.},
  \bibinfo{author}{Matsumoto, K.}, \bibinfo{author}{Nimmo, F.},
  \bibinfo{author}{Kuramoto, K.}, \bibinfo{author}{Namiki, N.},
  \bibinfo{year}{2016}.
\newblock \bibinfo{title}{{Tidal deformation of Ganymede: Sensitivity of Love
  numbers on the interior structure}}.
\newblock \bibinfo{journal}{J. Geophys. Res.} \bibinfo{volume}{121},
  \bibinfo{pages}{1362--1375}.
\newblock \DOIprefix\doi{10.1002/2016JE005071}.
\bibitem[{Kamata et~al.(2015)Kamata, Matsuyama and Nimmo}]{Kamata:2015}
\bibinfo{author}{Kamata, S.}, \bibinfo{author}{Matsuyama, I.},
  \bibinfo{author}{Nimmo, F.}, \bibinfo{year}{2015}.
\newblock \bibinfo{title}{{Tidal resonance in icy satellites with subsurface
  oceans}}.
\newblock \bibinfo{journal}{J. Geophys. Res. Planets} \bibinfo{volume}{120},
  \bibinfo{pages}{1528--1542}.
\newblock \DOIprefix\doi{10.1002/2015JE004821}.
\bibitem[{Kivelson et~al.(2000)Kivelson, Khurana, Russell, Volwerk, Walker and
  Zimmer}]{Kivelson:2000}
\bibinfo{author}{Kivelson, M.G.}, \bibinfo{author}{Khurana, K.K.},
  \bibinfo{author}{Russell, C.T.}, \bibinfo{author}{Volwerk, M.},
  \bibinfo{author}{Walker, R.J.}, \bibinfo{author}{Zimmer, C.},
  \bibinfo{year}{2000}.
\newblock \bibinfo{title}{{Galileo Magnetometer Measurements: A Stronger Case
  for a Subsurface Ocean at Europa}}.
\newblock \bibinfo{journal}{Science} \bibinfo{volume}{289},
  \bibinfo{pages}{1340--1343.}
\newblock \DOIprefix\doi{10.1126/science.289.5483.1340}.
\bibitem[{Kivelson et~al.(2002)Kivelson, Khurana and Volwerk}]{Kivelson:2002}
\bibinfo{author}{Kivelson, M.G.}, \bibinfo{author}{Khurana, K.K.},
  \bibinfo{author}{Volwerk, M.}, \bibinfo{year}{2002}.
\newblock \bibinfo{title}{{The Permanent and Inductive Magnetic Moments of
  Ganymede}}.
\newblock \bibinfo{journal}{Icarus} \bibinfo{volume}{157},
  \bibinfo{pages}{507--522}.
\newblock \DOIprefix\doi{10.1006/icar.2002.6834}.
\bibitem[{{Lamb}(1993)}]{Lamb:1993}
\bibinfo{author}{{Lamb}, H.}, \bibinfo{year}{1993}.
\newblock \bibinfo{title}{Hydrodynamics}.
\newblock \bibinfo{edition}{6} ed., \bibinfo{publisher}{Cambridge University
  Press, Cambridge}.
\bibitem[{{Lambeck}(1980)}]{Lambeck:1980earth}
\bibinfo{author}{{Lambeck}, K.}, \bibinfo{year}{1980}.
\newblock \bibinfo{title}{{The earth's variable rotation: Geophysical causes
  and consequences}}.
\newblock \bibinfo{publisher}{Cambridge University Press, Cambridge}.
\bibitem[{Longuet-Higgins(1968)}]{Longuet-Higgins:1968}
\bibinfo{author}{Longuet-Higgins, M.S.}, \bibinfo{year}{1968}.
\newblock \bibinfo{title}{{The Eigenfunctions of Laplace's Tidal Equations over
  a Sphere}}.
\newblock \bibinfo{journal}{Philosophical Transactions of the Royal Society A:
  Mathematical, Physical and Engineering Sciences} \bibinfo{volume}{262},
  \bibinfo{pages}{511--607}.
\newblock \DOIprefix\doi{10.1098/rsta.1968.0003}.
\bibitem[{Matsuyama(2014)}]{Matsuyama:2014}
\bibinfo{author}{Matsuyama, I.}, \bibinfo{year}{2014}.
\newblock \bibinfo{title}{{Tidal dissipation in the oceans of icy satellites}}.
\newblock \bibinfo{journal}{Icarus} \bibinfo{volume}{242},
  \bibinfo{pages}{11--18}.
\newblock \DOIprefix\doi{10.1016/j.icarus.2014.07.005}.
\bibitem[{McKinnon(2015)}]{McKinnon:2015}
\bibinfo{author}{McKinnon, W.B.}, \bibinfo{year}{2015}.
\newblock \bibinfo{title}{{Effect of Enceladus's rapid synchronous spin on
  interpretation of Cassini gravity}}.
\newblock \bibinfo{journal}{Geophys. Res. Lett.} \bibinfo{volume}{42},
  \bibinfo{pages}{2137--2143}.
\newblock \DOIprefix\doi{10.1002/2015GL063384}.
\bibitem[{Moore and Schubert(2000)}]{Moore:2000}
\bibinfo{author}{Moore, W.B.}, \bibinfo{author}{Schubert, G.},
  \bibinfo{year}{2000}.
\newblock \bibinfo{title}{{NOTE: The Tidal Response of Europa}}.
\newblock \bibinfo{journal}{Icarus} \bibinfo{volume}{147},
  \bibinfo{pages}{317--319}.
\newblock \DOIprefix\doi{10.1006/icar.2000.6460}.
\bibitem[{{Munk} and {MacDonald}(1960)}]{Munk:1960}
\bibinfo{author}{{Munk}, W.H.}, \bibinfo{author}{{MacDonald}, G.J.F.},
  \bibinfo{year}{1960}.
\newblock \bibinfo{title}{{The rotation of the earth; a geophysical
  discussion}}.
\newblock \bibinfo{publisher}{Cambridge University Press, Cambridge}.
\bibitem[{Nimmo and Bills(2010)}]{Nimmo:2010}
\bibinfo{author}{Nimmo, F.}, \bibinfo{author}{Bills, B.G.},
  \bibinfo{year}{2010}.
\newblock \bibinfo{title}{{Shell thickness variations and the long-wavelength
  topography of Titan}}.
\newblock \bibinfo{journal}{Icarus} \bibinfo{volume}{208},
  \bibinfo{pages}{896--904}.
\newblock \DOIprefix\doi{10.1016/j.icarus.2010.02.020}.
\bibitem[{Nimmo and Pappalardo(2016)}]{Nimmo:2016}
\bibinfo{author}{Nimmo, F.}, \bibinfo{author}{Pappalardo, R.T.},
  \bibinfo{year}{2016}.
\newblock \bibinfo{title}{{Ocean worlds in the outer solar system}}.
\newblock \bibinfo{journal}{J. Geophys. Res.} \bibinfo{volume}{121},
  \bibinfo{pages}{1378--1399}.
\newblock \DOIprefix\doi{10.1002/2016JE005081}.
\bibitem[{Nimmo et~al.(2014)Nimmo, Porco and Mitchell}]{Nimmo:2014}
\bibinfo{author}{Nimmo, F.}, \bibinfo{author}{Porco, C.},
  \bibinfo{author}{Mitchell, C.}, \bibinfo{year}{2014}.
\newblock \bibinfo{title}{{Tidally Modulated Eruptions on Enceladus: Cassini
  ISS Observations and Models}}.
\newblock \bibinfo{journal}{Astronomical Journal} \bibinfo{volume}{148},
  \bibinfo{pages}{46}.
\newblock \DOIprefix\doi{10.1088/0004-6256/148/3/46}.
\bibitem[{Nimmo et~al.(2007)Nimmo, Thomas, Pappalardo and Moore}]{Nimmo:2007}
\bibinfo{author}{Nimmo, F.}, \bibinfo{author}{Thomas, P.C.},
  \bibinfo{author}{Pappalardo, R.T.}, \bibinfo{author}{Moore, W.B.},
  \bibinfo{year}{2007}.
\newblock \bibinfo{title}{{The global shape of Europa: Constraints on lateral
  shell thickness variations}}.
\newblock \bibinfo{journal}{Icarus} \bibinfo{volume}{191},
  \bibinfo{pages}{183--192}.
\newblock \DOIprefix\doi{10.1016/j.icarus.2007.04.021}.
\bibitem[{Ojakangas(1989a)}]{Ojakangas:1989a}
\bibinfo{author}{Ojakangas, G.}, \bibinfo{year}{1989}a.
\newblock \bibinfo{title}{{Polar wander of an ice shell on Europa}}.
\newblock \bibinfo{journal}{Icarus} \bibinfo{volume}{81},
  \bibinfo{pages}{242--270}.
\newblock \DOIprefix\doi{10.1016/0019-1035(89)90053-5}.
\bibitem[{Ojakangas(1989b)}]{Ojakangas:1989}
\bibinfo{author}{Ojakangas, G.}, \bibinfo{year}{1989}b.
\newblock \bibinfo{title}{{Thermal state of an ice shell on Europa}}.
\newblock \bibinfo{journal}{Icarus} \bibinfo{volume}{81},
  \bibinfo{pages}{220--241}.
\newblock \DOIprefix\doi{10.1016/0019-1035(89)90052-3}.
\bibitem[{Ojakangas and Stevenson(1986)}]{Ojakangas:1986}
\bibinfo{author}{Ojakangas, G.W.}, \bibinfo{author}{Stevenson, D.J.},
  \bibinfo{year}{1986}.
\newblock \bibinfo{title}{{Episodic volcanism of tidally heated satellites with
  application to Io}}.
\newblock \bibinfo{journal}{Icarus} \bibinfo{volume}{66},
  \bibinfo{pages}{341--358}.
\newblock \DOIprefix\doi{10.1016/0019-1035(86)90163-6}.
\bibitem[{Peltier(1974)}]{Peltier:1974ki}
\bibinfo{author}{Peltier, W.R.}, \bibinfo{year}{1974}.
\newblock \bibinfo{title}{{The Impulse Response of a Maxwell Earth}}.
\newblock \bibinfo{journal}{Rev. Geophys. Space Phys.} \bibinfo{volume}{12},
  \bibinfo{pages}{649--669}.
\newblock \DOIprefix\doi{10.1029/RG012i004p00649}.
\bibitem[{Ross and Schubert(1990)}]{Ross:1990}
\bibinfo{author}{Ross, M.N.}, \bibinfo{author}{Schubert, G.},
  \bibinfo{year}{1990}.
\newblock \bibinfo{title}{{The coupled orbital and thermal evolution of
  Triton}}.
\newblock \bibinfo{journal}{Geophys. Res. Lett.} \bibinfo{volume}{17},
  \bibinfo{pages}{1749--1752}.
\newblock \DOIprefix\doi{10.1029/GL017i010p01749}.
\bibitem[{{Sabadini} and {Vermeersen}(2004)}]{Sabadini:2004a}
\bibinfo{author}{{Sabadini}, R.}, \bibinfo{author}{{Vermeersen}, B.},
  \bibinfo{year}{2004}.
\newblock \bibinfo{title}{Global Dynamics of the Earth: Applications of Normal
  Mode Relaxation Theory to Solid-Earth Geophysics}.
\newblock \bibinfo{publisher}{Kluwer Academic Publishers}.
\bibitem[{Sagan and Dermott(1982)}]{Sagan:1982}
\bibinfo{author}{Sagan, C.}, \bibinfo{author}{Dermott, S.F.},
  \bibinfo{year}{1982}.
\newblock \bibinfo{title}{{The tide in the seas of Titan}}.
\newblock \bibinfo{journal}{Nature} \bibinfo{volume}{300},
  \bibinfo{pages}{731--733}.
\newblock \DOIprefix\doi{10.1038/300731a0}.
\bibitem[{Saito(1974)}]{Saito:1974}
\bibinfo{author}{Saito, M.}, \bibinfo{year}{1974}.
\newblock \bibinfo{title}{{Some problems of static deformation of the earth.}}
\newblock \bibinfo{journal}{Journal of Physics of the Earth}
  \bibinfo{volume}{22}, \bibinfo{pages}{123--140}.
\newblock \DOIprefix\doi{10.4294/jpe1952.22.123}.
\bibitem[{Sasao et~al.(1980)Sasao, Okubo and Saito}]{Sasao:1980iv}
\bibinfo{author}{Sasao, T.}, \bibinfo{author}{Okubo, S.},
  \bibinfo{author}{Saito, M.}, \bibinfo{year}{1980}.
\newblock \bibinfo{title}{{A simple theory on the dynamical effects of a
  stratified fluid core upon nutational motion of the Earth}}, in:
  \bibinfo{booktitle}{Proceedings of the International Astronomical Union}, pp.
  \bibinfo{pages}{165--183}.
\newblock \DOIprefix\doi{10.1017/S0074180900032009}.
\bibitem[{Sasao and Wahr(1981)}]{Sasao:1981}
\bibinfo{author}{Sasao, T.}, \bibinfo{author}{Wahr, J.M.},
  \bibinfo{year}{1981}.
\newblock \bibinfo{title}{{An excitation mechanism for the free 'core
  nutation'}}.
\newblock \bibinfo{journal}{Geophysical Journal} \bibinfo{volume}{64},
  \bibinfo{pages}{729--746}.
\newblock \DOIprefix\doi{10.1111/j.1365-246X.1981.tb02692.x}.
\bibitem[{Saur et~al.(2015)Saur, Duling, Roth, Jia, Strobel, Feldman,
  Christensen, Retherford, McGrath, Musacchio, Wennmacher, Neubauer, Simon and
  Hartkorn}]{Saur:2015}
\bibinfo{author}{Saur, J.}, \bibinfo{author}{Duling, S.},
  \bibinfo{author}{Roth, L.}, \bibinfo{author}{Jia, X.},
  \bibinfo{author}{Strobel, D.F.}, \bibinfo{author}{Feldman, P.D.},
  \bibinfo{author}{Christensen, U.R.}, \bibinfo{author}{Retherford, K.D.},
  \bibinfo{author}{McGrath, M.A.}, \bibinfo{author}{Musacchio, F.},
  \bibinfo{author}{Wennmacher, A.}, \bibinfo{author}{Neubauer, F.M.},
  \bibinfo{author}{Simon, S.}, \bibinfo{author}{Hartkorn, O.},
  \bibinfo{year}{2015}.
\newblock \bibinfo{title}{{The search for a subsurface ocean in Ganymede with
  Hubble Space Telescope observations of its auroral ovals}}.
\newblock \bibinfo{journal}{Journal of Geophysical Research: Space Physics}
  \bibinfo{volume}{120}, \bibinfo{pages}{1715--1737}.
\newblock \DOIprefix\doi{10.1002/2014JA020778}.
\bibitem[{Showman et~al.(1997)Showman, Stevenson and Malhotra}]{Showman:1997}
\bibinfo{author}{Showman, A.P.}, \bibinfo{author}{Stevenson, D.J.},
  \bibinfo{author}{Malhotra, R.}, \bibinfo{year}{1997}.
\newblock \bibinfo{title}{{Coupled Orbital and Thermal Evolution of Ganymede}}.
\newblock \bibinfo{journal}{Icarus} \bibinfo{volume}{129},
  \bibinfo{pages}{367--383}.
\newblock \DOIprefix\doi{10.1006/icar.1997.5778}.
\bibitem[{Sohl et~al.(1995)Sohl, Sears and Lorenz}]{Sohl:1995}
\bibinfo{author}{Sohl, F.}, \bibinfo{author}{Sears, W.D.},
  \bibinfo{author}{Lorenz, R.D.}, \bibinfo{year}{1995}.
\newblock \bibinfo{title}{{Tidal dissipation on Titan.}}
\newblock \bibinfo{journal}{Icarus} \bibinfo{volume}{115},
  \bibinfo{pages}{278--294}.
\newblock \DOIprefix\doi{10.1006/icar.1995.1097}.
\bibitem[{Spencer et~al.(2013)Spencer, Howett, Verbiscer, Hurford, Segura and
  Spencer}]{Spencer:2013}
\bibinfo{author}{Spencer, J.R.}, \bibinfo{author}{Howett, C.J.A.},
  \bibinfo{author}{Verbiscer, A.}, \bibinfo{author}{Hurford, T.A.},
  \bibinfo{author}{Segura, M.}, \bibinfo{author}{Spencer, D.C.},
  \bibinfo{year}{2013}.
\newblock \bibinfo{title}{{Enceladus Heat Flow from High Spatial Resolution
  Thermal Emission Observations}}, in: \bibinfo{booktitle}{European Planetary
  Science Congress 2013}, pp. \bibinfo{pages}{EPSC2013--840--1}.
\bibitem[{Spencer et~al.(2006)Spencer, Pearl, Segura, Flasar, Mamoutkine,
  Romani, Buratti, Hendrix, Spilker and Lopes}]{Spencer:2006}
\bibinfo{author}{Spencer, J.R.}, \bibinfo{author}{Pearl, J.C.},
  \bibinfo{author}{Segura, M.}, \bibinfo{author}{Flasar, F.M.},
  \bibinfo{author}{Mamoutkine, A.}, \bibinfo{author}{Romani, P.},
  \bibinfo{author}{Buratti, B.J.}, \bibinfo{author}{Hendrix, A.R.},
  \bibinfo{author}{Spilker, L.J.}, \bibinfo{author}{Lopes, R.M.C.},
  \bibinfo{year}{2006}.
\newblock \bibinfo{title}{{Cassini Encounters Enceladus: Background and the
  Discovery of a South Polar Hot Spot}}.
\newblock \bibinfo{journal}{Science} \bibinfo{volume}{311},
  \bibinfo{pages}{1401--1405}.
\newblock \DOIprefix\doi{10.1126/science.1121661}.
\bibitem[{Spohn and Schubert(2003)}]{Spohn:2003iw}
\bibinfo{author}{Spohn, T.}, \bibinfo{author}{Schubert, G.},
  \bibinfo{year}{2003}.
\newblock \bibinfo{title}{{Oceans in the icy Galilean satellites of Jupiter?}}
\newblock \bibinfo{journal}{Icarus} \bibinfo{volume}{161},
  \bibinfo{pages}{456--467}.
\newblock \DOIprefix\doi{10.1016/S0019-1035(02)00048-9}.
\bibitem[{Takeuchi and Saito(1972)}]{Takeuchi:1972kh}
\bibinfo{author}{Takeuchi, H.}, \bibinfo{author}{Saito, M.},
  \bibinfo{year}{1972}.
\newblock \bibinfo{title}{{Seismic surface waves}}, in: \bibinfo{editor}{Bolt,
  B.A.} (Ed.), \bibinfo{booktitle}{Methods in Computational Physics}.
  \bibinfo{publisher}{Academic Press, New York,}, pp.
  \bibinfo{pages}{217--295}.
\newblock \DOIprefix\doi{10.1016/B978-0-12-460811-5.50010-6}.
\bibitem[{Thomas et~al.(2016)Thomas, Tajeddine, Tiscareno, Burns, Joseph,
  Loredo, Helfenstein and Porco}]{Thomas:2016ex}
\bibinfo{author}{Thomas, P.C.}, \bibinfo{author}{Tajeddine, R.},
  \bibinfo{author}{Tiscareno, M.S.}, \bibinfo{author}{Burns, J.A.},
  \bibinfo{author}{Joseph, J.}, \bibinfo{author}{Loredo, T.J.},
  \bibinfo{author}{Helfenstein, P.}, \bibinfo{author}{Porco, C.},
  \bibinfo{year}{2016}.
\newblock \bibinfo{title}{{Enceladus's measured physical libration requires a
  global subsurface ocean}}.
\newblock \bibinfo{journal}{Icarus} \bibinfo{volume}{264},
  \bibinfo{pages}{37--47}.
\newblock \DOIprefix\doi{10.1016/j.icarus.2015.08.037}.
\bibitem[{Tobie et~al.(2005)Tobie, Grasset, Lunine, Mocquet and
  Sotin}]{Tobie:2005a}
\bibinfo{author}{Tobie, G.}, \bibinfo{author}{Grasset, O.},
  \bibinfo{author}{Lunine, J.I.}, \bibinfo{author}{Mocquet, A.},
  \bibinfo{author}{Sotin, C.}, \bibinfo{year}{2005}.
\newblock \bibinfo{title}{{Titan's internal structure inferred from a coupled
  thermal-orbital model}}.
\newblock \bibinfo{journal}{Icarus} \bibinfo{volume}{175},
  \bibinfo{pages}{496--502}.
\newblock \DOIprefix\doi{10.1016/j.icarus.2004.12.007}.
\bibitem[{Tyler(2014)}]{Tyler:2014gf}
\bibinfo{author}{Tyler, R.}, \bibinfo{year}{2014}.
\newblock \bibinfo{title}{{Comparative estimates of the heat generated by ocean
  tides on icy satellites in the outer Solar System}}.
\newblock \bibinfo{journal}{Icarus} \bibinfo{volume}{243},
  \bibinfo{pages}{358--385}.
\newblock \DOIprefix\doi{10.1016/j.icarus.2014.08.037}.
\bibitem[{Tyler(2008)}]{Tyler:2008}
\bibinfo{author}{Tyler, R.H.}, \bibinfo{year}{2008}.
\newblock \bibinfo{title}{{Strong ocean tidal flow and heating on moons of the
  outer planets}}.
\newblock \bibinfo{journal}{Nature} \bibinfo{volume}{456},
  \bibinfo{pages}{770--772}.
\newblock \DOIprefix\doi{10.1038/nature07571}.
\bibitem[{Tyler(2009)}]{Tyler:2009}
\bibinfo{author}{Tyler, R.H.}, \bibinfo{year}{2009}.
\newblock \bibinfo{title}{{Ocean tides heat Enceladus}}.
\newblock \bibinfo{journal}{Geophys. Res. Lett.} \bibinfo{volume}{36},
  \bibinfo{pages}{L15205}.
\bibitem[{Tyler(2011)}]{Tyler:2011}
\bibinfo{author}{Tyler, R.H.}, \bibinfo{year}{2011}.
\newblock \bibinfo{title}{{Tidal dynamical considerations constrain the state
  of an ocean on Enceladus}}.
\newblock \bibinfo{journal}{Icarus} \bibinfo{volume}{211},
  \bibinfo{pages}{770--779}.
\newblock \DOIprefix\doi{10.1016/j.icarus.2010.10.007}.
\bibitem[{Wahr et~al.(2009)Wahr, Selvans, Mullen, Barr, Collins, Selvans and
  Pappalardo}]{Wahr:2009}
\bibinfo{author}{Wahr, J.}, \bibinfo{author}{Selvans, Z.A.},
  \bibinfo{author}{Mullen, M.E.}, \bibinfo{author}{Barr, A.C.},
  \bibinfo{author}{Collins, G.C.}, \bibinfo{author}{Selvans, M.M.},
  \bibinfo{author}{Pappalardo, R.T.}, \bibinfo{year}{2009}.
\newblock \bibinfo{title}{{Modeling stresses on satellites due to
  nonsynchronous rotation and orbital eccentricity using gravitational
  potential theory}}.
\newblock \bibinfo{journal}{Icarus} \bibinfo{volume}{200},
  \bibinfo{pages}{188--206}.
\newblock \DOIprefix\doi{10.1016/j.icarus.2008.11.002}.
\bibitem[{Wahr et~al.(2006)Wahr, Zuber, Smith and Lunine}]{Wahr:2006}
\bibinfo{author}{Wahr, J.M.}, \bibinfo{author}{Zuber, M.T.},
  \bibinfo{author}{Smith, D.E.}, \bibinfo{author}{Lunine, J.I.},
  \bibinfo{year}{2006}.
\newblock \bibinfo{title}{{Tides on Europa, and the thickness of Europa's icy
  shell}}.
\newblock \bibinfo{journal}{J. Geophys. Res.} \bibinfo{volume}{111},
  \bibinfo{pages}{E12005}.
\newblock \DOIprefix\doi{10.1029/2006JE002729}.
\bibitem[{Webb(1980)}]{Webb:1980}
\bibinfo{author}{Webb, D.J.}, \bibinfo{year}{1980}.
\newblock \bibinfo{title}{{Tides and tidal friction in a hemispherical ocean
  centred at the equator}}.
\newblock \bibinfo{journal}{Geophysical Journal International}
  \bibinfo{volume}{61}, \bibinfo{pages}{573--600}.
\newblock \DOIprefix\doi{10.1111/j.1365-246X.1980.tb04833.x}.
\bibitem[{Zimmer et~al.(2000)Zimmer, Khurana and Kivelson}]{Zimmer:2000}
\bibinfo{author}{Zimmer, C.}, \bibinfo{author}{Khurana, K.K.},
  \bibinfo{author}{Kivelson, M.G.}, \bibinfo{year}{2000}.
\newblock \bibinfo{title}{{Subsurface Oceans on Europa and Callisto:
  Constraints from Galileo Magnetometer Observations}}.
\newblock \bibinfo{journal}{Icarus} \bibinfo{volume}{147},
  \bibinfo{pages}{329--347}.
\newblock \DOIprefix\doi{10.1006/icar.2000.6456}.
\bibitem[{Zschau(1978)}]{Zschau:1978jp}
\bibinfo{author}{Zschau, J.}, \bibinfo{year}{1978}.
\newblock \bibinfo{title}{{Tidal Friction in the Solid Earth: Loading Tides
  Versus Body Tides}}, in: \bibinfo{booktitle}{Tidal Friction and the
  Earth{\textquoteright}s Rotation}. \bibinfo{publisher}{Springer, Berlin,
  Heidelberg}, \bibinfo{address}{Berlin, Heidelberg}, pp.
  \bibinfo{pages}{62--94}.
\newblock \DOIprefix\doi{10.1007/978-3-662-40203-0_7}.

\end{thebibliography}

\end{document}